\documentclass[aps,eqsecnum]{revtex4}
\usepackage{epsfig}
\usepackage{latexsym}
\usepackage{bm}

\begin{document}
\thispagestyle{empty}

%
%
\leftline{\epsfbox{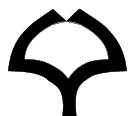}}
\vspace{-10.0mm}
{\baselineskip-4pt
\font\yitp=cmmib10 scaled\magstep2
\font\elevenmib=cmmib10 scaled\magstep1  \skewchar\elevenmib='177
\leftline{\baselineskip20pt
\hspace{12mm} 
\vbox to0pt
   { {\yitp\hbox{Osaka \hspace{1.5mm} University} }
     {\large\sl\hbox{{Theoretical Astrophysics}} }\vss}}

%
%
{\baselineskip0pt
\rightline{\large\baselineskip14pt\rm\vbox
        to20pt{\hbox{OU-TAP-217}
               \hbox{OCU-PHYS-203}
               \hbox{AP-GR-12}
               \hbox{KUNS-1864}
               \hbox{YITP-03-52}
\vss}}
}
\vskip15mm

\begin{center}

{\Large\bf Gauge Problem in the Gravitational Self-Force}

\vspace{3mm}

{\large \it - First Post Newtonian
Force under Regge-Wheeler Gauge -}

\bigskip

Hiroyuki Nakano$^1$, Norichika Sago$^{2,3}$ and Misao Sasaki$^4$

\smallskip

\smallskip
$^1${\em Department of Mathematics and Physics,~Graduate School of 
 Science,~Osaka City University,\\ Osaka 558-8585, Japan
}

\smallskip
$^2${\em Department of Earth and Space Science,~Graduate School of
 Science,~Osaka University,\\ Osaka 560-0043, Japan
}

\smallskip
$^3${\em Department of Physics,~Graduate School of Science,
Kyoto University,\\ Kyoto 606-8502,~Japan}\\

\smallskip
$^4${\em Yukawa Institute for Theoretical Physics,
Kyoto University,\\ Kyoto 606-8502, Japan}\\

\smallskip

\medskip

\today

\end{center}

\bigskip

We discuss the gravitational self-force on a particle in a black hole
space-time. For a point particle, the full (bare) self-force diverges.
It is known that the metric perturbation induced by a particle can be
divided into two parts, the direct part (or the S part) and the tail part
(or the R part), in the harmonic gauge, and the regularized self-force
is derived from the R part which is regular and satisfies the source-free
perturbed Einstein equations. In this paper, we consider a gauge
transformation from the harmonic gauge to the Regge-Wheeler gauge
in which the full metric perturbation can be calculated, and present
a method to derive the regularized self-force for a particle in circular
 orbit around a Schwarzschild black hole in the Regge-Wheeler gauge.
As a first application of this method, we then calculate the self-force
to first post-Newtonian order. We find the correction to the total
mass of the system due to the presence of the particle is correctly
reproduced in the force at the Newtonian order.

\section{Introduction}

Thanks to recent advances in technology, an era of gravitational
wave astronomy has arrived. There are already several large-scale
laser interferometric gravitational wave detectors that are in operation
in the world. Among them are LIGO \cite{LIGO}, GEO-600 \cite{GEO}
and TAMA300 \cite{TAMA}. VIRGO \cite{VIRGO} is expected to start
its operation soon. The primary targets for these ground-based detectors
are inspiralling compact binaries, and they are expected to be detected
in the near future.

On the other hand, there is a future space-based interferometric
detector project LISA \cite{LISA} that can detect gravitational waves
from solar-mass compact objects orbiting supermassive black holes.
There is also a future plan called DECIGO \cite{DECIGO}.
To extract out physical information of such binary systems from detected
gravitational wave signals, it is essential to know the
theoretical gravitational waveforms accurately.
The black hole perturbation approach is most suited for
this purpose. In this approach, one considers gravitational waves
emitted by a point particle that represents a compact
object orbiting a black hole, assuming the mass of the particle ($\mu$)
is much less than that of the black hole ($M$); $\mu\ll M$.

In the lowest order in the mass ratio $(\mu/M)^0$,
the orbit of the particle can be represented
a geodesic on the background geometry of a black hole.
Already in this lowest order,
by combining with the assumption of adiabatic orbital evolution,
this approach has been proved to be very powerful for
evaluating general relativistic corrections to the gravitational
waveforms, even for neutron star-neutron star
(NS-NS) binaries \cite{MSSTT}.

In the next order, the orbit deviates from the geodesic on
the black hole background because the spacetime is perturbed by
the particle.
We can interpret this deviation as the effect of the self-force
on the particle itself. Since it is essential to take account
of this deviation to predict the orbital evolution accurately,
we have to derive the equation of motion that includes
the self-force on the particle. The self-force is formally given
by the tail part (or the R part by Detweiler and
 Whiting~\cite{Detweiler:2002mi})
of the metric perturbation which is regular at the location of
the particle.

The gravitational self-force is, however, not easily obtainable.
There are two main reasons. First, the full (bare) metric perturbation
due to a point particle diverges at the location of the particle,
hence so does the self-force. As mentioned above, one has to identify
the R part of the metric perturbation to obtain a meaningful self-force.
However, the R part cannot be determined locally but depends on the
whole history of the particle. Therefore, one usually identifies
the divergent part which can be evaluated locally (called the S part)
to a necessary order and subtract it from the full metric perturbation.
This identification of the S part is sometimes called the subtraction
problem. Second, the regularized self-force is formally defined
only in the harmonic gauge because the form of the S part is
known only in the harmonic gauge, whereas the metric perturbation of a
black hole geometry can be calculated only in the ingoing or
outgoing radiation gauge in the Kerr background, or in the Regge-Wheeler
gauge in the Schwarzschild background. Hence, one has to find a
gauge transformation to express the full metric perturbation
and the divergent part in the same gauge.
This is called the gauge problem.

In this paper, as a first step toward a complete derivation
of the gravitational self-force, we consider a particle
orbiting a Schwarzschild black hole, and propose a method to
calculate the regularized self-force by solving the subtraction
and gauge problems simultaneously. Namely, we develop a method
to regularize the self-force in the Regge-Wheeler gauge.
The regularization is done by the ``mode decomposition
regularization'' \cite{MNS1}, which is effectively the same
in the present case as the ``mode-sum regularization'' developed
in \cite{Barack:2001gx,Barack:2002mh,Barack:2002bt}.

Recently, Barack and Ori~\cite{Barack:2001ph} proposed what they call
the intermediate gauge approach to the gauge problem.
Applying this method, the gravitational self-force for an orbit
plunging straight into a Schwarzschild black hole was calculated
by Barack and Lousto \cite{Barack:2002ku}.
It is noted that, although their approach is philosophically quite
different from our present approach, practically both approaches
turn out to give the same result as far as the
Regge-Wheeler gauge calculations are concerned.

As for the case of the Kerr background, the only known gauge in
 which the metric perturbation can be
evaluated is the radiation gauge formulated by
Chrzanowski \cite{Chrzanowski:wv}.
However, the Chrzanowski construction of the
metric perturbation becomes ill-defined in the
neighborhood of the particle, i.e., the Einstein equations are
not satisfied there \cite{Barack:2001ph}.
Some progress was made by Ori \cite{Ori:2002uv}
to obtain the correct, full metric perturbation in the Kerr
background. The regularization parameters in the mode-sum
regularization for Kerr case are calculated by
 Barack and Ori \cite{Barack:2002mh2}.

The paper is organized as follows.
In Sec.~\ref{sec:gauge}, we briefly review the situation of
the self-force problem and explain our strategy.
In Sec.~\ref{sec:presc}, we give the regularization prescription
under the Regge-Wheeler gauge condition.
In Sec.~\ref{sec:full}, we calculate the full metric perturbation
and the full force in the Regge-Wheeler gauge with the
Regge-Wheeler-Zerilli formalism.
In Sec.~\ref{sec:dir}, we evaluate the singular, divergent part
in the harmonic gauge by local analysis at the particle location
and expand it in the Fourier-harmonic form.
In Sec.~\ref{sec:trans}, we calculate the S part under the
Regge-Wheeler gauge condition by using
the gauge transformation.
By subtracting this S part from
the full force evaluated in Sec.~\ref{sec:full},
we obtain the regularized gravitational self-force
in Sec.~\ref{sec:result}.
Finally, we summarize our calculation and discuss the future work
in Sec.~\ref{sec:conclusion}.
Some details of the calculations as well as discussions on
the $\ell=0$ and $1$ modes are given in 
Appendices~\ref{app:mano} $\sim$ \ref{app:M}.

\section{Gauge problem}\label{sec:gauge}

We consider the linearized metric perturbation
\begin{eqnarray}
h_{\mu\nu} =
\tilde g_{\mu\nu}-g_{\mu\nu} \,,
\end{eqnarray}
where $g_{\mu\nu}$ and $\tilde g_{\mu\nu}$ is the background
and the perturbed metric, respectively.
Here we define the force due to the metric
perturbation as the part that gives rise to a deviation from the background
geodesic:
\begin{eqnarray}
\frac{d^2 z^{\alpha}}{d\tau^2}
+\Gamma_{\mu\nu}^{\alpha}\frac{dz^{\mu}}{d\tau}\frac{dz^{\nu}}{d\tau}
={1 \over \mu}\,F^{\alpha}[h] \,,
\end{eqnarray}
where $z^{\alpha}(\tau)$ is an orbit of the particle
parametrized by the background proper time
 (i.e., $g_{\mu\nu} (dz^\mu/d\tau)(dz^\nu/d\tau)=-1$).
 From the geodesic equation on $\tilde g_{\mu\nu}$,
we obtain
\begin{eqnarray}
F^{\alpha}[h] =
-\mu P_{\beta}^{\alpha}
(\bar{h}_{\beta\gamma;\delta}
-\frac{1}{2}g_{\beta\gamma}
 {\bar{h}^{\epsilon}}{}_{\epsilon;\delta}
-\frac{1}{2}\bar{h}_{\gamma\delta;\beta}
+\frac{1}{4}g_{\gamma\delta}
{\bar{h}^{\epsilon}}{}_{\epsilon;\beta}
)u^{\gamma}u^{\delta} \,,
\label{eq:formalF}
\end{eqnarray}
where ${P_{\alpha}}^{\beta}={\delta_{\alpha}}^{\beta}+u_{\alpha}u^{\beta}$,
$\bar{h}_{\alpha\beta}
=h_{\alpha\beta}-\frac{1}{2}g_{\alpha\beta}h_{\mu}{}^{\mu}$
and $u^{\alpha}=dz^\alpha/d\tau$.

In the case that the perturbation is produced by a point particle,
however, we face the problem that $h_{\mu\nu}$ diverges at the
location of the particle, and so does the force.
Therefore, we cannot naively apply the above calculation to obtain the
self-force of the particle.
Mino, Sasaki and Tanaka \cite{reaction} and Quinn and Wald \cite{QuiWal}
gave a formal answer to this problem by considering the metric perturbation
in the harmonic gauge. According to them, the metric perturbation in
the vicinity of the orbit can be divided into two parts: the direct part
and the tail part.
The direct part has support only on the past null cone of the field
point $x^{\mu}$ and diverges in the limit
$x^{\mu}\to z^{\mu}(\tau)$.
The tail part has support inside the past null cone and
gives the physical self-force  which is regular
at the location of the particle. But it is almost impossible
to calculate the tail part of the metric perturbation directly,
because it depends on the global structure of the space-time
as well as on the history of the particle motion.
In contrast, the direct part can be evaluated locally in terms of
geometrical quantities. Hence, instead of directly calculating the
tail part, we consider the subtraction of the direct part from the
full metric perturbation, where the latter can be calculated in
principle by the Regge-Wheeler-Zerilli or Teukolsky formalism
for black hole
perturbations~\cite{RegWhe,Zer,Teukolsky:1973ha,Sago:2002fe,Jhingan:2002kb}.

 From the fact that $F^{\alpha}$ is a linear differential operator
on $h_{\mu\nu}$ (with a suitable extension of $u^\mu$ off the particle
trajectory), we can calculate the self-force by subtracting
the direct part from the full force under the harmonic gauge as
\begin{eqnarray}
\lim_{x\to z(\tau)}F_{\alpha}[h^{\rm tail,H}(x)] =
\lim_{x\to z(\tau)}
\left(F_{\alpha}[h^{{\rm full,H}}(x)]
-F_{\alpha}[h^{{\rm dir,H}}(x)]\right) \,,
\label{eq:subt}
\end{eqnarray}
where the superscript H stands for the harmonic gauge.
When we perform this subtraction, the full metric perturbation
and the direct part must be evaluated in the harmonic gauge
because this division is meaningful only in this gauge.

However, it is difficult to obtain the full metric perturbation
directly in the harmonic gauge. In order to overcome this difficulty,
one possibility is to perform the gauge transformation
to the harmonic gauge from the gauge in which the full metric
perturbation is obtained.
In our previous paper~\cite{Sago:2002fe},
we investigated this problem for the
Schwarzschild case, namely, we formulated a method
to perform the gauge transformation from the Regge-Wheeler (RW) gauge
to the harmonic gauge.
We expressed the gauge transformation equations in the Fourier-harmonic
expanded form and derived a set of decoupled equations for the
coefficients of each mode.
Applications of this method are now under study.

Recently, Detweiler and Whiting found a slight but important
modification of the above division of the metric
 perturbation~\cite{Detweiler:2002mi}.
The new direct part, called the S part, $h^{\rm S,H}_{\mu\nu}$,
is constructed to be an inhomogeneous
solution of the linearized Einstein equations (in the harmonic gauge)
as
\begin{eqnarray}
\bar{h}^{{\rm full/S,H}}_{\mu\nu;\alpha}{}^{;\alpha}
+2R_{\mu}{}^{\alpha}{}_{\nu}{}^{\beta}\bar{h}^{{\rm full/S,H}}_{\alpha\beta}
=-16\pi T_{\mu\nu} \,.
\end{eqnarray}
The new tail part, called the R part, $h^{\rm R,H}_{\mu\nu}$,
is then a homogeneous solution.
Since the S and R parts are both the solutions of the
Einstein equations, we can define the S and R parts in another
gauge, which are also the solutions of the Einstein equations,
by performing the gauge transformation of each part.
Therefore, we can consider the subtraction procedure
under some other convenient gauge by transforming
the S part from the harmonic gauge to the desired gauge.
Thus, another, perhaps more promising possibility is to
formulate a method to derive the S part in the Regge-Wheeler or radiation
gauge, where we have formalisms to evaluate the full metric perturbation,
and to obtain the regularized self-force by subtracting
the S part in this gauge. In this paper, we focus on the
Schwarzschild case and consider the subtraction in the
Regge-Wheeler gauge.

To subtract the S part, we adopt the mode decomposition
regularization \cite{MNS1}.
In this method, the subtraction procedure (\ref{eq:subt}) is done
at each harmonic mode.
The full force is obtained in the form of the Fourier-harmonic
expansion. The Fourier (frequency) integral can be easily done
in the case of circular orbits.
On the other hand, the S part is known only in the vicinity
of the particle. Hence, one has to extend it over the sphere
to obtain its harmonic coefficients. This procedure introduces
some ambiguity in the harmonic expansion of the S part.
In particular, each harmonic mode obtained by this
extension has no physical significance by itself. The physical
significance is recovered only after we sum over all the modes.
Because of this ambiguity, we have to treat the
$\ell=0$ and $1$ modes with special care, as will be
shown later.

\section{Self-force in the Regge-Wheeler gauge} \label{sec:presc}

The Schwarzschild metric is given in the standard Schwarzschild
coordinates as
\begin{equation}
g_{\mu\nu} dx^{\mu}dx^{\nu} =
-f(r)dt^2+f(r)^{-1}dr^2+r^2(d\theta^2+\sin^2\theta d\phi^2) \,,
\quad \quad f(r) = 1-\frac{2M}{r} \,.
\end{equation}
We denote the location of the particle at its proper time $\tau=\tau_0$
as
\begin{eqnarray}
\{z^\alpha_{0}\} = \{z^\alpha(\tau_{0})\} = \{t_0,r_0,\theta_0,\phi_0\}\,.
\label{eq:ploc}
\end{eqnarray}

Formally, the gravitational self-force
acting on the particle is given by the tail part
in the harmonic gauge, as expressed in the left-hand
side of Eq.~(\ref{eq:subt}).
Using the notions of the S and R parts introduced by
Detweiler and Whiting~\cite{Detweiler:2002mi},
it may be rewritten as
\begin{eqnarray}
F_\alpha^{\rm H}(\tau) &=& \lim_{x \to z(\tau)}
F_\alpha \left[h^{\rm R,H}_{\mu\nu} \right](x)
\nonumber \\
&=&
\lim_{x \to z(\tau)}
F_\alpha \left[h^{\rm full,H}_{\mu\nu}-h^{\rm S,H}_{\mu\nu} \right](x)
\nonumber \\
&=&
\lim_{x \to z(\tau)}
\left(F_\alpha \left[h^{\rm full,H}_{\mu\nu} \right](x)
-F_\alpha \left[h^{\rm S,H}_{\mu\nu} \right](x)
\right)
\,, \label{eq:force2}
\end{eqnarray}
where
$h^{\rm S,H}_{\mu\nu}$ and $h^{\rm R,H}_{\mu\nu}$ denote the S
and R parts, respectively, of the metric perturbation
in the harmonic gauge. The S part can be calculated by the local
coordinate expansion~\cite{MNS1}.

Now, we consider the gauge transformation from the harmonic
gauge to the RW gauge defined by
\begin{eqnarray}
x^{{\rm H}}_{\mu} &\to&
x^{{\rm RW}}_{\mu} = x^{{\rm H}}_{\mu}
+ \xi_{\mu}^{{\rm H} \to {\rm RW}} \,,
\label{displace} \\
h_{\mu\nu}^{{\rm H}} &\to&
h_{\mu\nu}^{{\rm RW}} =
h_{\mu\nu}^{{\rm H}} - 2\,
\nabla\kern1.3ex\xi^{{\rm H}\to{\rm RW}}\kern-9.4ex{}_{(\mu~~\nu)}\,,
\label{gauge-trans}
\end{eqnarray}
where $\xi_{\mu}^{{\rm H} \to {\rm RW}}$
is the generator of the gauge transformation.
Then the self-force in the RW gauge is
given by
\begin{eqnarray}
F_\alpha^{\rm RW}(\tau)
&=&\lim_{x \to z(\tau)}
F_\alpha\left[h^{\rm R,RW}\right]
=\lim_{x \to z(\tau)}
F_\alpha \left[h^{\rm R,H} - 2\,\nabla\xi^{{\rm H} \to {\rm RW}}
\left[h^{\rm R,H}\right] \right](x)
\nonumber\\
&=&
\lim_{x \to z(\tau)}
F_\alpha \left[h^{\rm full,H}-h^{\rm S,H}
- 2\,\nabla\xi^{{\rm H} \to {\rm RW}}
\left[h^{\rm full,H}-h^{\rm S,H}\right]
\right](x)
\nonumber \\
&=&
\lim_{x \to z(\tau)}
F_\alpha \left[h^{\rm full,H}
- 2\,\nabla\xi^{{\rm H} \to {\rm RW}}
\left[h^{\rm full,H}\right]
-h^{\rm S,H}
+ 2\,\nabla\xi^{{\rm H} \to {\rm RW}}
\left[h^{\rm S,H}\right]
\right](x)
\nonumber \\
&=&
\lim_{x \to z(\tau)}
\Bigl(
F_\alpha \left[h^{\rm full,RW}\right](x)
-F_\alpha \left[h^{\rm S,H}-2\,\nabla\xi^{{\rm H} \to {\rm RW}}
\left[h^{\rm S,H}\right]
\right](x)
\Bigr)\,,
\label{eq:RWsf}
\end{eqnarray}
where we have omitted the spacetime indices of $h_{\mu\nu}$
and $\nabla_{(\mu}\xi_{\nu)}$ for notational simplicity.
The full metric perturbation $h^{\rm full,RW}_{\mu\nu}$
can be calculated by using the Regge-Wheeler-Zerilli formalism,
while the S part $h^{\rm S,H}_{\mu\nu}$ can be obtained with
sufficient accuracy by the local analysis near the particle
location. Thus the remaining issue is if we can
unambiguously determine the gauge transformation
\begin{eqnarray}
\xi_\alpha^{{\rm S},{\rm H} \to {\rm RW}}
=\xi_\alpha^{{\rm H} \to {\rm RW}}\left[h_{\mu\nu}^{\rm S,H}\right]\,.
\label{eq:Sgtrans}
\end{eqnarray}
Note that the self-force (\ref{eq:RWsf}) is almost identical
to the expression obtained in the intermediate
gauge approach~\cite{Barack:2001ph}, if we replace
the S and R parts by the direct and tail parts,
respectively. The only difference is that the S and R parts are
now solutions of the inhomogeneous and homogeneous Einstein
equations, respectively. Hence the S part in the
RW gauge is (at least formally) well-defined provided that
the gauge transformation of the S part, Eq.~(\ref{eq:Sgtrans})
is unique. As will be shown later in
Eqs.~(\ref{eq:GGT4}),
this turns out to be indeed the case.
Therefore one may identify
the self-force (\ref{eq:RWsf}) to be actually the one evaluated
in the RW gauge~\cite{Minopc}, not in some intermediate gauge.

\section{Full Metric Perturbation and its Force}\label{sec:full}

In this section, we consider the full metric perturbation
and its self-force in the case of a circular orbit.
First, the metric perturbation is calculated
by the Regge-Wheeler-Zerilli formalism in which a Fourier-harmonic expansion
is used because of the symmetry of the background spacetime.
Next, we derive the self-force by acting force operators
and represent it in terms of $\ell$ mode coefficients
after summing over $\omega$ and $m$ for
the Fourier-harmonic series.

\subsection{Regge-Wheeler-Zerilli formalism}

On the Schwarzschild background,
the metric perturbation $h_{\mu\nu}$ can be expanded in terms of
tensor harmonics as
\begin{eqnarray}
\bm{h} &=& \sum_{\ell m} \left[
f(r)H_{0\ell m}(t,r)\bm{a}^{(0)}_{\ell m}
-i\sqrt{2}H_{1\ell m}(t,r)\bm{a}^{(1)}_{\ell m}
+\frac{1}{f(r)}H_{2\ell m}(t,r)\bm{a}_{\ell m}
\right. \nonumber \\
& &
-\frac{i}{r}\sqrt{2\ell(\ell+1)}h^{(e)}_{0\ell m}(t,r)\bm{b}^{(0)}_{\ell m}
+\frac{1}{r}\sqrt{2\ell(\ell+1)}h^{(e)}_{1\ell m}(t,r)\bm{b}_{\ell m}
\nonumber \\
& &
+\sqrt{\frac{1}{2}\ell(\ell+1)(\ell-1)(\ell+2)}G_{\ell m}(t,r)\bm{f}_{\ell m}
+\left(\sqrt{2}K_{\ell m}(t,r)
 -\frac{\ell(\ell+1)}{{\sqrt{2}}}G_{\ell m}(t,r)\right)\bm{g}_{\ell m}
\nonumber \\
& &
\left.
-\frac{\sqrt{2\ell(\ell+1)}}{r}h_{0\ell m}(t,r)\bm{c}^{(0)}_{\ell m}
+\frac{i\sqrt{2\ell(\ell+1)}}{r}h_{1\ell m}(t,r)\bm{c}_{\ell m}
+\frac{\sqrt{2\ell(\ell+1)(\ell-1)(\ell+2)}}{2r^2}h_{2\ell m}(t,r)\bm{d}_{\ell m}
\right] \,, \label{eq:hharm}
\end{eqnarray}
where $\bm{a}_{\ell m}^{(0)}$, $\bm{a}_{\ell m}~\cdots$ are the
tensor harmonics introduced by Zerilli~\cite{Zer}.
The energy-momentum tensor of a point particle takes the form,
\begin{eqnarray}
T^{\mu\nu}&=& \mu \int^{+\infty}_{-\infty} \delta^{(4)}(x-z(\tau))
{dz^{\mu} \over d\tau}{dz^{\nu} \over d\tau}d\tau \nonumber \\
&=& \mu {1 \over u^t}u^{\mu}u^{\nu}
{\delta(r-r_0(t)) \over r^2} \delta ^{(2)}
({\bf\Omega}-{\bf\Omega}_0(t)) \,,
\end{eqnarray}
where the orbit has been expressed as
\begin{eqnarray}
x^{\mu}=z^{\mu}(\tau)=\{t_0(\tau),r_0(\tau),\theta_0(\tau),\phi_0(\tau)\}\,,
\end{eqnarray}
with $\tau$ being regarded as a function of time determined
by $t=T(\tau)$.
The RW gauge is defined by the conditions on the
metric perturbation as
\begin{eqnarray}
h_2^{\rm RW}
=h_{0}^{(e){\rm RW}}=h_{1}^{(e){\rm RW}}=G^{\rm RW}=0\,.
\label{eq:RWcond}
\end{eqnarray}
The Regge-Wheeler and Zerilli equations are obtained
by plugging the metric perturbation (\ref{eq:hharm}) in
the linearized Einstein equations and Fourier
decomposing them. (Recently,
the Regge-Wheeler-Zerilli formalism is
improved by Jhingan and Tanaka~\cite{Jhingan:2002kb}.)

For odd parity waves that are defined by the parity $(-1)^{\ell+1}$
under the transformation $(\theta,\phi)\to(\pi-\theta,\phi+\pi)$,
we introduce a new radial function
$R_{\ell m \omega}^{{\rm (odd)}}(r)$
in terms of which the two radial functions of the metric perturbation
are expressed as
\begin{eqnarray}
h_{1\ell m \omega}^{\rm RW}
&=&{r^2  \over (r-2M)}R_{\ell m \omega}^{{\rm (odd)}} \,,
\cr
h_{0\ell m \omega}^{\rm RW}
&=&{i \over \omega}{d \over dr^*}(r R_{\ell m \omega}^{{\rm (odd)}})
-{8 \pi r(r-2M) \over
\omega [{1\over2}\ell(\ell+1)(\ell-1)(\ell+2)]^{1/2}}D_{\ell m\omega} \,.
\label{eq:h0}
\end{eqnarray}
The new radial function $R_{\ell m \omega}^{{\rm (odd)}}(r)$
 satisfies the Regge-Wheeler equation,
\begin{eqnarray}
{d^2 \over dr^{*2}} R_{\ell m \omega}^{{\rm (odd)}}
&+&[\omega ^2 -V_{\ell}(r)]
R_{\ell m \omega}^{{\rm (odd)}}
\nonumber \\
&=&{8\pi i \over [{1\over2}\ell(\ell+1)(\ell-1)(\ell+2)]^{1/2}}{r-2M \over r^2}
\nonumber \\ &&
\times \left(-r^2{d \over dr}[(1-{2M \over r})D_{\ell m \omega}]
+(r-2M)[(\ell-1)(\ell+2)]^{1/2}Q_{\ell m \omega} \right) \,,
\end{eqnarray}
where
$r^*=r+2M \log (r/2M-1)$, and the potential $V_{\ell}$ is
given by
\begin{eqnarray}
V_{\ell}(r)
=\left( 1-{2M \over r} \right)
\left( {\ell(\ell+1) \over r^2}-{6M \over r^3} \right) \,.
\end{eqnarray}
The source term $Q_{\ell m \omega}$ vanishes
in the case of a circular orbit and
\begin{eqnarray}
D_{\ell m\omega}(r) =
\left[{1\over 2}\ell(\ell+1)(\ell-1)(\ell+2)\right]^{-1/2}
\,\mu\,{(u^{\phi})^2\over u^t}\,\delta(r-r_0)\,m\,
\partial_{\theta}\,Y_{\ell m}^*(\theta_0,\,\phi_0) \,\,,
\end{eqnarray}
where the orbit is given by
\begin{eqnarray}
z^\alpha(\tau)=
\left\{u^t \tau ,\, r_0 ,\, {\pi\over 2} ,\, u^\phi \tau \right\}
\,;\quad
u^t = \sqrt{r_0 \over r_0-3M} \,, \quad
u^\phi ~=~ {1\over r_0}\sqrt{M\over r_0-3M}=\Omega u^t \,,
\label{eq:orbit}
\end{eqnarray}
where $\Omega=\sqrt{M/r_0^3}$ is the orbital frequency.
The orbit is assumed to be on the equatorial plane
without loss of generality.

For even parity waves with the parity $(-1)^{\ell}$,
we introduce a new radial function $R_{\ell m \omega}^{{\rm (Z)}}(r)$
in terms of which the four radial functions
of the metric perturbation are expressed as
\begin{eqnarray}
K_{\ell m \omega}^{\rm RW}
&=&{\lambda (\lambda +1)r^2+3\lambda Mr+6M^2 \over r^2(\lambda r+3M)}
R_{\ell m \omega}^{{\rm (Z)}}
+{r-2M \over r}{d  \over dr} R_{\ell m \omega}^{{\rm (Z)}}
\cr
&& -{r(r-2M) \over \lambda r+3M}\tilde C_{1\ell m \omega}
+{i (r-2M)^2 \over r(\lambda r+3M)}\tilde C_{2\ell m \omega} \,,
\cr
H_{1\ell m \omega}^{\rm RW}&=&-i \omega
{\lambda r^2-3\lambda Mr-3M^2 \over
(r-2M)(\lambda r+3M)}R_{\ell m \omega}^{{\rm (Z)}}
-i \omega r{d \over dr} R_{\ell m \omega}^{{\rm (Z)}} 
\nonumber \\
&& +{i \omega r^3 \over \lambda r+3M}\tilde C_{1\ell m \omega}
+{\omega r (r-2M) \over r(\lambda r+3M)}\tilde C_{2\ell m \omega} \,,
\cr
H_{0\ell m \omega}^{\rm RW}
&=&{\lambda r(r-2M)-\omega ^2 r^4 +M(r-3M) \over (r-2M)(\lambda r+3M)}
K_{\ell m \omega}^{\rm RW}
+{M(\lambda +1)-\omega ^2 r^3 \over i \omega r(\lambda r+3M)}
H_{1\ell m \omega}^{\rm RW}
+\tilde B_{\ell m \omega} \,, \cr
H_{2\ell m \omega}^{\rm RW}&=&H_{0\ell m \omega}^{\rm RW}
-16 \pi r^2 [{1\over2}\ell(\ell+1)(\ell-1)(\ell+2)]^{-1/2}F_{\ell m \omega}
\,, \label{eq:H2}
\end{eqnarray}
where
\begin{eqnarray}
\lambda={1 \over 2}(\ell-1)(\ell+2) \,,
\end{eqnarray}
and the source terms are given by
\begin{eqnarray}
\tilde B_{\ell m \omega}&=&{8\pi r^2(r-2M) \over \lambda r+3M}\{A_{\ell m \omega}
+[{1\over2}\ell(\ell+1)]^{-1/2}B_{\ell m \omega}\}
-{4\pi \sqrt{2} \over \lambda r +3M}
{Mr \over \omega}A_{\ell m \omega}^{(1)} \,, \cr
\tilde C_{1\ell m \omega}&=&{8\pi \over \sqrt{2}\omega}A_{\ell m \omega}^{(1)}
+{1 \over r}\tilde B_{\ell m \omega}
-16\pi r [{1 \over 2}\ell(\ell+1)(\ell-1)(\ell+2)]^{-1/2}F_{\ell m \omega} \,,
\cr
\tilde C_{2\ell m \omega}&=&-{8\pi r^2 \over i \omega}
{[{1\over2}l(l+1)]^{-1/2} \over r-2M}B_{\ell m \omega}^{(0)}
-{ir \over r-2M}\tilde B_{\ell m \omega}
+{16\pi ir^3 \over r-2M}
[{1 \over 2}\ell(\ell+1)(\ell-1)(\ell+2)]^{-1/2}F_{\ell m \omega}
\,.
\end{eqnarray}
Here the harmonic coefficients of the source terms $A_{\ell m \omega}$, 
$A_{\ell m \omega}^{(1)}$ and
$B_{\ell m \omega}$ vanish in the circular case and 
\begin{eqnarray}
B_{\ell m \omega}^{(0)}&=&
\left[{\ell(\ell+1)\over 2}\right]^{-1/2}
\mu \,u^{\phi} \left(1-{2M\over r}\right){1\over r}
\delta(r-r_0)\,m\,Y_{\ell m}^*(\theta_0,\,\phi_0)
\,,
\cr
F_{\ell m \omega}&=&
{1\over 2}\left[{\ell(\ell+1)(\ell-1)(\ell+2)\over 2}\right]^{-1/2}
\mu \,{(u^{\phi})^2  \over u^t}
\delta(r-r_0)\,(\ell(\ell+1)-2m^2)\,Y_{\ell m}^*(\theta_0,\,\phi_0)
\,.
\end{eqnarray}

The new radial function $R_{\ell m \omega}^{{\rm (Z)}}(r)$
obeys the Zerilli equation,
\begin{eqnarray}
{d^2 \over dr^{*2}} R_{\ell m \omega}^{{\rm (Z)}}
+[\omega ^2 -V_{\ell}^{{\rm (Z)}}(r)]R_{\ell m \omega}^{{\rm (Z)}}
=S_{\ell m \omega}^{{\rm (Z)}} \,,
\label{eq:elederi}
\end{eqnarray}
where
\begin{eqnarray}
V_{\ell}^{{\rm (Z)}}(r)=\left( 1-{2M \over r} \right)
{2\lambda ^2(\lambda +1)r^3+6\lambda ^2 Mr^2+18\lambda M^2r+18M^3
\over r^3(\lambda r+3M)^2} \,,
\end{eqnarray}
and
\begin{eqnarray}
S_{\ell m \omega}^{{\rm (Z)}}&=&-i{r-2M \over r}{d \over dr}
\left[{(r-2M)^2 \over r(\lambda r+3M)}\left({ir^2 \over r-2M}\tilde C_{1\ell m \omega}
+\tilde C_{2\ell m \omega} \right)\right]
\nonumber \\ &&
+i{(r-2M)^2 \over r(\lambda r+3M)^2}
\left[{\lambda (\lambda +1)r^2+3\lambda Mr+6M^2 \over r^2}\tilde C_{2\ell m \omega}
+i{\lambda r^2-3\lambda Mr-3M^2 \over (r-2M)}\tilde C_{1\ell m \omega}\right] \,.
\end{eqnarray}
The Zerilli equation can be transformed to the Regge-Wheeler equation
by the Chandrasekhar transformation if desired,
as shown in Appendix~\ref{app:mano}. 
However, here, we treat the original Zerilli equation.

\subsection{Full Metric Perturbation}

The homogeneous solutions of the Regge-Wheeler equation are discussed
in detail by Mano et al.~\cite{Mano2} and in Appendix \ref{app:mano}.
By constructing the retarded Green function from the homogeneous
solutions with appropriate boundary conditions, namely, the two
independent solutions with the in-going and up-going wave boundary
conditions, we can solve the Regge-Wheeler and Zerilli equations to
obtain the full metric perturbation in the RW gauge.
Here, we consider the radial functions up to the first post-Newtonian
(1PN) order.

The radial function for the odd part of the metric perturbation
metric perturbation is obtained as
\begin{eqnarray}
R_{\ell m \omega}^{{\rm (odd)}}(r) =
\cases{
{\displaystyle \frac {16\,i\,\pi \,\mu \,\Omega ^{2}\,m\,r}
{(2\,\ell + 1)\,\ell\,(\ell + 1)\,(\ell + 2)}}
\,\left( {\displaystyle \frac {r}{r_0}} \right)^{\ell}
\partial_{\theta}Y_{\ell m}^*(\theta_0,\,\phi_0)
& for $r<r_0$ \,, \cr
- {\displaystyle \frac {16\,i\,\pi \,\mu \,\Omega ^{2}\,m\,r_0}
{(2\,\ell + 1)\,(\ell - 1)\,\ell\,(\ell + 1)}}
\,\left( {\displaystyle \frac {r_0}{r}} \right)^{\ell}
\partial_{\theta}Y_{\ell m}^*(\theta_0,\,\phi_0)
& for $r>r_0$ \,, \cr
}
\label{eq:Rodd}
\end{eqnarray}
where $\Omega=u^{\phi}/u^t$.
For the even part, the radial function
is obtained as
\begin{eqnarray}
R_{\ell m \omega}^{{\rm (Z)}} &=&
{8\, \Omega \,m\,\pi \,u^t\,\mu
\over (2\,\ell + 1)\,(\ell + 2)\,(\ell + 1)
\omega }
\biggl( \left( - \,{\displaystyle \frac {r^{3}}{(2\,\ell + 3)\,r_0}}  +
{\displaystyle \frac {(\ell^{2} - \ell + 4)\,r_0\,r}
{\ell\,(2\,\ell - 1)\,(\ell - 1)}} \right) \,\omega ^{2}
\nonumber \\
&&\quad
+ 2{r \over r_0}
+ 2\,{\displaystyle \frac {(\ell^{2} - 2\,\ell - 1)\,M\,r}{(\ell - 1)\,r_0^2}}
\mbox{} - {\displaystyle \frac {2\,(\ell^{4} + \ell^{3} - 6\,\ell^{2}
- 4\,\ell - 4)\,M}{\ell\,(\ell - 1)\,(\ell + 2)\,r_0}} \biggr)
\nonumber \\
 &&\quad \times
\,\left( {\displaystyle \frac {r}{r_0}} \right)^{\ell}
Y_{\ell m}^*(\theta_0,\,\phi_0)
\qquad \qquad {\rm for} \quad r<r_0 \,,
\nonumber\\
R_{\ell m \omega}^{{\rm (Z)}} &=&
{8\, \Omega \,m\,\pi \,u^t\,\mu
\over (2\,\ell + 1)\,\ell \,(\ell - 1)
\omega }
\biggl( \left( \,{\displaystyle \frac {r^{2}}{2\,\ell - 1}}  -
{\displaystyle \frac {(\ell^{2} + 3\,\ell + 6)\,r_0^{2}}
{(\ell + 1)\,(2\,\ell + 3)\,(\ell + 2)}} \right) \,\omega ^{2}
\nonumber \\
&&\quad
+ 2
- 2\,{\displaystyle \frac {(\ell^{2} + 4\,\ell + 2)\,M}{(\ell + 2)\,r_0}}
\mbox{}
+ {\displaystyle \frac {2\,(\ell^{4} + 3\,\ell^{3} - 3\,\ell^{2}
- 7\,\ell - 6)\,M}{(\ell + 1)\,(\ell - 1)\,(\ell + 2)\,r}} \biggr)
\nonumber \\
&&\quad \times
\,\left( {\displaystyle \frac {r_0}{r}} \right)^{\ell}
Y_{\ell m}^*(\theta_0,\,\phi_0)
\qquad \qquad {\rm for} \quad r>r_0 \,,
\label{eq:Reven}
\end{eqnarray}
The metric perturbation in the RW gauge is obtained from
Eqs.~(\ref{eq:h0}) and (\ref{eq:H2}).

\subsection{Full Force}

Formally, the force derived from the full metric perturbation
is given by
\begin{eqnarray}
F_{\rm full,RW}^{\mu}(z) = -{\mu \over 2}
\left(g^{\mu\nu}+u^{\mu}u^{\nu}\right)
\left(2 h^{{\rm full,RW}}_{\nu\alpha;\beta}
 - h^{{\rm full,RW}}_{\alpha\beta;\nu}\right)
u^{\alpha}u^{\beta}\,.
\end{eqnarray}
If we decompose the above into harmonic modes, each mode becomes
finite at the location of the particle though the sum over the modes
diverges. We therefore apply the
`mode decomposition regularization' method, in which
the force is decomposed into harmonic modes and
subtract the harmonic-decomposed S part mode by mode before
the coincidence limit $x\to z(\tau)$ is taken.

Since the orbit under consideration is circular, the source term
contains the factor $\delta(\omega-m\Omega)$, and the frequency integral
can be trivially performed. Hence we can calculate the harmonic coefficients
of the full metric perturbation in the time-domain. This is a great
advantage of the circular orbit case, since the S part can be given
only in the time-domain.
We also note that the $\theta$-component of the force vanishes because
of the symmetry, and $F^{\phi}(z)=[(r_0-2M)/(r_0^3 \Omega)] F^t$
for a circular orbit.

The even and odd parity parts of the full self-force are
expressed in terms of the metric perturbation as
\begin{eqnarray}
F^{t\,{\rm RW}}_{{\rm (even)}} &=& \sum_{\ell m}
{\displaystyle \frac {i\,\mu\,m\,\Omega\,r_0}{2(r_0-3\,M)(r_0-2\,M)}}
\,\left( (r_0-2\,M)\,H_{0 \ell m, m\Omega}^{\rm RW}(r_0)
 +M\, K_{\ell m, m\Omega}^{\rm RW}(r_0)\right)
\,Y_{\ell m}(\theta_0 , \,\phi_0 )
\,, 
\cr
F^{r\,{\rm RW}}_{{\rm (even)}} &=& \sum_{\ell m}
{\displaystyle \frac {\mu \,(r_0 - 2\,M)}{2\,r_0^{2}\,(r_0 - 3\,M)}}
\Biggl(2\,M \,H_{0 \ell m, m\Omega}^{\rm RW}(r_0)
+ 2\,\,M\,K_{\ell m, m\Omega}^{\rm RW}(r_0)
\nonumber \\ &&
+r_0\,(r_0- 2\,M)
{\frac {d }{d r}}\,H_{0 \ell m, m\Omega}^{\rm RW}(r_0)
 + r_0\,M\,{\frac {d }{d r}}
\,K_{\ell m, m\Omega}^{\rm RW}(r_0)\Biggr)
\,Y_{\ell m}(\theta_0 , \,\phi_0 )
\,, \cr 
F^{t\,{\rm RW}}_{{\rm (odd)}} &=& \sum_{\ell m} 
 {\displaystyle \frac {i\,\mu \,M
\,m}{r_0\,(r_0 - 3\,M)
\,(r_0 - 2\,M)}} \,h_{0\ell m, m\Omega}^{\rm RW}(r_0)\,
\partial_{\theta}\,Y_{\ell m}(\theta_0 , \,\phi_0 )
\,, \cr
F^{r\,{\rm RW}}_{{\rm (odd)}} &=& \sum_{\ell m}
{\displaystyle \frac {\mu \,(r_0 - 2\,M)\,\Omega}{r_0 - 3\,M}}
\,({\frac {d }{d r}}\,h_{0\ell m, m\Omega}^{\rm RW}(r_0))
\,\partial_{\theta}\,Y_{\ell m}(\theta_0 , \,\phi_0 ) \,,
\label{eq:Ffull}
\end{eqnarray}
It is understood that the derivatives appearing in the above expressions
are taken before the coincidence limit. It may be noted that
there is no contribution from the components $H^{RW}_1$ and $H^{RW}_2$
to the even force and no contribution from $h^{RW}_1$ to the odd force
for a circular orbit. 

{}Inserting the metric perturbation under the RW gauge 
to the above, and performing the summation over $m$,
we find
\begin{eqnarray}
\left. F^t_{{\rm full,RW}}\right|_{\ell}
&=& 0 \,, \cr
\left. F^{r(+)}_{{\rm full,RW}}\right|_{\ell} &=&
- {\displaystyle \frac {(\ell + 1)\,\mu ^{2}}{r_0^{2}}}
+ {\displaystyle \frac {1}{2}} \,
{\displaystyle \frac {\mu ^{2}\,(12\,\ell^{3} + 25\,\ell^{2} + 4\,\ell - 21)\,M}
{r_0^{3}\,(2\,\ell + 3)\,(2\,\ell - 1)}}
\,, \cr
\left. F^{r(-)}_{{\rm full,RW}}\right|_{\ell} &=&
 {\displaystyle \frac {\ell\,\mu ^{2}}{r_0^{2}}}
- {\displaystyle \frac {1}{2}}
\,{\displaystyle \frac {\mu ^{2}\,(12\,\ell^{3} + 11\,\ell^{2} - 10\,\ell + 12)\,M}
{(2\,\ell - 1)\,(2\,\ell + 3)\,r_0^{3}}}
\,, \cr
\left. F^{\theta}_{{\rm full,RW}}\right|_{\ell}
&=& 0 \,, \cr
\left. F^{\phi}_{{\rm full,RW}}\right|_{\ell}
&=& 0
 \,.
\end{eqnarray}
We see that the only non-vanishing component is the radial component
as expected because there is no radiation reaction effect at 1PN order.
In the above, the indices $(+)$ and $(-)$ denote that the coincidence limit
is taken from outside ($r>r_0)$ of the orbit and inside ($r<r_0$)
of the orbit, respectively, and the vertical bar suffixed with $\ell$,
\begin{eqnarray*}
{\cdots}~\Bigl|_{\ell} \,,
\end{eqnarray*}
denotes the coefficient of the $\ell$ mode in the coincidence limit.
The formulas for the summation over $m$ are shown in Appendix \ref{app:M}.

We note that the above result is valid for $\ell \geq 2$.
Although the $\ell=0$ and $1$ modes do not contribute to the
self-force formally, because of our inability to
know the exact form of the S part, it turns out that we do need
to calculate the contributions from the $\ell=0$ and $1$ modes.
These modes are treated in Appendix \ref{app:L01}.

\section{S part of the metric perturbation and Force}\label{sec:dir}

In this section, we calculate the S part of metric perturbation
and its self-force (S-force) by using the local coordinate expansion.
The S part of the metric perturbation in the harmonic gauge
is given covariantly as
\begin{eqnarray}
\bar{h}_{\mu\nu}^{\rm S,H} &=& 4 \mu \left[{
\bar g_{\mu\alpha}(x,z_{\rm ret})\bar g_{\nu\beta}(x,z_{\rm ret})
u^\alpha(\tau_{\rm ret})u^\beta(\tau_{\rm ret})
\over \sigma_{;\gamma}(x,z_{\rm ret})u^\gamma(\tau_{\rm ret})}
\right]
\nonumber \\ &&
+ 2 \mu (\tau_{{\rm adv}}-\tau_{{\rm ret}})
\bar g_{\mu}{}^{\alpha}(x,z_{\rm ret})
\bar g_{\nu}{}^{\beta}(x,z_{\rm ret})
R_{\gamma \alpha \delta \beta}(z_{\rm ret})
u^\gamma(\tau_{{\rm ret}})u^\delta(\tau_{{\rm ret}})
+O(y^2) \,,
\label{eq:spart}
\end{eqnarray}
where $z_{\rm ret}=z(\tau_{\rm ret})$,
 $\tau_{\rm ret}$ is the retarded proper time
defined by the past light cone condition of the field point $x$,
$\tau_{\rm adv}$ is the advanced proper time defined
by the future light cone condition of the field point $x$,
$\bar g_{\mu\alpha}$ is the parallel displacement bi-vector, and
$y$ is the expansion parameter of the local expansion,
which may be taken to be the difference of the coordinates between
$x$ and $z_0$, $y^\mu=x^\mu-z_0^\mu$.
Details of the local expansion are given in \cite{MNS1}.
The difference between the S part and the direct part
appears in the terms of $O(y)$, i.e., the second term on the
right-hand side of Eq.~(\ref{eq:spart}).
In the local coordinate expansion of the S part,
it is convenient to use the quantities
\begin{eqnarray}
&&\epsilon:=
\left(r_0^2+r^2-2\,r_0\,r\,\cos\Theta\cos\Phi\right)^{1/2}\,,
\nonumber\\
&&T:=t-t_0\,,\quad R:=r-r_0\,,
\nonumber\\
&&\Theta:=\theta-{\pi\over2}\,,\quad\Phi:=\phi-\phi_0\,.
\label{eq:xi}
\end{eqnarray}

\subsection{S part of the metric perturbation}

Using the variables defined in Eqs.~(\ref{eq:xi}),
it is straightforward to calculate the S part to 1PN order.
Here we note that, in general, we have to evaluate the S part
up through the accuracy of $O(y)$,
because the force is given by first derivatives of the metric components.
The result takes the form,
\begin{eqnarray}
{h_{\mu\nu}^{\rm S,H}} &=&
\mu \sum_{m,n,p,q,r}c_{m,n,p,q,r}
\frac{T^mR^n\Theta^p\Phi^q}{\epsilon^r}
\,,
\end{eqnarray}
where $m$, $n$, $p$, $q$ and $r$ are positive intergers.
The explicit expressions for the
components are shown in Appendix \ref{app:OS},
Eqs.~(\ref{eq:SlocalH}).

\subsection{Tensor harmonics expansion of the S part}

In the previous subsection, we calculated
the S part of the metric perturbation
in the local coordinates expansion.
In order to use them in the mode decomposition regularization,
it is necessary to expand them in terms of tensor spherical harmonics,
which involves an extension of the locally expanded S part to
a quantity defined over the sphere. Since the only requirement is
to recover the local behavior near the orbit correctly, there exists
much freedom in the way of extending the locally known S part
to a globally defined (but only approximate) S part on the
whole sphere. To guarantee the accuracy of $h^{\rm S,H}_{\mu\nu}$
up through $O(y)$ in the local expansion, because the leading term
diverges as $1/y$, a spherical extension must be accurate enough to
recover the behavior at $O(y^2)$ beyond the leading order.
Below, using one of such extensions as given in Appendix~\ref{app:CHD},
we derive the harmonic coefficients of the S part.

Once we fix the method of spherical extension, it is possible
in principle to calculate the harmonic coefficients of the
extended S part exactly. However, it is neither necessary
nor quite meaningful because the extension is only approximate.
In fact, corresponding to the fact that all the terms in positive
powers of $y$ vanish in the coincidence limit, it is known 
that all the terms of $O(1/L^2)$ or higher,
where $L=\ell+1/2$, vanish when summed over $\ell$ \cite{MNS1}
in the harmonic gauge. It should be noted, however, this result
is obtained by expanding the force in the
scalar spherical harmonics. In our present analysis, we employ
the tensor spherical harmonic expansion. So, the meaning of
the index $\ell$ is slightly different. Nevertheless, the
same is found to be true. Namely, by expanding the S-part of
the metric perturbation in the tensor spherical harmonics,
the S-force in the harmonic gauge is found to have the form,
\begin{eqnarray}
F^{\mu(\pm)}_{\rm S,H}|_\ell=\pm A^\mu L+B^\mu +D^\mu_\ell\,,
\end{eqnarray}
where $A^\mu$ and $B^\mu$ are independent of $\ell$,
and the $\pm$ denotes that the limit to $r_0$ is taken 
from the greater or smaller value of $r$, and
\begin{eqnarray}
D^\mu_\ell=\frac{d^\mu}{L^2-1}+
\frac{e^\mu}{(L^2-1)(L^2-4)}+
\frac{f^\mu}{(L^2-1)(L^2-4)(L^2-9)}+
\cdots.
\end{eqnarray}
Then the summation of $D^\mu_\ell$ over $\ell$ 
(from $\ell=0$ to $\infty$) vanishes.
For convenience, let us call this the standard form.
As we shall see later, the standard form of the S-force 
is found to persist also in the RW gauge. 

For the moment, let us assume the standard form of the
S-force both in the harmonic gauge and the RW gauge.
Then we may focus our discussion on the divergent terms.
When we calculate the S-force in the RW gauge,
we first transform the metric perturbation from the harmonic
gauge to the RW gauge, and then take appropriate linear combinations
of their first derivatives. We then find that
the harmonic coefficients 
$h_{2\ell m}^{\rm S,H}$, $h_{0\ell m}^{\rm (e) S,H}$ and
$h_{1\ell m}^{\rm (e) S,H}$
are differentiated two times, and $G_{\ell m}^{\rm S,H}$
is differentiated three times,
while the rest are differentiated once, to obtain the S-force.
So, it is necessary and sufficient to perform the Taylor expansion
of the harmonic coefficients up to $O(X^2)$ for
$h_{2\ell m}^{\rm S,H}$, $h_{0\ell m}^{\rm (e) S,H}$ and
$h_{1\ell m}^{\rm (e) S,H}$,
and up to $O(X^3)$ for $G_{\ell m}^{\rm S,H}$,
and the rest up to $O(X)$, where $X=T$ or $R$.

To the accuracy mentioned above,
the harmonic coefficients of the S part are found in the form,
\begin{eqnarray}
h_{0\ell m}^{\rm S,H}(t,\,r)
&=& {2 \over L}\,\pi\, \mu \biggl[
\frac {4\,i\,T\,m\,
r_0\,(L^{2} - 2)\,(u^{\phi})^2}
{{\cal L}^{(2)}\,(L^2 - 1)} 
 +\cdots\,\biggr]\partial_{\theta}\,Y_{\ell m}^*(\theta_0,\,\phi_0) 
\,,
\cr
h_{1\ell m}^{\rm S,H}(t,\,r)
&=& {2 \over L}\,\pi\,\mu \biggl[
{\displaystyle \frac {-2\,i\,r_0\,
m\,(2\,r_0 + R)\,(u^{\phi})^2}
{{\cal L}^{(2)}\,(L^2 - 1)}} 
\biggr] \partial_{\theta}\,Y_{\ell m}^*(\theta_0,\,\phi_0) \,,
\cr
h_{2\ell m}^{\rm S,H}(t,\,r)
&=& {2 \over L}\,\pi\,\mu \biggl[
-  \frac {1}{6} \frac{r_0\,
m\,(72\,r_0\,R\,L^{4} + 48\,r_0\,R\,L^{5} 
+\cdots)}{{\cal L}^{(4)}\,(L^2 - 1)\,(L^2 - 4)}
(u^{\phi})^2\biggr]\partial_{\theta}\,Y_{\ell m}^*(\theta_0,\,\phi_0)
\,,\qquad \mbox{etc.},
\label{eq:harmonicS}
\end{eqnarray}
where we have defined
\begin{eqnarray}
{\cal L}^{(2)} &=& \ell(\ell+1) = \left(L^2-{1\over 4}\right)
\,,
\nonumber\\
{\cal L}^{(4)} &=& \ell(\ell+1)(\ell-1)(\ell+2) 
= \left(L^2-{1\over 4}\right)\left(L^2-{9\over 4}\right)\,.
\end{eqnarray}
The explicit expressions for the coefficients are
given in Appendix \ref{app:OS}, Eqs.~(\ref{eq:ScoeffH}).
Shown there are the coefficients in the case when we approach
the orbit from inside ($r<r_0$).
The results in the case of approaching from outside ($r>r_0$)
are obtained in the same manner. For readers' convenience, 
these are placed at the web page: 
{http://www2.yukawa.kyoto-u.ac.jp/\~{}misao/BHPC/}. 

Now we consider the S-force in the harmonic gauge. 
It is noted that the $t$, $\theta$ and $\phi$-components of the S-force
vanish after summing over $m$ modes. 
The $r$-component of the S-force is derived as 
\begin{eqnarray}
\left. F^{r(-)}_{{\rm S,H}}\right|_{\ell} 
&=& \sum_m {2\,\pi\,\mu^2 \over L}\biggl[
\biggl(
{2\,L-1 \over 2\,r_0^2}
+{M\,(10\,L^3+11\,L^2-10\,L-17) \over 4\,r_0^3\,(L^2-1)}
\nonumber \\ && \qquad \qquad
-{M\,(64\,L^5+28\,L^4-320\,L^3-695\,L^2+256\,L+442)\,m^2 \over 
16\,r_0^3\,{\cal L}^{(2)}\,(L^2-1)(L^2-4)} 
\nonumber \\ && \qquad \qquad
-{M\,(156\,L^2-179)\,m^4 \over 4\,r_0^3\,
{\cal L}^{(2)}\,(L^2-1)(L^2-4)(L^2-9)} 
\biggr) \,|Y_{\ell m}(\theta_0 , \,\phi_0 )|^2
\nonumber \\ && \qquad
+\biggl(
{13\,M\,m^2 \over r_0^3\,{\cal L}^{(2)}\,(L^2-1)(L^2-4)}
\nonumber \\ && \qquad \qquad
-{M\,(2\,L-1)(2\,L^2+2\,L-1) \over r_0^3\,{\cal L}^{(2)}\,(L^2-1)}
\biggr)
|\partial_{\theta}\,Y_{\ell m}(\theta_0 , \,\phi_0 )|^2
\biggr] \,.
\label{eq:bmFr}
\end{eqnarray}
The formulas for summation over $m$ are summarized
in Appendix \ref{app:M}. For example, we have
\begin{eqnarray}
\sum_m {2\,\pi \over L} m^2 \,|Y_{\ell m}(\pi/2, \,0 )|^2
&=& {{\cal L}^{(2)} \over 2} \,, \cr 
\sum_m {2\,\pi \over L}|\partial_{\theta}
\,Y_{\ell m}(\pi/2 , \,0 )|^2
&=& {{\cal L}^{(2)} \over 2} \,.
\end{eqnarray}
Using these formulas, we obtain
\begin{eqnarray}
\left. F^t_{{\rm S,H}}\right|_{\ell}
&=& 0 \,, \cr
\left. F^{r(\pm)}_{{\rm S,H}}\right|_{\ell}
&=& 
\mp {\displaystyle \frac {1}{2}}
\,{\displaystyle \frac {\mu ^{2}\,(2\,r_0 - 3\,M)}
{r_0^{3}}} \,L
- {\displaystyle \frac {1}{8}}
\,{\displaystyle \frac {\mu ^{2}\,(4\,r_0 - 7\,M)}{r_0^{3}}}
\nonumber \\ && 
+{\mu^2 \,M \,(172\,L^4-14784\,L^2+299)
\over 128\,r_0^3 (L^2-1)(L^2-4)(L^2-9)}
\nonumber \\ 
&=& \mp {\displaystyle \frac {1}{2}}
\,{\displaystyle \frac {\mu ^{2}\,(2\,r_0 - 3\,M)}
{r_0^{3}}} \,L
- {\displaystyle \frac {1}{8}}
\,{\displaystyle \frac {\mu ^{2}\,(4\,r_0 - 7\,M)}{r_0^{3}}}
+O({1 \over L^2})
\,, \cr
\left. F^{\theta}_{{\rm S,H}}\right|_{\ell}
&=& 0 \,, \cr
\left. F^{\phi}_{{\rm S,H}}\right|_{\ell}
&=& 0 \,.
\end{eqnarray}
This is indeed of the standard form.
In particular, the factor ${\cal L}^{(2)}$ which is
present in the denominators before summing over $m$ turns
out to be cancelled by the same factor that arises from
summation over $m$. If it were present in the final result,
we would not be able to conclude that the summation of
$D^\mu_\ell$ over $\ell$ vanishes.
We note that, apart from the fact that the 
denominator of the $D^\mu_\ell$ term takes the standard form,
the numerical coefficients appearing in the numerator should not be
taken rigorously. This is because our calculation is accurate only
to $O(y^0)$ of the S-force, while the numerical coefficients depend
on the $O(y)$ behavior of it (an example is shown in Appendix \ref{app:y2}).
It is also noted that the $O(1/L)$-terms are absent
in the S-force, implying the absence of logarithmic divergence.

It is important to note that $\ell$ in the above runs from $0$ 
to $\infty$. Although there are some tensor harmonics that do not
exist for $\ell=0$ and/or $\ell=1$, we may regard that 
the corresponding harmonic coefficients contribute
to the $B^\mu$ and $D^\mu_\ell$ terms of the S-force
individually, with $B^\mu+D^\mu_\ell=0$. That is, 
we set the contributions to $A^\mu$ to zero and adjust
the $D^\mu_\ell$ term in such a way that
$D^\mu_\ell=-B^\mu$ for these special coefficients
while keeping the standard form for $D^\mu_\ell$.

\section{S Part in the Regge-Wheeler Gauge} \label{sec:trans}

Now, we transform the S part of the metric perturbation
from the harmonic gauge to the RW gauge.
The gauge transformation functions are
given in the tensor-harmonic expansion form as
\begin{eqnarray}
\xi_{\mu}^{{\rm (odd)}} &=& \sum_{\ell m}
\Lambda_{\ell m}^{{\rm S,H} \to {\rm RW}}(t,\,r)
\left\{ 0, 0, \frac{-1}{\sin\theta}\partial_{\phi}Y_{\ell m}(\theta,\phi),
\sin\theta\partial_{\theta}Y_{\ell m}(\theta,\phi) \right\} \,,
\cr
\xi_{\mu}^{{\rm (even)}} &=& \sum_{\ell m}
\biggl\{M_{0\ell m}^{{\rm S,H} \to {\rm RW}}(t,\,r)
Y_{\ell m}(\theta,\phi),
M_{1\ell m}^{{\rm S,H} \to {\rm RW}}(t,\,r)Y_{\ell m}(\theta,\phi),
\nonumber \\ && \qquad \qquad
M_{2\ell m}^{{\rm S,H} \to {\rm RW}}(t,\,r)
\partial_{\theta}Y_{\ell m}(\theta,\phi),
M_{2\ell m}^{{\rm S,H} \to {\rm RW}}(t,\,r)
\partial_{\phi}Y_{\ell m}(\theta,\phi)
 \biggr\}
 \,.
\label{eq:Gtransgen}
\end{eqnarray}
There are one degree of gauge freedom for the odd part and
three for the even part.
To satisfy the RW gauge condition~(\ref{eq:RWcond}),
we obtain the equations for the gauge functions
that are found to be rather simple:
\begin{eqnarray}
h_{2\ell m}^{{\rm S,H}}(t,\,r) &=&
-2\,i\,\Lambda_{\ell m}^{{\rm S,H} \to {\rm RW}}(t,\,r)\,,
\nonumber \\
\cr 
h_{0\ell m}^{{\rm (e)S,H}}(t,\,r) &=&
-M_{0\ell m}^{{\rm S,H} \to {\rm RW}}(t,\,r)
- \partial_t  M_{2\ell m}^{{\rm S,H} \to {\rm RW}}(t,\,r)
\,,
\nonumber\\
h_{1\ell m}^{{\rm (e)S,H}}(t,\,r) &=&
-M_{1\ell m}^{{\rm S,H} \to {\rm RW}}(t,\,r)
- r^2 \,\partial_r  \left(
{M_{2\ell m}^{{\rm S,H} \to {\rm RW}}(t,\,r)
\over r^2}\right)
\,,
\nonumber\\
G_{\ell m}^{{\rm S,H}}(t,\,r) &=&
-{2 \over r^2} M_{2\ell m}^{{\rm S,H} \to {\rm RW}}(t,\,r)\,.
\label{eq:ggtran}
\end{eqnarray}
We therefore find
\begin{eqnarray}
\Lambda_{\ell m}^{{\rm S,H} \to {\rm RW}}(t,\,r)
&=&{i\over2}h_{2\ell m}^{{\rm S,H}}(t,\,r)\,,
\label{eq:GGT1} \\ 
\cr
M_{2\ell m}^{{\rm S,H} \to {\rm RW}}(t,\,r)
&=&-{r^2\over2}G_{\ell m}^{{\rm S,H}}(t,\,r)\,,
\cr
M_{0\ell m}^{{\rm S,H} \to {\rm RW}}(t,\,r)
&=&-h_{0\ell m}^{{\rm (e)S,H}}(t,\,r)
- \partial_t  M_{2\ell m}^{{\rm S,H} \to {\rm RW}}(t,\,r)\,,
\cr
M_{1\ell m}^{{\rm S,H} \to {\rm RW}}(t,\,r)
&=&-h_{1\ell m}^{{\rm (e)S,H}}(t,\,r)
- r^2 \,\partial_r
\left({M_{2\ell m}^{{\rm S,H} \to {\rm RW}}(t,\,r)
\over r^2}\right)\,.
\label{eq:GGT4}
\end{eqnarray}
We note that it is not necessary to calculate any integration with respect to 
$t$ or $r$. It is also noted that the gauge functions are determined uniquely.
This is because the RW gauge is a gauge in which there is no
residual gauge freedom (for $\ell\geq2$).

Then the S part of the metric perturbation in the RW gauge
is expressed in terms of those in the harmonic gauge as
follows. The odd parity components are found as
\begin{eqnarray}
h_{0\ell m}^{\rm S,RW}(t,\,r)
&=& h_{0\ell m}^{\rm S,H}(t,\,r)
+ \partial_t \,\Lambda_{\ell m}^{{\rm S,H} \to {\rm RW}}(t,\,r) \,,
\nonumber\\
h_{1\ell m}^{\rm S,RW}(t,\,r)
&=& h_{1\ell m}^{\rm S,H}(t,\,r)
+ r^2 \,\partial_r \,
\left(
{\Lambda_{\ell m}^{{\rm S,H} \to {\rm RW}}(t,\,r)
\over r^2}\right) \,,
\label{eq:GTodd}
\end{eqnarray}
and the even parity components are found as
\begin{eqnarray}
H_{0\ell m}^{\rm S,RW}(t,\,r)
&=& H_{0\ell m}^{\rm S,H}(t,\,r)
+{2\,r \over r-2\,M}
\left[ \partial_t
\,M_{0\ell m}^{{\rm S,H} \to {\rm RW}}(t,\,r)
-{M(r-2\,M) \over r^3}
M_{1\ell m}^{{\rm S,H} \to {\rm RW}}(t,\,r)\right] \,,
\nonumber\\
H_{1\ell m}^{\rm S,RW}(t,\,r)
&=& H_{1\ell m}^{\rm S,H}(t,\,r)
+
\left[ \partial_t
\,M_{1\ell m}^{{\rm S,H} \to {\rm RW}}(t,\,r)
+\partial_r \,M_{0\ell m}^{{\rm S,H} \to {\rm RW}}(t,\,r)
-{2\,M \over r(r-2\,M) }
M_{0\ell m}^{{\rm S,H} \to {\rm RW}}(t,\,r)\right] \,,
\nonumber\\
H_{2\ell m}^{\rm S,RW}(t,\,r)
&=& H_{2\ell m}^{\rm S,H}(t,\,r)
+{2(r-2\,M) \over r}
\left[ \partial_r
\,M_{1\ell m}^{{\rm S,H} \to {\rm RW}}(t,\,r)
+{M \over r(r-2\,M)}
M_{1\ell m}^{{\rm S,H} \to {\rm RW}}(t,\,r)\right] \,,
\nonumber\\
K_{\ell m}^{\rm S,RW}(t,\,r)
&=& K_{\ell m}^{\rm S,H}(t,\,r)
+{2(r-2\,M) \over r^2}
M_{1\ell m}^{{\rm S,H} \to {\rm RW}}(t,\,r)
\,,
\label{eq:GTeven}
\end{eqnarray}
where the gauge functions are given by Eqs.~(\ref{eq:GGT1})
 and (\ref{eq:GGT4}).

\subsection{Gauge transformation and the S part in the RW gauge}

Inserting the results obtained in the previous section
to Eqs.~(\ref{eq:GGT1}) and (\ref{eq:GGT4}),
we obtain the gauge functions that transform the S part from the
harmonic gauge to the RW gauge.  
They are shown in Appendix \ref{app:OS}, Eqs.~(\ref{eq:Gfunctions}).
It may be noted that the gauge functions do not
contribute to the metric at the Newtonian order.
In other words, both the RW gauge and the harmonic gauge reduce
to the same (Newtonian) gauge in the Newtonian limit.

The S part of the metric perturbation in the RW gauge is now
found in the form,
\begin{eqnarray}
h_{0\ell m}^{\rm S,RW}(t,\,r)
&=& {2 \over L}\,\pi\,\mu \biggl[
{\displaystyle \frac 
{4\,i\,T \,m\,r_0\,(L^{2} - 2)\,(u^{\phi})^2}
{{\cal L}^{(2)}\,(L^2-1)}}  
+\cdots\biggr] 
\partial_{\theta}\,Y_{\ell m}^*(\theta_0,\,\phi_0) \,,
\cr
h_{1\ell m}^{\rm S,RW}(t,\,r)
&=& {2 \over L}\,\pi\,\mu \biggl[
-\frac {i\,m\,r_0}{3} \,\frac{( - 60\,r_0\,L^{3} 
+ 174\,r_0\,L^{2} + \cdots)(u^{\phi})^2 }
{{\cal L}^{(4)}\,(L^2-1)\,(L^2-4)} \biggr] 
\partial_{\theta}\,Y_{\ell m}^*(\theta_0,\,\phi_0)
\,,\qquad\mbox{etc.}.
\label{eq:rws6}
\end{eqnarray}
The explicit expressions are given in Appendix \ref{app:OS},
Eqs.~(\ref{eq:RWSmetric}).
\subsection{S force}

Next we calculate the S part of the self-force.
Of course, it diverges in the coincidence limit. However,
as we noted several times,
in the mode decomposition regularization in which the
regularization is done for each harmonic mode,
the harmonic coefficients of the S part are finite.

The calculation is straightforward. Expanding the formula for
the self-force~(\ref{eq:formalF}) in terms of the tensor harmonics,
we obtain
\begin{eqnarray}
F^{t\,{\rm RW}}_{{\rm (even)}} &=& \sum_{\ell m}
{\displaystyle \frac {\mu\,r_0}{2\,( r_0 - 3\,M)^{2} ( r_0 - 2\,M)  }}
\biggl(
- r_0\,M \,(\partial_{t}
\,H_{0\ell m}^{\rm RW}(t_0,\,r_0))
+ 2\,M^{2}\,(\partial_{t}\,H_{0\ell m}^{\rm RW}(t_0,\,r_0))
\nonumber \\ && \quad \quad
+ i\,m\,r_0^{2}\,\Omega \,H_{0\ell m}^{\rm RW}(t_0,\,r_0)
- 6\,i\,m\,M\,r_0\,\Omega \,H_{0\ell m}^{\rm RW}(t_0,\,r_0)
+ 8\,i\,m\,M^{2}\,\Omega \,H_{0\ell m}^{\rm RW}(t_0,\,r_0)
\nonumber \\ && \quad \quad
- i\,m\,r_0\,M\,\Omega \,K_{\ell m}^{\rm RW}(t_0,\,r_0)
+ 2\,i\,m\,M^{2}\,\Omega \,K_{\ell m}^{\rm RW}(t_0,\,r_0)
+ 5\,M^{2}\,(\partial_{t}\,K_{\ell m}^{\rm RW}(t_0,\,r_0))
\nonumber \\ && \quad \quad
- 2\,r_0\,M\,(\partial_{t}
\,K_{\ell m}^{\rm RW}(t_0,\,r_0))
\biggr)
\,Y_{\ell m}(\theta_0 , \,\phi_0 )
\,, \cr
F^{r\,{\rm RW}}_{{\rm (even)}} &=& \sum_{\ell m}
-{\displaystyle \frac {\mu \,( r_0 - 2\,M)}{2\,r_0^{2}\,( r_0 - 3\,M) }}
\biggl(
2\,r_0^{2}\,(\partial_{t}\,H_{1\ell m}^{\rm RW}(t_0,\,r_0))
- 2\,M\,H_{0\ell m}^{\rm RW}(t_0,\,r_0)
\nonumber \\ && \quad \quad
+ 2\,i\,m\,r_0^{2}\,\Omega \,H_{1\ell m}^{\rm RW}(t_0,\,r_0)
- 2\,M\,K_{\ell m}^{\rm RW}(t_0,\,r_0) 
- r_0^{2}\,(\partial_{r}\,H_{0\ell m}^{\rm RW}(t_0,\,r_0))
\nonumber \\ && \quad \quad
+ 2\,r_0\,M \,(\partial_{r}
\,H_{0\ell m}^{\rm RW}(t_0,\,r_0))
- r_0\,M
\,(\partial_{r}\,K_{\ell m}^{\rm RW}(t_0,\,r_0))
\biggr)
\,Y_{\ell m}(\theta_0 , \,\phi_0 )
\,, \cr
F^{t\,{\rm RW}}_{{\rm (odd)}} &=& \sum_{\ell m}
{\displaystyle \frac { - i\,\mu\,\Omega \,r_0^{2}}{(r_0 - 3\,M)^{2}\,(r_0 - 2\,M)}}
\,\left(
\Omega \,m\,M\,h_{0 \ell m}^{\rm RW}(t_0, \, r_0)
 - i\,(r_0-2\,M)\,(\partial_{t}\,h_{0 \ell m}^{\rm RW}(t_0, \, r_0)) 
\right)
\nonumber \\ && \qquad \times 
\partial_{\theta}\,Y_{\ell m}(\theta_0 , \,\phi_0 )
\,, \cr
F^{r\,{\rm RW}}_{{\rm (odd)}} &=& \sum_{\ell m}
{\displaystyle \frac { \mu\,\Omega \,(r_0 - 2\,M) }{r_0 - 3\,M}}
\,\left( 
(\partial_{r}
\,h_{0 \ell m}^{\rm RW}(t_0, \, r_0))
- \,(\partial_{t}\,h_{1 \ell m}^{\rm RW}(t_0, \, r_0)) 
-i\,\Omega \,m  \, h_{1 \ell m}^{\rm RW}(t_0, \, r_0)
\right)
\nonumber \\ && \qquad \times
\partial_{\theta}\,Y_{\ell m}(\theta_0 , \,\phi_0 )
\,.
\end{eqnarray}
Substituting the S part of the metric components in the RW gauge
as shown in Eqs.~(\ref{eq:rws6}),
given explicitly in Eqs.~(\ref{eq:RWSmetric}), into the above, 
we find that the $t$, $\theta$ and $\phi$-components of the S-force
vanish after summing over $m$. 
The $r$-component of the S-force inside the particle trajectory 
is derived as
\begin{eqnarray}
\left. F^{r(-)}_{{\rm S,RW}}\right|_{\ell} 
&=& \sum_m {2\,\pi\,\mu^2 \over L}\biggl[
\biggl(
{2\,L-1 \over 2\,r_0^2}
+{M\,(10\,L^3+11\,L^2-10\,L-17) \over 4\,r_0^3\,(L^2-1)}
\nonumber \\ && \qquad \qquad
-{M\,(64\,L^5+28\,L^4-320\,L^3-695\,L^2+256\,L+442)\,m^2 \over 
16\,r_0^3\,{\cal L}^{(2)}\,(L^2-1)(L^2-4)} 
\nonumber \\ && \qquad \qquad
-{M\,(156\,L^2-179)\,m^4 \over 4\,r_0^3\,
{\cal L}^{(2)}\,(L^2-1)(L^2-4)(L^2-9)} 
\biggr) \,|Y_{\ell m}(\theta_0 , \,\phi_0 )|^2
\nonumber \\ && \qquad
+\biggl(
{13\,M\,m^2 \over r_0^3\,{\cal L}^{(2)}\,(L^2-1)(L^2-4)}
\nonumber \\ && \qquad \qquad
-{M\,(2\,L-1)(2\,L^2+2\,L-1) \over r_0^3\,{\cal L}^{(2)}\,(L^2-1)}
\biggr)
|\partial_{\theta}\,Y_{\ell m}(\theta_0 , \,\phi_0 )|^2
\biggr] \,.
\end{eqnarray}
Summing the above over $m$, we obtain
\begin{eqnarray}
\left. F^t_{{\rm S,RW}}\right|_{\ell}
&=& 0 \,, \cr
\left. F^{r(\pm)}_{{\rm S,RW}}\right|_{\ell}
&=& \mp {\displaystyle \frac {1}{2}}
\,{\displaystyle \frac {\mu ^{2}\,(2\,r_0 - 3\,M)}
{r_0^{3}}} \,L
- {\displaystyle \frac {1}{8}}
\,{\displaystyle \frac {\mu ^{2}\,(4\,r_0 - 7\,M)}{r_0^{3}}}
\nonumber \\ && 
+{\mu^2 \,M \,(172\,L^4-14784\,L^2+299)
\over 128\,r_0^3 (L^2-1)(L^2-4)(L^2-9)}
\nonumber \\ 
&=& \mp {\displaystyle \frac {1}{2}}
\,{\displaystyle \frac {\mu ^{2}\,(2\,r_0 - 3\,M)}
{r_0^{3}}} \,L
- {\displaystyle \frac {1}{8}}
\,{\displaystyle \frac {\mu ^{2}\,(4\,r_0 - 7\,M)}{r_0^{3}}}
+O({1 \over L^2})
\,, \cr
\left. F^{\theta}_{{\rm S,RW}}\right|_{\ell}
&=& 0 \,, \cr
\left. F^{\phi}_{{\rm S,RW}}\right|_{\ell}
&=& 0 \,.
\end{eqnarray}
We now see that the S-force in the RW gauge also has the standard
form as in the case of the harmonic gauge and there is no
$O(1/L)$ term. 
Note that, again with the same reason as we explained at the
end of the previous section,
the final formula above should be regarded as valid for
all $\ell$ from $0$ to $\infty$.

\section{Regularized gravitational self-force} \label{sec:result}

In the previous two sections,
we have calculated the full and S parts of the self-force
in the RW gauge. Now we are ready to evaluate the regularized
self-force. But there is one more issue to be discussed,
namely, the treatment of the $\ell=0$ and $1$ modes.

The full metric perturbation and its self-force are
derived by the Regge-Wheeler-Zerilli formalism. This means
they contain only the harmonic modes with $\ell \geq 2$.
If we could know the exact S part, then the knowledge of
the modes $\ell\geq2$ would be sufficient to derive the
regular, R part of the self-force, because the R part of
the metric perturbation is known to satisfy the homogeneous
Einstein equations~\cite{Detweiler:2002mi}, and
because there are no non-trivial homogeneous solutions
in the $\ell=0$ and $1$ modes. To be more precise, apart of
the gauge modes that are always present,
the $\ell=0$ homogeneous solution corresponds
to a shift of the black hole mass and the $\ell=1$ odd parity to
adding a small angular momentum to the black hole,
both of which should be put to zero in the absence of 
an orbiting particle. As for the $\ell=1$ even mode,
it is a pure gauge that corresponds to a dipolar shift
of the coordinates.
In other words, apart from possible gauge mode contributions,
the $\ell=0$ and $1$ modes of the full force should be exactly
cancelled by those of the S part.
In reality, however, what we have in hand is only an approximate
S part. In particular, its individual harmonic coefficients do not
have physical meaning. Let us denote the harmonic coefficients of
the approximate S-force by $F_\ell^{\rm S,Ap}$, while
the exact S-force and the full force by $F_\ell^{\rm S}$
and $F_\ell^{\rm full}$, respectively.
Then the R-force $F^R$ may be expressed as
\begin{eqnarray}
F^{R} &=& \sum_{\ell \geq 2}
\left(F^{\rm full}_{\ell} -F^{\rm S}_{\ell}\right)
\nonumber \\
&=&
\sum_{\ell \geq 0}
\left(F^{\rm full}_{\ell} -F^{\rm S}_{\ell}\right)
\nonumber \\
&=&
\sum_{\ell \geq 0}
\left(F^{\rm full}_{\ell} -F^{\rm S,Ap}_{\ell}\right)
-\sum_{\ell\geq0} D_{\ell}
\nonumber \\
&=&
\sum_{\ell \geq 2}
\left(F^{\rm full}_{\ell} -F^{\rm S,Ap}_{\ell}\right)
+\sum_{\ell =0,1}
\left(F^{\rm full}_{\ell} -F^{\rm S,Ap}_{\ell}\right)
\,,
\label{eq:Fldecomp}
\end{eqnarray}
where $D_\ell=F_\ell^{\rm S}-F_\ell^{\rm S,Ap}$, and
the last line follows from the fact that
$F_\ell^{\rm S,Ap}$ are assumed to be obtained from a
sufficiently accurate spherical extension of the local
behavior of the S part to guarantee $\sum_{\ell\geq0}D_\ell=0$.
Thus, it is necessary to evaluate the $\ell=0$ and $1$ modes
of the full force to evaluate the self-force correctly.

First, we consider the contributions of $\ell \geq 2$ to
the self-force. As noted before, for the 1PN calculation,
the only $r$-component of the full and S part of the self-force
is non-zero.
The $\ell$ mode coefficients
corresponding to the first term in the last line of
Eq.~(\ref{eq:Fldecomp}) are derived as
\begin{eqnarray}
\left. F^{r}_{{\rm RW}}\right|_{\ell} &=&
\left. F^{r}_{{\rm full,RW}}\right|_{\ell}
-\left. F^{r}_{{\rm S,RW}}\right|_{\ell}
\nonumber \\ &=&
-{45\,\mu^2\, M\over 8 (2\ell-1) (2\ell +3) \,r_0^3} \,.
\end{eqnarray}
Summing over $\ell$ modes, we obtain
\begin{eqnarray}
F^r_{{\rm RW}}(\ell \geq 2) = -{3\,\mu^2\, M \over 4 \,r_0^3} \,.
\label{eq:FRW2}
\end{eqnarray}

Next, we consider the $\ell=0$ and $1$ modes. 
Detailed analyses are given in Appendix~\ref{app:L01}. 
It is noted that the $\ell=0$ and $\ell=1$ odd modes, 
which describe the perturbation in the total mass and angular momentum,
respectively, of the system due to the presence of the particle,
are determinable in the harmonic gauge, with the retarded boundary
condition. On the other hand, we were unable to solve for 
the $\ell=1$ even mode in the harmonic gauge.
Since it is locally a gauge mode describing a shift of the center
of mass coordinates, this gives rise to an ambiguity in the
final result of the self-force. Nevertheless,
we were able to resolve this ambiguity at Newtonian order,
and hence to obtain an unambiguous interpretation of
the resulting self-force.

The correction to the regularized self-force that
arises from the $\ell=0$ and $1$ modes,
corresponding to the second term in the last line of
Eq.~(\ref{eq:Fldecomp}), is found as
\begin{eqnarray}
\delta F^r_{{\rm RW}}(\ell =0,\,1) = {2\,\mu^2 \over r_0^2}
-{41\,\mu^2 M \over 4 r_0^3} \,.
\label{eq:FRW01}
\end{eqnarray}

Finally, adding Eqs.~(\ref{eq:FRW2}) and (\ref{eq:FRW01}),
we obtain the regularized gravitational self-force to
the 1PN order as
\begin{eqnarray}
F^r_{{\rm RW}} = {2\,\mu^2 \over r_0^2}
-{11\,\mu^2 M \over r_0^3} \,.
\end{eqnarray}
Since there will be no effect of the gravitational radiation at the
1PN order, i.e., the $t-$ and $\phi-$components are zero,
the above force describes the correction to the radius of the orbit
that deviates from the geodesic on the unperturbed background.
It is noted that the first term proportional to $\mu^2$ is just
the correction to the total mass of the system at the Newtonian
order, where $r_0$ is interpreted as the distance from the center of
mass of the system to the particle.
 
\section{Conclusion} \label{sec:conclusion}

In this paper, we proposed a new method to derive
the regularized gravitational self-force on a point particle
in circular orbit around the Schwarzschild black hole,
and, as a demonstration, we derived the regularized 
self-force analytically to the first post-Newtonian (1PN) order.

The regularization of the gravitational self-force may be
divided to the two problems, the subtraction problem and
the gauge problem. To regularize and subtract the divergent
part, we employed the `mode decomposition regularization', in
which everything is expanded in the spherical harmonics and
the regularization is performed at each $\ell$ mode.
As for the gauge problem, utilizing the recent discovery by
Detweiler and Whiting that the regularized force may be derived
from the R part of the metric perturbation that satisfies the
source-free Einstein equations, we considered the regularized
force in the Regge-Wheeler gauge.

In the present paper, actual calculations were done only
for circular orbit and to the 1PN order. 
However, there remains a problem for the even parity $\ell=1$ mode. 
In this metric perturbation approach, there inevitably remains
ambiguity of the gauge in the resulting self-force. To circumvent this
difficulty, the only way seems to be to regularize at the level of 
the Weyl scalar $\psi_4$
or the Hertz potential $\Psi$, which are free from the $\ell=0$ and $1$ modes
As another problem, to make our method applicable to general cases, 
it is therefore necessary to extend to general orbits and to higher PN orders.
Some progress in this direction based on analytical methods
is under way~\cite{BHPC}.
It will also be necessary to incorporate numerical techniques
if we are to treat completely general orbits.
Some development is done by Fujita et al. \cite{Fujita}.

Our final goal is to derive the self-force on the Kerr background.
Recently, Mino~\cite{Minon} has proposed a new approach
to the radiation reaction problem by using the radiative Green function.
In his method, assuming the validity of the adiabatic approximation,
the radiation reaction to the conserved quantities including the
Carter constant can be calculated from the radiative Green function,
which is free from any singular, divergent behavior.
This is a great computational advantage.
However, this method cannot treat the self-force for completely
general orbit because of the assumption of adiabaticity. It is
therefore still necessary to derive the self-force in the general case.
One possibility is to consider the regularization of the Weyl scalar
$\Psi_4$ and construct the R part of the metric perturbation in the
radiation gauge by using the Chrzanowski method.
Investigations in this direction is also in progress~\cite{psireg}.

\acknowledgements

We would like to thank W.~Hikida, S.~Jhingan,
Y.~Mino, H.~Tagoshi, T.~Tanaka and M.~Shibata
for useful discussions.
HN and MS would like to thank all the participants of 
the Radiation Reaction Focus Session and
the 5th Capra Ranch Meeting at Penn State for invaluable discussions.
We have also benefitted a lot from 
the YITP workshop on ``Gravitational Wave physics'' (YITP-W-02-22),
and from the 6th Capra Meeting
and the Post-Capra Meeting (YITP-W-03-02) held at Yukawa Institute,
Kyoto University. Discussions with S.~Detweiler,
A.~Ori and E.~Poisson on the $\ell=0$ and 1 modes at the
Post-Capra meeting were particularly useful for us to complete
the present work.
HN is supported by the JSPS Research Fellowships
for Young Scientists, No.~5919. 
This work was supported in part by Monbukagaku-sho Grant-in-Aid
for Scientific Research, Nos.~1047214 and 12640269, and by
the Center for Gravitational Wave Physics, PSU,
which is funded by the National Science Foundation under
Cooperative Agreement PHY 0114375.

\begin{appendix}

\section{Mano et al. Analysis}\label{app:mano}

In this Appendix, we summarize the analysis of Mano et al.~\cite{Mano2}
which we use in order to derive the full metric perturbation
 for $\ell\geq 2$ modes.

\subsection{Homogeneous solutions}

We investigate
the analytic expression of the Regge-Wheeler functions,
and generate these functions in an explicit manner
up to $O(v^2)$ corrections relative to the leading order
in the slow-motion expansion, i.e., first post-Newtonian order.
(More detail analysis is given in \cite{SaNa}.)
Here $v$ is a characteristic velocity of the particle.
The Regge-Wheeler equation is
\begin{eqnarray}
\left[
{d \over dr} \left(1-{2M \over r}\right) {d \over dr}
+\left(1-{2M \over r}\right)^{-1}\left(\omega^2 - V_{\ell}(r)\right)
\right]R_{\ell m \omega}^{{\rm (even/odd)}}(r)
= \left(1-{2M \over r}\right)^{-1} S_{\ell m \omega}^{{\rm (even/odd)}}(r) \,.
\end{eqnarray}
The source term
$S_{\ell m\omega}^{\rm (even)}$ is expressed in terms of
the source terms of the Zerilli equations~\cite{Zer} as
\begin{eqnarray}
S_{\ell m\omega}^{\rm (even)} =
\left(\lambda(\lambda+1)+{9M^2(r-2M)\over r^2(\lambda r+3M)}\right)
S_{\ell m\omega}^{{\rm (Z)}}
-3M \left(1-{2M\over r}\right){d\over dr}S_{\ell m\omega}^{{\rm (Z)}} \,,
\end{eqnarray}
and the Zerilli function, $R_{\ell m\omega}^{\rm (Z)}$ is
derived from $R_{\ell m \omega}^{\rm (even)}$ as
\begin{eqnarray}
R_{\ell m\omega}^{\rm (Z)} = {1 \over (\lambda^2(\lambda+1)^2+9\,\omega^2 M^2)}
\left[
\left(\lambda(\lambda+1)+{9M^2(r-2M)\over r^2(\lambda r+3M)}\right)
R_{\ell m\omega}^{\rm (even)}
+3M \left(1-{2M\over r}\right){d\over dr}R_{\ell m\omega}^{\rm (even)}
\right]
\,.
\end{eqnarray}
So, we may focus on the Regge-Wheeler function.
The Regge-Wheeler equation is rewritten as
\begin{eqnarray}
{d^2 \over dz^2}X(z)+\left[
{1\over z-\epsilon}-{1\over z}\right]{d \over dz}X(z)
+\left[1+{2\epsilon\over z-\epsilon}
+{\epsilon^2\over (z-\epsilon)^2}-{\ell(\ell+1)\over z(z-\epsilon)}
+{3\epsilon\over z^2(z-\epsilon)}\right]X(z)&=&
\left(1-{\epsilon \over z}\right)^{-2} S(z) \,.
\end{eqnarray}
Here $z=\omega r$ and $\epsilon=2M \omega$, and
we use the symbol $X(z)$ for $R_{\ell m \omega}^{{\rm (even/odd)}}(r)$,
$S(z)$ for $S_{\ell m \omega}^{{\rm (even/odd)}}(r)$.
In the post-Newtonian expansion, both $z$ and $\epsilon$
are assumed to be small, while only $\epsilon$
is considered to be small in the post-Minkowskian expansion.
We note that $z \sim O(v)$ and $\epsilon \sim O(v^3)$
in the post-Newtonian expansion.

First, we consider a homogeneous Regge-Wheeler function
in the form of a series of the Coulomb wave functions,
$X_C{}^\nu$. (See (3.4) and (3.6) in Ref.~\cite{Mano2}.)
\begin{eqnarray}
X_C{}^\nu (z) &=&
\left(1-{\epsilon\over z}\right)^{-i\epsilon}
\sum_{n=-\infty}^{\infty} i^n
{\Gamma(n+\nu-1-i\epsilon)\Gamma(n+\nu+1-i\epsilon)\over
\Gamma(n+\nu+1+i\epsilon)\Gamma(n+\nu+3+i\epsilon)}
a_n{}^\nu F_{\nu+n}(z) \,, \label{eq:xc001}
\cr
F_{n+\nu} (z) &=&
e^{-iz} (2z)^{n+\nu} z{\Gamma(n+\nu+1+i\epsilon)\over\Gamma(2n+2\nu+2)}
{}_1 F_1(n+\nu+1+i\epsilon;2n+2\nu+2;2iz) \,, \label{eq:xc002}
\end{eqnarray}
where ${}_1 F_1$ is the confluent hypergeometric function, and
the expansion coefficients $a_n{}^\nu$ are determined
by the three-term recurrence relation,
(See (2.5) and below in Ref.~\cite{Mano2}.)
\begin{eqnarray}
&&\alpha_n{}^\nu a_{n+1}{}^\nu +\beta_n{}^\nu a_n{}^\nu
+\gamma_n{}^\nu a_{n-1}{}^\nu = 0 \,,
\cr
&&\alpha_n{}^\nu = -i\epsilon
{(n+\nu-1+i\epsilon)(n+\nu-1-i\epsilon)(n+\nu+1-i\epsilon)
\over (n+\nu+1)(2n+2\nu +3)} \,,
\cr
&&\beta_n{}^\nu = (n+\nu)(n+\nu+1)-\ell(\ell+1)
+2\epsilon^2+{\epsilon^2 (4+\epsilon^2)\over (n+\nu)(n+\nu+1)} \,,
\cr
&&\gamma_n{}^\nu = i\epsilon
{(n+\nu+2+i\epsilon)(n+\nu+2-i\epsilon)(n+\nu+i\epsilon)
\over (n+\nu)(2n+2\nu -1)} \,,
\end{eqnarray}
and $\nu$, which is called the renormalized angular momentum, is
determined by requiring the convergence of the series expansion 
in $X_C{}^\nu$. Replacing $\nu$ by $-\nu-1$, one obtains the other
independent solution $X_C{}^\nu$.
 It is important to note that the renormalized
angular momentum in the post-Minkowskian expansion becomes
\begin{eqnarray}
\nu = \ell +O(\epsilon^2)=\ell+O(v^6)\,.
\label{eq:ang001}
\end{eqnarray}
Hence $\nu=\ell$ to 1PN order.

The post-Minkowskian expansion of the coefficients $a_n{}^\nu$
is also discussed in Ref.~\cite{Mano2}.
With the normalization $a_0{}^\nu=1$, they are
found for $\ell\geq 2$,
\begin{eqnarray}
a_n{}^\nu &\sim& O(\epsilon^{|n|})
\qquad \qquad (n\geq -\ell+2) \,,
\cr
a_{-\ell+1}{}^\nu &\sim& O(\epsilon^{\ell+1}) \,,
\cr
a_{-\ell}{}^\nu &\sim& O(\epsilon^{\ell+2}) \,,
\cr
a_{-\ell-1}{}^\nu &\sim& O(\epsilon^{\ell+2}) \,,
\cr
a_n{}^\nu &\sim& O(\epsilon^{|n|+1})
\qquad \quad (-\ell-2\geq n\geq -2\ell) \,,
\cr
a_n{}^\nu &\sim& O(\epsilon^{|n|-1})
\qquad \quad (-2\ell-1\geq n) \,.
\end{eqnarray}
The post-Minkowskian expansion of the coefficients $a_n{}^{-\nu-1}$
can be obtained by using the symmetry,
\begin{eqnarray}
a_n{}^\nu = a_{-n}{}^{-\nu-1} \,.
\end{eqnarray}
(See (2.13) in Ref.\cite{Mano2}.)

The leading terms in the Regge-Wheeler functions
in the slow-motion expansion become
\begin{eqnarray}
X_C{}^\nu &\sim& O(z^{\ell+1}\epsilon^0) \,, \cr
X_C{}^{-\nu-1} &\sim& O(z^{-\ell}\epsilon^0) \,.
\end{eqnarray}
Then, for instance, if we consider 1PN order,
it is sufficient to take account of the
$a_0^\nu$ and $a_{-1}^\nu$ terms in 
$X_C{}^\nu$ and $X_C{}^{-\nu-1}$ ($\ell \geq 2$).

In Ref.~\cite{Mano2}, the homogeneous Regge-Wheeler functions
with the in-going and up-going boundary conditions
are derived in the form of linear combinations
of $X_C{}^\nu$ and $X_C{}^{-\nu-1}$.
The in-going boundary condition is that
waves are purely in-going at the black hole horizon, and
the up-going boundary condition is that
waves are purely out-going at the infinity.
\begin{eqnarray}
X_{{\rm in}}{}^\nu &=& K_\nu X_C{}^\nu +K_{-\nu-1} X_C{}^{-\nu-1} \,,
\label{eq:xin001} \cr
&& K_\nu = -{\pi 2^{-\nu} \epsilon^{-\nu-1} \over
\Gamma(\nu+1+i\epsilon)\Gamma(\nu-1+i\epsilon)
\Gamma(\nu+3+i\epsilon)\sin\pi(\nu+i\epsilon)} \nonumber \\
&& \qquad \qquad \times
\sum_{n=0}^\infty {\Gamma(n+\nu-1+i\epsilon)\Gamma(n+2\nu+1) \over
n!\Gamma(n+\nu+3-i\epsilon)}a_n{}^\nu \nonumber \\
&& \qquad \qquad \times
\left[\sum_{n=-\infty}^0
{\Gamma(n+\nu-1-i\epsilon)\Gamma(n+\nu+1-i\epsilon) \over
(-n)!\Gamma(n+\nu+1+i\epsilon)\Gamma(n+\nu+3+i\epsilon)\Gamma(n+2\nu+2)}
a_n{}^\nu \right]^{-1} \,,
\\
X_{{\rm up}}{}^\nu &=&
{1\over e^{2i\pi\nu}+{\sin\pi(\nu+i\epsilon)\over\sin\pi(\nu-i\epsilon)}}
\left[{\sin\pi(\nu+i\epsilon)\over \sin\pi(\nu-i\epsilon)}X_C{}^\nu
-ie^{i\pi\nu}X_C^{-\nu-1}\right] \,. \label{eq:xup001}
\end{eqnarray}

The leading order of $K_\nu$ and $K_{-\nu-1}$ for $\ell\geq 2$ becomes
\begin{eqnarray}
K_\nu &\sim& O(\epsilon^{-\ell-2}) \,, \cr
K_{-\nu-1} &\sim& O(\epsilon^{\ell-2}) \,.
\end{eqnarray}
Then we find
\begin{eqnarray}
{K_{-\nu-1}X_C^{-\nu-1} \over K_\nu X_C{}^\nu} \sim
O(\epsilon^{2\ell}z^{-2\ell-1})
=O(v^{4\ell-1}) \,.
\end{eqnarray}
Therefore, we may replace $X_{{\rm in}}{}^\nu$ by $X_C{}^\nu$
to 3PN order.
As for $X_{\rm up}{}^\nu$, we find
\begin{eqnarray}
{ {\sin\pi(\nu+i\epsilon)\over\sin\pi(\nu-i\epsilon)}X_C{}^\nu
\over -ie^{i\pi\nu}X_C^{-\nu-1}} \sim
O(z^{2\ell+1})=O(v^{2\ell+1}) \,.
\end{eqnarray}
Thus, we may replace $X_{{\rm up}}{}^\nu$ by $X_C{}^{-\nu-1}$
to 2PN order.

For convenience, we define the homogeneous solutions
$\tilde X_C{}^\nu$ and $\tilde X_C^{-\nu-1}$ normalized as
\begin{eqnarray}
\tilde X_C{}^\nu(z) &=&
{\Gamma(\nu+3+i\epsilon)\Gamma(2\nu+2)\over
\Gamma(\nu-1-i\epsilon)\Gamma(\nu+1-i\epsilon)}X_C{}^\nu
\cr
&=& z^{\ell+1}\left(1+O(v)\right) \,,
\\
\tilde X_C{}^{-\nu-1}(z) &=&
{\Gamma(-\nu+2+i\epsilon)\Gamma(-2\nu)\over
\Gamma(-\nu-2-i\epsilon)\Gamma(-\nu-i\epsilon)}X_C^{-\nu-1}
\cr
&=& z^{-\ell}\left(1+O(v)\right) \,.
\end{eqnarray}
These are expanded to $O(v^2)$ as
\begin{eqnarray}
\tilde X_C{}^\nu(z) &=& z(2z)^\nu
\,\Biggl(1
-{1\over 2}{z^2\over 3+2\ell}
-{1\over 2}{(\ell-2)(\ell+2)\epsilon\over \ell z}
+O(v^3)\Biggr) \,, \\
\tilde X_C{}^{-\nu-1}(z) &=& z(2z)^{-\nu-1}
\,
\Biggl(1
+{1\over 2}{z^2\over 2\ell-1}
+{1\over 2}{(\ell -1)(\ell +3)\epsilon\over (\ell +1)z}
+O(v^3)\Biggr) \,,
\end{eqnarray}
where $\nu=\ell+O(v^6)$.

To summarize, for the in-going homogeneous solution normalized as
\begin{eqnarray}
\tilde X_{{\rm in}}{}^\nu(z) &=&
\tilde X_C{}^\nu + \beta_\nu \tilde X_C{}^{-\nu-1} \,,
\end{eqnarray}
all the coefficients $\beta_\nu$ can be set to zero
up through 3PN order,
while, for the up-going solution normalized as
\begin{eqnarray}
\tilde X_{{\rm up}}{}^\nu(z) &=&
\tilde X_C{}^{-\nu-1}+\gamma_\nu \tilde X_C{}^\nu
 \,,
\end{eqnarray}
we may put $\gamma_\nu= 0$ up through 2PN order.

\subsection{Retarded Green function}

Using the Regge-Wheeler functions
$X_{{\rm in}}{}^\nu$ and $X_{{\rm up}}{}^\nu$,
the retarded Green function is constructed as
\begin{eqnarray}
G_{{\rm ret}}{}^\nu(z,z')
&=& {1\over W(X_{{\rm in}}{}^\nu(z'),X_{{\rm up}}{}^\nu(z'))}
\left(X_{{\rm in}}{}^\nu(z) X_{{\rm up}}{}^\nu(z') \theta(z'-z)
+X_{{\rm up}}{}^\nu(z) X_{{\rm in}}{}^\nu(z') \theta(z-z')\right)
\label{eq:totalgreen1} \,, 
\end{eqnarray}
where $W(X_1,X_2)$ is the Wronskian,
\begin{eqnarray}
W(X_1(z'),X_2(z')) &\equiv&
-\left(1-{\epsilon \over z'}\right)
\left(X_1(z'){d \over dz'}X_2(z') -X_2(z'){d\over dz'}X_1(z')\right)
=\mbox{const.} \,.
\end{eqnarray}
This Green function satisfies
\begin{eqnarray}
&&\left\{\partial_z^2 +\left[
{1\over z-\epsilon}-{1\over z}\right]
\partial_z 
+ \left[1+{2\epsilon\over z-\epsilon}
+{\epsilon^2\over (z-\epsilon)^2}-{\ell(\ell+1)\over z(z-\epsilon)}
+{3\epsilon\over z^2(z-\epsilon)}\right]\right\}
G_{{\rm ret}}{}^\nu(z,z')
\nonumber \\
&&\hspace{20ex}
=- \left(1-{\epsilon \over z}\right)^{-1} \delta(z-z') \,.
\end{eqnarray}
Then the Regge-Wheeler function with the source term
$S_{\ell m \omega}^{{\rm (even/odd)}}(r)$ 
is given by
\begin{eqnarray}
R_{\ell m \omega}^{{\rm (even/odd)}}(r) = -
\int_{2M}^{\infty} dr' G_{{\rm ret}}{}^\nu(r, r')
{1\over \omega}
\left(1-{2 M \over r'}\right)^{-1} S_{\ell m \omega}^{{\rm (even/odd)}}(r')
\,.
\end{eqnarray}
Here we are only interested in the Green function accurate
to 1PN order. A numerical method to construct an accurate Green
function based on this Mano-Suzuki-Takasugi method is discussed
in Ref.~\cite{Fujita}.

\section{Spherical extension of S part}\label{app:CHD}

In this Appendix,
we consider the tensor harmonic decomposition of the S part.
First, we give the decomposition of $\epsilon^n$
where
\begin{eqnarray}
\epsilon=\left(r^2+r_0^2-2r_0r\cos{\bf\Omega\cdot\Omega}_0\right)^{1/2},
\end{eqnarray}
and ${\bf\Omega_0}$ is taken to be on the equatorial plane,
 $(\pi/2,\phi_0)$. Extending ${\bf\Omega}$ over the whole
sphere, we have
\begin{eqnarray}
{\epsilon}^p = \sum_{\ell m} {4\pi \over 2\ell+1} 
E_{\ell}^{(p)}(r,r_0)
Y_{\ell m}({\bf \Omega})Y_{\ell m}^*({\bf \Omega}_0) \,,
\label{eq:sphext}
\end{eqnarray}
where the detail derivation as well as the coefficients
$E_\ell^{(p)}$ are given in Appendix D of \cite{MNS1}.

In terms of the coefficients $E_{\ell}^{(p)}$,
the formulas needed to decompose the S part are derived as
\begin{eqnarray}
{\displaystyle \frac {1}{\epsilon }} &=&
\sum_{\ell m} {4\pi \over 2\ell+1}E_{\ell}^{(-1)}
Y_{\ell m}({\bf \Omega})Y_{\ell m}^*({\bf \Omega}_0)
\,, \cr
{\displaystyle \frac {\Phi }{\epsilon }} &=& \sum_{\ell m}
 {4\pi \over 2\ell+1}
{\displaystyle
\frac {i\,m\,E_{\ell}^{(1)}}{r_0\,r}}
Y_{\ell m}({\bf \Omega})Y_{\ell m}^*({\bf \Omega}_0)
\,, \cr
{\displaystyle \frac {\Phi ^{2}}{\epsilon }} &=& \sum_{\ell m}
 {4\pi \over 2\ell+1}\Biggl[
- {\displaystyle \frac {E_{\ell}^{(1)}}{r_0^{2}}}
- {\displaystyle \frac {1}{3}} \,{\displaystyle \frac {m^{2}\,E_{\ell}^{(3)}}
{r_0^{4}}} \Biggr]
Y_{\ell m}({\bf \Omega})Y_{\ell m}^*({\bf \Omega}_0)
\,, \cr
{\displaystyle \frac {1}{\epsilon ^{3}}} &=& \sum_{\ell m}
 {4\pi \over 2\ell+1}
E_{\ell}^{(-3)}
Y_{\ell m}({\bf \Omega})Y_{\ell m}^*({\bf \Omega}_0)
\,, \cr
{\displaystyle \frac {\Phi }{\epsilon ^{3}}} &=& \sum_{\ell m}
 {4\pi \over 2\ell+1}\Biggl[
- {\displaystyle \frac {1}{2}} \,{\displaystyle \frac
{i\,m\,E_{\ell}^{(1)}}{r_0^{3}\,r}}
- {\displaystyle \frac {1}{2}} \,{\displaystyle \frac
{i\,m\,( - R^{2} + 2\,r_0^{2})\,E_{\ell}^{(-1)}}{r_0^{3}\,r}}
- {\displaystyle \frac {1}{9}}
\,{\displaystyle \frac {i\,m^{3}\,E_{\ell}^{(3)}}{r_0^{5}\,r}} \Biggr]
Y_{\ell m}({\bf \Omega})Y_{\ell m}^*({\bf \Omega}_0)
\,, \cr
{\displaystyle \frac {\Phi ^{2}}{\epsilon ^{3}}} &=& \sum_{\ell m}
 {4\pi \over 2\ell+1}\Biggl[
{\displaystyle \frac {2}{45}} \,{\displaystyle \frac {m^{4}\,E_{\ell}^{(5)}}
{r_0^{7}\,r}}  + {\displaystyle \frac {1}{2}}
\,{\displaystyle \frac {(2\,r_0^{2}\,r + 2\,m^{2}\,r_0^{3}
- 2\,R^{2}\,m^{2}\,r - r_0^{3})\,E_{\ell}^{(1)}}{r_0^{5}\,r^{2}}}
\cr & & \mbox{}
+ {\displaystyle \frac {1}{2}} \,{\displaystyle
\frac {( - 2\,R^{2}\,r + 2\,r_0^{2}\,r + R^{2}\,r_0)\,E_{\ell}^{(-1)}}
{r_0^{3}\,r^{2}}}  + {\displaystyle \frac {m^{2}\,E_{\ell}^{(3)}}
{r_0^{5}\,r}} \Biggr]
Y_{\ell m}({\bf \Omega})Y_{\ell m}^*({\bf \Omega}_0)
\,, \cr
{\displaystyle \frac {\Phi ^{3}}{\epsilon ^{3}}} &=& \sum_{\ell m}
 {4\pi \over 2\ell+1}\Biggl[
3\,{\displaystyle \frac {i\,m\,E_{\ell}^{(1)}}{r_0^{4}}}
+ {\displaystyle \frac {1}{3}} \,{\displaystyle \frac {i\,m^{3}\,E_{\ell}^{(3)}}
{r_0^{6}}} \Biggr]
Y_{\ell m}({\bf \Omega})Y_{\ell m}^*({\bf \Omega}_0)
\,, \cr
{\displaystyle \frac {\Phi ^{4}}{\epsilon ^{3}}} &=& \sum_{\ell m}
 {4\pi \over 2\ell+1}\Biggl[
- {\displaystyle \frac {1}{15}} \,{\displaystyle \frac {m^{4}\,E_{\ell}^{(5)}}
{r_0^{8}}}  - 3\,{\displaystyle \frac {E_{\ell}^{(1)}}{r_0^{4}}}
- 2\,{\displaystyle \frac {m^{2}\,E_{\ell}^{(3)}}{r_0^{6}}} \Biggr]
Y_{\ell m}({\bf \Omega})Y_{\ell m}^*({\bf \Omega}_0)
\,, \cr
{\displaystyle \frac {1}{\epsilon ^{5}}} &=& \sum_{\ell m}
 {4\pi \over 2\ell+1}
E_{\ell}^{(-5)}
Y_{\ell m}({\bf \Omega})Y_{\ell m}^*({\bf \Omega}_0)
\,, \cr
{\displaystyle \frac {\Phi ^{2}}{\epsilon ^{5}}} &=& \sum_{\ell m}
 {4\pi \over 2\ell+1}\Biggl[
{\displaystyle \frac {1}{6}} \,{\displaystyle \frac
{( - 3\,R^{2}\,r + 2\,R\,r_0\,r
- 4\,R\,r_0^{2} + 2\,r_0^{2}\,r)\,E_{\ell}^{(-3)}}{r_0^{4}\,r}}
- {\displaystyle \frac {1}{3}} \,{\displaystyle \frac {m^{2}\,E_{\ell}^{(1)}}
{r_0^{6}}}
\cr & & \mbox{}
- {\displaystyle \frac {1}{6}} \,{\displaystyle \frac
{( - r_0^{2}\,r^{2} - 4\,R\,m^{2}\,r_0^{3} - 4\,R^{2}\,m^{2}\,r^{2}
+ 2\,m^{2}\,r_0^{2}\,r^{2})\,E_{\ell}^{(-1)}}{r_0^{6}\,r^{2}}}
- {\displaystyle \frac {2}{27}} \,
{\displaystyle \frac {m^{4}\,E_{\ell}^{(3)}}{r_0^{8}}} \Biggr]
Y_{\ell m}({\bf \Omega})Y_{\ell m}^*({\bf \Omega}_0)
\,, \cr
{\displaystyle \frac {\Phi ^{4}}{\epsilon ^{5}}} &=& \sum_{\ell m}
 {4\pi \over 2\ell+1}\Biggl[
2\,{\displaystyle \frac {m^{2}\,E_{\ell}^{(1)}}{r_0^{10}}}
+ {\displaystyle \frac {E_{\ell}^{(-1)}}{r_0^{4}}}
+ {\displaystyle \frac {1}{9}} \,{\displaystyle \frac {m^{4}\,E_{\ell}^{(3)}}
{r_0^{8}}} \Biggr]
Y_{\ell m}({\bf \Omega})Y_{\ell m}^*({\bf \Omega}_0)
\,, \cr
{\displaystyle \frac {\Theta }{\epsilon }} &=& \sum_{\ell m}
 {4\pi \over 2\ell+1}\Biggl[
- {\displaystyle \frac {E_{\ell}^{(1)}}{r_0^{2}}} \Biggr]
Y_{\ell m}({\bf \Omega})\partial_{\theta}Y_{\ell m}^*({\bf \Omega}_0)
\,, \cr
{\displaystyle \frac {\Theta }{\epsilon ^{3}}} &=& \sum_{\ell m}
 {4\pi \over 2\ell+1}\Biggl[
- {\displaystyle \frac {1}{2}} \,{\displaystyle \frac {E_{\ell}^{(1)}}
{r_0^{3}\,r}}
+ {\displaystyle \frac {1}{6}} \,{\displaystyle \frac
{( - R^{2} + 6\,r_0^{2})\,E_{\ell}^{(-1)}}
{r_0^{3}\,r}}
- {\displaystyle \frac {1}{9}} \,{\displaystyle \frac {m^{2}\,E_{\ell}^{(3)}}
{r_0^{5}\,r}} \Biggr]
Y_{\ell m}({\bf \Omega})\partial_{\theta}Y_{\ell m}^*({\bf \Omega}_0)
\,, \cr
{\displaystyle \frac {\Theta \,\Phi }{\epsilon ^{3}}} &=& \sum_{\ell m}
 {4\pi \over 2\ell+1}
{\displaystyle \frac {i\,m\,E_{\ell}^{(1)}}{r_0^{2}\,r^{2}}}
Y_{\ell m}({\bf \Omega})\partial_{\theta}Y_{\ell m}^*({\bf \Omega}_0)
\,, \cr
{\displaystyle \frac {\Theta \,\Phi ^{2}}{\epsilon ^{3}}} &=& \sum_{\ell m}
 {4\pi \over 2\ell+1}\Biggl[
- {\displaystyle \frac {E_{\ell}^{(1)}}{r_0^{4}}}
- {\displaystyle \frac {1}{3}} \,{\displaystyle \frac {m^{2}\,E_{\ell}^{(3)}}
{r_0^{6}}} \Biggr]
Y_{\ell m}({\bf \Omega})\partial_{\theta}Y_{\ell m}^*({\bf \Omega}_0)
\,, \cr
{\displaystyle \frac {\Theta \,\Phi }{\epsilon ^{5}}} &=& \sum_{\ell m}
 {4\pi \over 2\ell+1}\Biggl[
- {\displaystyle \frac {1}{6}} \,{\displaystyle \frac
{i\,m\,(2\,r_0^{2} + 3\,r^{2} - 4\,r_0\,r)\,E_{\ell}^{(1)}}
{r_0^{4}\,r^{4}}}  + {\displaystyle \frac {1}{9}}
\,{\displaystyle \frac {i\,m\,( - 3\,r_0^{3} + 2\,R^{2}\,r)\,E_{\ell}^{(-1)}}
{r_0^{5}\,r^{2}}}
\cr & &
- {\displaystyle \frac {1}{18}}
\,{\displaystyle \frac {i\,m^{3}\,E_{\ell}^{(3)}}{r_0^{6}\,r^{2}}} \Biggr]
Y_{\ell m}({\bf \Omega})\partial_{\theta}Y_{\ell m}^*({\bf \Omega}_0)
\,, \cr
{\displaystyle \frac {\Theta \,\Phi ^{3}}{\epsilon ^{5}}} &=& \sum_{\ell m}
 {4\pi \over 2\ell+1}\Biggl[
{\displaystyle \frac {i\,m\,E_{\ell}^{(1)}}{r_0^{3}\,r^{3}}}
 + {\displaystyle \frac {1}{9}} \,{\displaystyle \frac
{i\,m^{3}\,E_{\ell}^{(3)}}{r_0^{4}\,r^{4}}} \Biggr]
Y_{\ell m}({\bf \Omega})\partial_{\theta}Y_{\ell m}^*({\bf \Omega}_0)
\,.
\end{eqnarray}
Note that these formulas are valid only in the sense of the
spherical extension given by Eq.~(\ref{eq:sphext}).

\section{$\bm{O(y^2)}$-correction} \label{app:y2}

In this Appendix, as an example to clarify how the standard form
is recovered and why it is necessary to include
the $\ell=0$, $1$ modes even if some of the tensor harmonics
are identically zero for these modes,
we consider the S part of the
metric components $h_{t\theta}$ and $h_{t\phi}$
and analyze the contribution of their $O(y^2)$ terms to 
the self-force in the harmonic gauge. 

These metric components give rise to 
the coefficient $h_{0\,\ell m}^{\rm (e)}$
of the vector harmonic proportional to
$(\partial_\theta Y_{\ell m},\partial_\phi Y_{\ell m})$.
Note that this vanishes identically for $\ell=0$.
Since the contribution of the $O(y^2)$ terms
to the self-force is zero, 
the sum of $h_{0\,\ell m}^{\rm (e)}$ over $\ell$
should vanish. We show that it indeed has the
standard form for general $\ell$. However, to
guarantee that the sum over $\ell$ is zero, it
is necessary to include the contribution from
$\ell=0$ as well. This implies $B^\mu+D^\mu_\ell=0$
for $\ell=0$ as discussed at the end of Section~\ref{sec:dir}.

The local expansion of the S part of 
the metric components $h_{t\theta}$ and $h_{t\phi}$
takes the form
\begin{eqnarray}
{}^{(2)}{h_{t\theta}^{\rm S,H}}
=
\left[
{\displaystyle \frac {\Phi^{2n+2}\Theta}{\epsilon^{2n+1} }},\,
{\displaystyle \frac {\Phi^{2n+1}R\,\Theta}{\epsilon^{2n+1} }}
\right] \,, 
\quad
{}^{(2)}{h_{t\phi}^{\rm S,H}} 
=
\left[
{\displaystyle \frac {\Phi^{2n+3}}{\epsilon^{2n+1} }},\,
{\displaystyle \frac {\Phi^{2n+2}R}{\epsilon^{2n+1} }}
\right] \,,
\label{eq:y2term}
\end{eqnarray}
where we have retained only terms that may contribute
to the self-force, and the superscript ${}^{(2)}$ means $O(y^2)$. 
The tensor harmonic coefficients $h_{0\,\ell m}^{(e)}$ 
are given by
\begin{eqnarray}
h_{0\,\ell m}^{\rm (e) S,H}(t,\,r) 
= {-1 \over \ell(\ell+1)} \int \left(
h_{t\theta} \partial_{\theta}Y_{\ell m}^*(\theta, \,\phi )
+h_{t\phi} \partial_{\phi}Y_{\ell m}^*(\theta, \,\phi )
\right) d{\bf \Omega} \,,
\end{eqnarray}
where $\ell\neq0$. For the $O(y^2)$ terms of the form
(\ref{eq:y2term}), we have
\begin{eqnarray}
{}^{(2)}h_{0\ell m}^{\rm (e) S,H}(t,\,r) 
= {-1 \over \ell(\ell+1)} \int 
\left[
{\displaystyle \frac {\Phi^{2n+2}}{\epsilon^{2n+1} }},\,
{\displaystyle \frac {\Phi^{2n+1}R}{\epsilon^{2n+1} }}
\right]
Y_{\ell m}^*(\theta, \,\phi ) d{\bf \Omega} \,.
\label{eq:y2h0}
\end{eqnarray}
The force is given by
\begin{eqnarray}
F^{r(\pm)}_{{\rm S,H}}[h_{0}^{\rm (e) S,H}]
&=& \sum_{\ell m} {\mu\,(r_0-2\,M) \over 2\,r_0^3\,(r_0-3\,M)} 
\left(-2\,i\,m\,\Omega\,\partial_r h_{0\ell m}^{\rm (e) S,H}(t_0,\,r_0) \right)
Y_{\ell m}(\pi/2, \,0 ) \,;
\quad
\Omega = {u^{\phi} \over u^t} \,. 
\label{eq:fh0e}
\end{eqnarray}

Here, since the terms of interest are already of $O(y^2)$,
we may use the leading order formulas for the spherical
extension of the local coordinate expansion~\cite{MNS1}. 
We have
\begin{eqnarray}
\epsilon^{2n-1} &\sim& 
\sum_{\ell m}{2\pi \over L}
{\kappa_n \over (L^2-1)(L^2-2^2)\cdot\cdot\cdot(L^2-n^2)}
\left({r_< \over r_>}\right)^{\ell}Y_{\ell m}(\theta, \,\phi )
Y_{\ell m}^*(\pi/2, 0 )
\,, \\
{\Phi \over \epsilon} &\sim& \partial_{\phi} \epsilon \,,
\end{eqnarray}
where $n\geq1$, $r_{>}={\rm max}\{r,r_0\}$, $r_{<}={\rm min}\{r,r_0\}$, 
$L=\ell+1/2$ and $\kappa_n$ is a constant 
independent of $L$. 
Therefore, Eq.~(\ref{eq:y2h0}) is evaluated as
\begin{eqnarray}
{-1 \over \ell(\ell+1)} \int 
{\displaystyle \frac {\Phi^{2n+2}}{\epsilon^{2n+1} }}
Y_{\ell m}^*(\theta, \,\phi ) d{\bf \Omega} &\sim& 
{1 \over {\cal L}^{(2)}} \int 
\partial_{\phi}^{2n+2}\epsilon^{2n+3} 
Y_{\ell m}^*(\theta, \,\phi ) d{\bf \Omega} \nonumber \\ 
&& \hspace{-2cm} 
\sim {2\pi \over L}{1 \over {\cal L}^{(2)}}
{m^{2n+2} \over (L^2-1)(L^2-2^2)\cdot\cdot\cdot(L^2-(n+2)^2)}
\left({r_< \over r_>}\right)^{\ell}
Y_{\ell m}^*(\pi/2, 0 ) \,, \cr
{-1 \over \ell(\ell+1)} \int 
{\displaystyle \frac {\Phi^{2n+1}R}{\epsilon^{2n+1} }}
Y_{\ell m}^*(\theta, \,\phi ) d{\bf \Omega} &\sim& 
{1 \over {\cal L}^{(2)}} \int 
R\,\partial_{\phi}^{2n+1}\epsilon^{2n+1} 
Y_{\ell m}^*(\theta, \,\phi ) d{\bf \Omega} \nonumber \\ 
&& \hspace{-2cm} 
\sim {2\pi \over L}{1 \over {\cal L}^{(2)}}
{m^{2n+1}(r-r_0) \over (L^2-1)(L^2-2^2)\cdot\cdot\cdot(L^2-(n+1)^2)}
\left({r_< \over r_>}\right)^{\ell}
Y_{\ell m}^*(\pi/2, 0 ) \,,
\end{eqnarray}
where $n\geq0$ and
\begin{eqnarray}
{\cal L}^{(2)} = \ell(\ell+1) = \left(L^2-{1\over 4}\right)
\,. 
\end{eqnarray}
Using Eq.~(\ref{eq:fh0e}), and retaining only the terms
that will remain after summing over $m$, we have
\begin{eqnarray}
F^{r(\pm)}_{{\rm S,H}}[h_{0}^{\rm (e) S,H}]
&\sim& \sum_{\ell m} {2\pi \over L}{1 \over {\cal L}^{(2)}}
{m^{2n+2} \over (L^2-1)(L^2-2^2)\cdot\cdot\cdot(L^2-(n+1)^2)}
|Y_{\ell m}(\pi/2, \,0 )|^2 \,.
\end{eqnarray}
The $m$ summation gives
\begin{eqnarray}
\sum_m {2\,\pi \over L} {m^{2n+2} \over {\cal L}^{(2)}}
\,|Y_{\ell m}(\pi/2, \,0 )|^2
&=& \sum_{k=0}^{n} \alpha_k L^{2k} \,. 
\end{eqnarray}
Thus, the $O(y^2)$ terms contribute to the $D^\mu_\ell$ term
in the form of the standard form, and the sum over $\ell$
vanishes provided we include the $\ell=0$ term in the summation.
Since the $O(y^2)$ terms do not contribute to the force
anyway, it then follows that we may adjust the numerators
of the $D^\mu_\ell$ term so as to give $D^\mu_0=-B^\mu$.

\section{Calculation of the S part} \label{app:OS}

In this Appendix, we show the S part of the metric perturbation 
and its gauge transformaton. 
The S part of the metric perturbation 
under the harmonic gauge are obtained in the local coordinate 
expansion as 
\begin{eqnarray}
{h_{tt}^{\rm S,H}} &=&
\mu \Biggl[
2\,{\displaystyle \frac {1}{\epsilon }}
+ \biggl(
 + 2\,{\displaystyle \frac {\Phi \,T\,{r_0}^{2}}{\epsilon ^{3}}}
+ {\displaystyle \frac {\Phi \,T\,R^{2}}{\epsilon ^{3}}}
- {\displaystyle \frac {\Phi \,T}{\epsilon }}
+ {\displaystyle \frac {2}{3}}
\,{\displaystyle \frac {\Phi ^{3}\,T\,{r_0}^{2}}{\epsilon ^{3}}}
+ 2\,{\displaystyle \frac {\Phi \,T\,{r_0}\,R}{\epsilon ^{3}}}
\biggr)\,{u^{\phi}}
\nonumber \\
&&
+ \biggl(
-{\displaystyle \frac {1}{2}}
\,{\displaystyle \frac {\epsilon }{{r_0}^{3}}}
- {\displaystyle \frac {R\,T^{2}}{\epsilon ^{3}\,{r_0}^{2}}}
- 4\,{\displaystyle \frac {1}{\epsilon \,{r_0}}}
- 2\,{\displaystyle \frac {R^{2}}{\epsilon ^{3}\,{r_0}}}
+ {\displaystyle \frac {R^{3}}{\epsilon ^{3}\,{r_0}^{2}}}
- {\displaystyle \frac {R^{4}}{\epsilon ^{3}\,{r_0}^{3}}}
- {\displaystyle \frac {1}{2}}
\,{\displaystyle \frac {R^{2}\,T^{2}}{\epsilon ^{3}\,{r_0}^{3}}}
+ {\displaystyle \frac {1}{2}}
\,{\displaystyle \frac {T^{2}}{\epsilon \,{r_0}^{3}}}
\nonumber \\
 & & \quad
- {\displaystyle \frac {5}{2}}
\,{\displaystyle \frac {R^{2}}{\epsilon \,{r_0}^{3}}}
+ 4\,{\displaystyle \frac {R}{\epsilon \,{r_0}^{2}}}
\biggr)M
\nonumber \\
&&
- \biggl(
 - 2\,{\displaystyle \frac {{r_0}^{4}\,\Phi ^{4}\,T^{2}}
{\epsilon ^{5}}}
+ {\displaystyle \frac {{r_0}^{2}\,T^{2}}{\epsilon ^{3}}}
- {\displaystyle \frac {{r_0}^{2}\,\Phi ^{2}}{\epsilon }}
- 3\,{\displaystyle \frac {{r_0}^{4}\,\Phi ^{2}\,T^{2}}
{\epsilon ^{5}}}
- 6\,{\displaystyle \frac {{r_0}^{2}\,\Phi ^{2}\,R^{2}\,T^{2}}
{\epsilon ^{5}}}
\nonumber \\
 & & \quad
+ {\displaystyle \frac {2}{3}}
\,{\displaystyle \frac {{r_0}^{4}\,\Phi ^{4}}{\epsilon ^{3}}}
- 4\,{\displaystyle \frac {{r_0}^{2}}{\epsilon }}
- 6\,{\displaystyle \frac {{r_0}^{3}\,\Phi ^{2}\,T^{2}\,R}{\epsilon ^{5}}}
+ 3\,{\displaystyle \frac {{r_0}^{2}\,\Phi ^{2}\,T^{2}}{\epsilon ^{3}}}
+ 2\,{\displaystyle \frac {{r_0}^{3}\,\Phi ^{2}\,R}{\epsilon ^{3}}}
\nonumber \\
 & & \quad
+ 2\,{\displaystyle \frac {{r_0}^{2}\,\Phi ^{2}\,R^{2}}{\epsilon ^{3}}}
+ {\displaystyle \frac {{r_0}^{4}\,\Phi ^{2}}{\epsilon ^{3}}}
\biggr)({u^{\phi}})^{2} \Biggr] + O(y^2)
 \,, 
\cr
{h_{tr}^{\rm S,H}} = {h_{rt}^{\rm S,H}} &=&
\mu \Biggl[
-4\,{\displaystyle \frac {\Phi \,{r_0}}{\epsilon }}\,{u^{\phi}}
+\biggl(
4\,{\displaystyle \frac {T}{\epsilon \,{r_0}^{2}}}
- 4\,{\displaystyle \frac {T\,R}{\epsilon \,{r_0}^{3}}}
\biggr)\,M
+ \biggl(
-4\,{\displaystyle \frac {{r_0}^{3}\,\Phi ^{2}\,T}{\epsilon ^{3}}}
- 4\,{\displaystyle \frac {{r_0}^{2}\,\Phi ^{2}\,T\,R}{\epsilon ^{3}}}
\biggr)\,({u^{\phi}})^{2} \Biggr] + O(y^2)
 \,,
\cr
{h_{t\theta}^{\rm S,H}} = {h_{\theta t}^{\rm S,H}} &=&
\mu \Biggl[
- 2\,{\displaystyle \frac {T\,M\,\Theta }{\epsilon \,{r_0}}}
+ 4\,{\displaystyle \frac {{r_0}^{2}\,\Phi \,\Theta }{\epsilon }}\,{u^{\phi}}
+ 4\,{\displaystyle \frac {{r_0}^{4}\,\Phi ^{2}\,T\,\Theta }
{\epsilon ^{3}}} ({u^{\phi}})^{2} \Biggr] + O(y^2)
\,,
\cr
{h_{t\phi}^{\rm S,H}} = {h_{\phi t}^{\rm S,H}} &=&
\mu \Biggl[
\biggl(
  2\,\epsilon  - 2\,{\displaystyle \frac {R^{2}}{\epsilon }}
- 4\,{\displaystyle \frac {{r_0}\,R}{\epsilon }}
- 4\,{\displaystyle \frac {{r_0}^{2}}{\epsilon }}
\biggr)\,{u^{\phi}}
- 2\,{\displaystyle \frac {\Phi \,T}{\epsilon \,{r_0}}}\,M
\nonumber \\
 & & +
\biggl(
 4\,{\displaystyle \frac {{r_0}^{2}\,\Phi \,T}{\epsilon }}
- 8\,{\displaystyle \frac {{r_0}^{2}\,\Phi \,T\,R^{2}}{\epsilon ^{3}}}
- 8\,{\displaystyle \frac {{r_0}^{3}\,\Phi \,T\,R}{\epsilon ^{3}}}
- 4\,{\displaystyle \frac {{r_0}^{4}\,\Phi \,T}{\epsilon ^{3}}}
- {\displaystyle \frac {4}{3}}
\,{\displaystyle \frac {{r_0}^{4}\,\Phi ^{3}\,T}{\epsilon ^{3}}}
\biggr)({u^{\phi}})^{2} \Biggr] + O(y^2)
\,,
\cr
{h_{rr}^{\rm S,H}} &=&
\mu \Biggl[
 2\,{\displaystyle \frac {1}{\epsilon }}
+ \biggl(
  2\,{\displaystyle \frac {\Phi \,T\,{r_0}^{2}}{\epsilon ^{3}}}
+ {\displaystyle \frac {\Phi \,T\,R^{2}}{\epsilon ^{3}}}
- {\displaystyle \frac {\Phi \,T}{\epsilon }}
+ {\displaystyle \frac {2}{3}}
\,{\displaystyle \frac {\Phi ^{3}\,T\,{r_0}^{2}}{\epsilon ^{3}}}
+ 2\,{\displaystyle \frac {\Phi \,T\,{r_0}\,R}{\epsilon ^{3}}}
\biggr)\,{u^{\phi}}
\nonumber \\ &&
+ \biggl(
-{\displaystyle \frac {17}{2}} \,{\displaystyle \frac {\epsilon }{{r_0}^{3}}}
- 4\,{\displaystyle \frac {R}{\epsilon \,{r_0}^{2}}}
- {\displaystyle \frac {R\,T^{2}}{\epsilon ^{3}\,{r_0}^{2}}}
+ 4\,{\displaystyle \frac {1}{\epsilon \,{r_0}}}
- 2\,{\displaystyle \frac {R^{2}}{\epsilon ^{3}\,{r_0}}}
+ {\displaystyle \frac {R^{3}}{\epsilon ^{3}\,{r_0}^{2}}}
- {\displaystyle \frac {R^{4}}{\epsilon ^{3}\,{r_0}^{3}}}
- {\displaystyle \frac {1}{2}}
\,{\displaystyle \frac {R^{2}\,T^{2}}{\epsilon ^{3}\,{r_0}^{3}}}
\nonumber \\
 & & \quad + {\displaystyle \frac {1}{2}}
\,{\displaystyle \frac {T^{2}}{\epsilon \,{r_0}^{3}}}
+ {\displaystyle \frac {11}{2}} \,{\displaystyle \frac {R^{2}}
{\epsilon \,{r_0}^{3}}}
\biggr)M
\nonumber \\ &&
- \biggl(
- 2\,{\displaystyle \frac {{r_0}^{4}\,\Phi ^{4}\,T^{2}}{\epsilon ^{5}}}
- 5\,{\displaystyle \frac {{r_0}^{2}\,\Phi ^{2}}{\epsilon }}
- 3\,{\displaystyle \frac {{r_0}^{4}\,\Phi ^{2}\,T^{2}}{\epsilon ^{5}}}
- 6\,{\displaystyle \frac {{r_0}^{2}\,\Phi ^{2}\,R^{2}\,T^{2}}{\epsilon ^{5}}}
+ {\displaystyle \frac {{r_0}^{2}\,T^{2}}{\epsilon ^{3}}}
\nonumber \\
 & & \quad
+ {\displaystyle \frac {2}{3}}
\,{\displaystyle \frac {{r_0}^{4}\,\Phi ^{4}}{\epsilon ^{3}}}
- 6\,{\displaystyle \frac {{r_0}^{3}\,\Phi ^{2}\,T^{2}\,R}{\epsilon ^{5}}}
+ 3\,{\displaystyle \frac {{r_0}^{2}\,\Phi ^{2}\,T^{2}}{\epsilon ^{3}}}
+ 2\,{\displaystyle \frac {{r_0}^{3}\,\Phi ^{2}\,R}{\epsilon ^{3}}}
+ 2\,{\displaystyle \frac {{r_0}^{2}\,\Phi ^{2}\,R^{2}}{\epsilon ^{3}}}
\nonumber \\
 & & \quad
+ {\displaystyle \frac {{r_0}^{4}\,\Phi ^{2}}{\epsilon ^{3}}}
\biggr)({u^{\phi}})^{2} \Biggr] + O(y^2)
 \,,
\cr
{h_{r\theta}^{\rm S,H}} = {h_{\theta r}^{\rm S,H}} &=& O(y^2)
\,,
\cr
{h_{r\phi}^{\rm S,H}} = {h_{\phi r}^{\rm S,H}} &=&
\mu \biggl(
4\,{\displaystyle \frac {{r_0}^{2}\,\Phi \,R}{\epsilon }}
+4\,{\displaystyle \frac {{r_0}^{3}\,\Phi }{\epsilon }}
\biggr)({u^{\phi}})^{2} + O(y^2)
\,,
\cr
{h_{\theta\theta}^{\rm S,H}} &=&
\mu \Biggl[
 2\,{\displaystyle \frac {{r_0}^{2}}{\epsilon }}
+ 4\,{\displaystyle \frac {{r_0}\,R}{\epsilon }}
+ 2\,{\displaystyle \frac {R^{2}}{\epsilon }}
\nonumber \\ &&
+ \biggl(
-{\displaystyle \frac {{r_0}^{2}\,\Phi \,T}{\epsilon }}
+ 7\,{\displaystyle \frac {{r_0}^{2}\,\Phi \,T\,R^{2}}{\epsilon ^{3}}}
+ 6\,{\displaystyle \frac {{r_0}^{3}\,\Phi \,T\,R}{\epsilon ^{3}}}
+ {\displaystyle \frac {2}{3}}
\,{\displaystyle \frac {{r_0}^{4}\,\Phi ^{3}\,T}{\epsilon ^{3}}}
+ 2\,{\displaystyle \frac {{r_0}^{4}\,\Phi \,T}{\epsilon ^{3}}}
\biggr)\,{u^{\phi}}
\nonumber \\ &&
+ \biggl(
-{\displaystyle \frac {5}{2}}
\,{\displaystyle \frac {R^{2}\,T^{2}}{\epsilon ^{3}\,{r_0}}}
+ {\displaystyle \frac {1}{2}} \,{\displaystyle \frac {T^{2}}
{\epsilon \,{r_0}}}
- {\displaystyle \frac {R^{4}}{\epsilon ^{3}\,{r_0}}}
- 2\,{\displaystyle \frac {{r_0}\,R^{2}}{\epsilon ^{3}}}
- {\displaystyle \frac {R\,T^{2}}{\epsilon ^{3}}}
- 3\,{\displaystyle \frac {R^{3}}{\epsilon ^{3}}}
+ {\displaystyle \frac {7}{2}} \,{\displaystyle \frac {\epsilon }{{r_0}}}
+ {\displaystyle \frac {3}{2}}
\,{\displaystyle \frac {R^{2}}{\epsilon \,{r_0}}}
\biggr)\,M
\nonumber \\
 & &
+ \biggl(
 12\,{\displaystyle \frac {{r_0}^{5}\,\Phi ^{2}\,T^{2}\,R}{\epsilon ^{5}}}
+ 2\,{\displaystyle \frac {{r_0}^{6}\,\Phi ^{4}\,T^{2}}{\epsilon ^{5}}}
+ 3\,{\displaystyle \frac {{r_0}^{6}\,\Phi ^{2}\,T^{2}}{\epsilon ^{5}}}
- {\displaystyle \frac {{r_0}^{6}\,\Phi ^{2}}{\epsilon ^{3}}}
- {\displaystyle \frac {2}{3}}
\,{\displaystyle \frac {{r_0}^{6}\,\Phi ^{4}}{\epsilon ^{3}}}
\nonumber \\
 & & \quad
+ {\displaystyle \frac {{r_0}^{4}\,\Phi ^{2}}{\epsilon }}
- {\displaystyle \frac {{r_0}^{2}\,R^{2}\,T^{2}}{\epsilon ^{3}}}
- 2\,{\displaystyle \frac {{r_0}^{3}\,R\,T^{2}}{\epsilon ^{3}}}
- {\displaystyle \frac {{r_0}^{4}\,T^{2}}{\epsilon ^{3}}}
- 4\,{\displaystyle \frac {{r_0}^{5}\,\Phi ^{2}\,R}{\epsilon ^{3}}}
\nonumber \\
& & \quad
+ 21\,{\displaystyle \frac {{r_0}^{4}\,\Phi ^{2}\,R^{2}\,T^{2}}
{\epsilon ^{5}}}
- 3\,{\displaystyle \frac {{r_0}^{4}\,\Phi^{2}\,T^{2}}{\epsilon ^{3}}}
- 7\,{\displaystyle \frac {{r_0}^{4}\,\Phi ^{2}\,R^{2}}{\epsilon ^{3}}}
\biggr)({u^{\phi}})^{2} \Biggr] + O(y^2)
\,,
\cr
{h_{\theta\phi}^{\rm S,H}} = {h_{\phi\theta}^{\rm S,H}} &=&
-4\,\mu\,{\displaystyle \frac {{r_0}^{4}\,\Phi \,\Theta }{\epsilon }}
({u^{\phi}})^{2} + O(y^2)\,,
\cr
{h_{\phi\phi}^{\rm S,H}} &=&
\mu \Biggl[
-2\,\epsilon
+ 2\,{\displaystyle \frac {{r_0}^{2}}{\epsilon }}
+ 2\,{\displaystyle \frac {{r_0}^{2}\,\Phi ^{2}}{\epsilon }}
+ 4\,{\displaystyle \frac {{r_0}\,R}{\epsilon }}
+ 4\,{\displaystyle \frac {R^{2}}{\epsilon }}
\nonumber \\ &&
+ \biggl(
 {\displaystyle \frac {8}{3}}
\,{\displaystyle \frac {{r_0}^{4}\,\Phi ^{3}\,T}{\epsilon ^{3}}}
- 3\,{\displaystyle \frac {{r_0}^{2}\,\Phi \,T}{\epsilon }}
+ 9\,{\displaystyle \frac {{r_0}^{2}\,\Phi \,T\,R^{2}}{\epsilon ^{3}}}
+ 6\,{\displaystyle \frac {{r_0}^{3}\,\Phi \,T\,R}{\epsilon ^{3}}}
+ 2\,{\displaystyle \frac {{r_0}^{4}\,\Phi \,T}{\epsilon ^{3}}}
\biggr)\,{u^{\phi}}
\nonumber \\ &&
+\biggl(
-2\,{\displaystyle \frac {{r_0}\,\Phi ^{2}\,R^{2}}{\epsilon ^{3}}}
- {\displaystyle \frac {5}{2}}
\,{\displaystyle \frac {R^{2}\,T^{2}}{\epsilon ^{3}\,{r_0}}}
+ {\displaystyle \frac {1}{2}}
\,{\displaystyle \frac {T^{2}}{\epsilon \,{r_0}}}
- 3\,{\displaystyle \frac {R^{4}}{\epsilon ^{3}\,{r_0}}}
- 2\,{\displaystyle \frac {{r_0}\,R^{2}}{\epsilon ^{3}}}
- {\displaystyle \frac {R\,T^{2}}{\epsilon ^{3}}}
- 3\,{\displaystyle \frac {R^{3}}{\epsilon ^{3}}}
\nonumber \\
 & & \quad
+ {\displaystyle \frac {7}{2}}
\,{\displaystyle \frac {\epsilon }{{r_0}}}
+ {\displaystyle \frac {7}{2}}
\,{\displaystyle \frac {R^{2}}{\epsilon \,{r_0}}}
\biggr)M
\nonumber \\ &&
+ \biggl(
 4\,{\displaystyle \frac {{r_0}^{4}}{\epsilon }}
+ {\displaystyle \frac {{r_0}^{2}\,T^{2}}{\epsilon }}
- 4\,{r_0}^{2}\,\epsilon
+ 8\,{\displaystyle \frac {{r_0}^{3}\,R}{\epsilon }}
+ 8\,{\displaystyle \frac {{r_0}^{2}\,R^{2}}{\epsilon }}
\nonumber \\
 & & \quad
- {\displaystyle \frac {{r_0}^{4}\,T^{2}}{\epsilon ^{3}}}
- 4\,{\displaystyle \frac {{r_0}^{5}\,\Phi ^{2}\,R}{\epsilon ^{3}}}
+ 24\,{\displaystyle \frac {{r_0}^{4}\,\Phi ^{2}\,R^{2}\,T^{2}}
{\epsilon ^{5}}}
- 7\,{\displaystyle \frac {{r_0}^{4}\,\Phi ^{2}\,T^{2}}{\epsilon ^{3}}}
- 8\,{\displaystyle \frac {{r_0}^{4}\,\Phi ^{2}\,R^{2}}{\epsilon ^{3}}}
\nonumber \\
 & & \quad
+ 12\,{\displaystyle \frac {{r_0}^{5}\,\Phi^{2}\,T^{2}\,R}{\epsilon ^{5}}}
+ 5\,{\displaystyle \frac {{r_0}^{6}\,\Phi ^{4}\,T^{2}}{\epsilon ^{5}}}
+ 3\,{\displaystyle \frac {{r_0}^{6}\,\Phi ^{2}\,T^{2}}{\epsilon ^{5}}}
- {\displaystyle \frac {{r_0}^{6}\,\Phi ^{2}}{\epsilon ^{3}}}
- {\displaystyle \frac {5}{3}}
\,{\displaystyle \frac {{r_0}^{6}\,\Phi ^{4}}{\epsilon ^{3}}}
\nonumber \\
 & & \quad
+ 2\,{\displaystyle \frac {{r_0}^{4}\,\Phi ^{2}}{\epsilon }}
- 2\,{\displaystyle \frac {{r_0}^{2}\,R^{2}\,T^{2}}{\epsilon ^{3}}}
- 2\,{\displaystyle \frac {{r_0}^{3}\,R\,T^{2}}{\epsilon ^{3}}}
\biggr)({u^{\phi}})^{2} \Biggr]+ O(y^2)
\,.
\label{eq:SlocalH}
\end{eqnarray}

The harmonic coefficients of the above S part are calculated as 
\begin{eqnarray}
h_{0\ell m}^{\rm S,H}(t,\,r)
&=& {2 \over L}\,\pi\, \mu \biggl[
{\displaystyle \frac {4\,i\,T\,m\,
r_0\,(L^{2} - 2)\,(u^{\phi})^2}
{{\cal L}^{(2)}\,(L^2 - 1)}}  
\nonumber \\ && \qquad
- (8\,r_0 - 6\,r_0\,m^{2} - 18\,r_0\,L^{2}
+ 4\,r_0\,L^{4} - 4\,R + 16\,R\,L - 13\,R\,m^{2} 
 \nonumber \\ && \qquad
- 7\,R\,L^{2} - 20\,R\,L^{3} + 2\,R\,L^{4} + 4\,R\,L^{5})u^{\phi}
\nonumber \\ && \qquad
/({\cal L}^{(2)}\,(L^2- 1)\,(L^2 - 4)) 
\biggr] \partial_{\theta}\,Y_{\ell m}^*(\theta_0,\,\phi_0) 
\,,
\cr
h_{1\ell m}^{\rm S,H}(t,\,r)
&=& {2 \over L}\,\pi\,\mu \biggl[
{\displaystyle \frac {-2\,i\,r_0\,
m\,(2\,r_0 + R)\,(u^{\phi})^2}
{{\cal L}^{(2)}\,(L^2 - 1)}} 
\biggr] \partial_{\theta}\,Y_{\ell m}^*(\theta_0,\,\phi_0) \,,
\cr
h_{2\ell m}^{\rm S,H}(t,\,r)
&=& {2 \over L}\,\pi\,\mu \biggl[
- {\displaystyle \frac {1}{6}} r_0\,
m\,(72\,r_0\,R\,L^{4} + 48\,r_0\,R\,L^{5} 
- 240\,r_0\,R\,L^{3} - 288\,r_0\,R\,m^{2} + 4108\,R^{2} 
\nonumber \\ && \qquad
\mbox{} + 1056\,r_0^{2} - 240\,R^{2}\,L^{3} - 1488\,r_0\,R 
+ 1392\,R^{2}\,m^{2} + 24\,R^{2}\,L^{6} - 1147\,R^{2}\,L^{2} 
\nonumber \\ && \qquad
\mbox{} + 192\,R^{2}\,L - 66\,R^{2}\,L^{4} + 48\,R^{2}\,L^{5} 
+ 84\,r_0\,R\,L^{2} + 192\,r_0\,R\,L 
- 456\,r_0^{2}\,L^{2} 
\nonumber \\ && \qquad
\mbox{} - 48\,R^{2}\,m^{2}\,L^{2} + 48\,r_0^{2}\,L^{4} 
+ 288\,r_0^{2}\,m^{2})(u^{\phi})^2
\nonumber \\ && \qquad 
/({\cal L}^{(4)}\,(L^2 - 1)\,(L^2 - 4)) 
\biggr]\partial_{\theta}\,Y_{\ell m}^*(\theta_0,\,\phi_0) \,,
\cr
H_{0\ell m}^{\rm S,H}(t,\,r)
&=& {2 \over L}\,\pi\,\mu \biggl[
{\displaystyle \frac {1}{4}}  
( - 504\,r_0 + 144\,r_0\,m^{2} + 12\,r_0
\,L^{6} + 614\,r_0\,L^{2} - 62\,R\,m^{2}\,L^{2} + 2\,R\,m^{2}\,L^{4} 
\nonumber \\ && \qquad
- 170\,r_0\,L^{4} - 529\,R\,L^{2} - 10\,R\,L^{6}
 - 168\,R\,L^{5} + 588\,R\,L^{3} + 143\,R\,L^{4}
 \nonumber \\ && \qquad 
+ 396\,R\,m^{2} + 40\,R\,m^{4} - 432\,R\,L + 12\,R\,L^{7} + 468\,R - 52\,
r_0\,m^{2}\,L^{2} 
\nonumber \\ && \qquad
+ 4\,r_0\,m^{2}\,L^{4})(u^{\phi})^2
/((L^2 - 1)\,(L^2 - 4)\,(L^2 - 9)) 
\nonumber \\ && \qquad
- {\displaystyle \frac {2\,i\,m\,T\,u^{\phi}}
{r_0}}  
\nonumber \\ && \qquad
+ {\displaystyle \frac {1}{4}} 
\,{\displaystyle \frac {(2\,r_0 + R - 8\,R\,L + 8\,R\,L^{3})\,M}
{r_0^{3}\,(L^2 - 1)}}  + {\displaystyle \frac 
{(2\,r_0 - R + 2\,R\,L)}{r_0^{2}}}  
\biggr] Y_{\ell m}^*(\theta_0,\,\phi_0) \,,
\cr
H_{1\ell m}^{\rm S,H}(t,\,r)
&=& {2 \over L}\,\pi\,\mu \biggl[
- 2 {\displaystyle \frac {T\,( - 108 + 111\,L^{2}
 + 2\,L^{6} - 29\,L^{4} + 44\,m^{2}\,L^{2} - 2\,m^{2}\,L^{4} 
- 234\,m^{2} - 20\,m^{4})\,(u^{\phi})^2}{(L^2 - 1)\,(L^2 - 4)
\,(L^2 - 9)}}  
\nonumber \\ &&
 \qquad+ {\displaystyle \frac {2\,i\,m\,
( - R + 2\,r_0)\,u^{\phi}}{(L^2-1)\,r_0}}  
+ {\displaystyle \frac {4\,T\,M}{r_0^{3}}} 
\biggr] Y_{\ell m}^*(\theta_0,\,\phi_0) \,,
\cr
H_{2\ell m}^{\rm S,H}(t,\,r)
&=& {2 \over L}\,\pi\,\mu \biggl[
 - {\displaystyle \frac {1}{4}} 
 ( - 648\,r_0 - 576\,r_0\,m^{2} + 4\,r_0\,L^{6} 
+ 378\,r_0\,L^{2} + 134\,R\,m^{2}\,L^{2} - 2\,R\,m^{2}\,L^{4} 
\nonumber \\ &&
 \qquad - 70\,r_0\,L^{4} + 241\,R\,L^{2} + 2\,R\,L^{6} 
- 56\,R\,L^{5} + 196\,R\,L^{3} - 39\,R\,L^{4} - 1044\,R\,m^{2} 
\nonumber \\ &&
 \qquad - 40\,R\,m^{4} - 144\,R\,L + 4\,R\,L^{7} - 468\,R 
+ 100\,r_0\,m^{2}\,L^{2} - 4\,r_0\,m^{2}\,L^{4})(u^{\phi})^2
\nonumber \\ &&
 \qquad /((L^2 - 1)\,(L^2 - 4)\,(L^2 - 9))
\mbox{} 
- {\displaystyle \frac {2\,i\,m\,T\,u^{\phi}}{r_0}}  
\nonumber \\ &&
 \qquad + {\displaystyle \frac {1}{4}} 
\,{\displaystyle \frac {(2\,r_0 + R - 8\,R\,L + 8\,R\,L^{3})\,M}
{r_0^{3}\,(L^2 - 1)}}  
+ {\displaystyle \frac {(2\,r_0 - R + 2\,R\,L)}
{r_0^{2}}} 
\biggr] Y_{\ell m}^*(\theta_0,\,\phi_0) \,,
\cr
h_{0\ell m}^{\rm (e) S,H}(t,\,r)
&=& {2 \over L}\,\pi\,\mu \biggl[
T\, (72\,r_0 
+ 392\,R\,m^{2}\,L^{3} - 288\,R\,m^{2}\,L - 112\,R\,m^{2}\,L^{5} 
- 120\,r_0\,m^{4} 
\nonumber \\ &&
 \qquad  + 48\,R\,m^{4}\,L^{2} + 8\,R\,m^{2}\,L^{7} + 4\,R\,m^{2}\,L^{6} 
+ 8\,r_0\,m^{2}\,L^{6} - 828\,r_0\,m^{2} 
- 8\,r_0\,L^{6} 
\nonumber \\ && 
 \qquad  - 314\,r_0\,L^{2} + 172\,R\,m^{2}\,L^{2} 
- 50\,R\,m^{2}\,L^{4} + 106\,r_0\,L^{4} - 941\,R\,L^{2} 
- 20\,R\,L^{6} 
\nonumber \\ &&
 \qquad  + 277\,R\,L^{4} - 414\,R\,m^{2} - 612\,R\,m^{4} + 612\,R 
+ 344\,r_0\,m^{2}\,L^{2} - 100\,r_0\,m^{2}\,L^{4})
(u^{\phi})^2
\nonumber \\ && \qquad 
/
(2\,{\cal L}^{(2)}\,(L^2-1)\,(L^2-4)\,(L^2-9))
\nonumber \\ && 
 \qquad + i(8\,R^{2}\,L^{6}  + 16\,r_0\,R\,L^{5} + 16\,r_0^{2}\,L^{4} 
+ 8\,r_0\,R\,L^{4} - 46\,R^{2}\,L^{4} - 80\,r_0\,R\,L^{3} 
+ 55\,R^{2}\,L^{2} 
\nonumber \\ && 
 \qquad  - 56\,r_0^{2}\,L^{2} - 52\,r_0\,R\,L^{2} 
+ 64\,r_0\,R\,L - 32\,r_0^{2} + 80\,r_0\,R 
+ 16\,r_0\,R\,m^{2} + 4\,R^{2} 
\nonumber \\ &&
 \qquad + 40\,R^{2}\,m^{2})m\,u^{\phi}
\nonumber \\ &&
 \qquad /
(4\,r_0\,{\cal L}^{(2)}\,(L^2-1)\,(L^2-4)) 
+ {\displaystyle \frac {T\,
( - R + 2\,r_0)\,M}{r_0^{3}\,{\cal L}^{(2)}}} 
\biggr] Y_{\ell m}^*(\theta_0,\,\phi_0) \,,
\cr
h_{1\ell m}^{\rm (e) S,H}(t,\,r)
&=& {2 \over L}\,\pi\,\mu \biggl[
- (20\,r_0\,R\,L^{4} 
+ 3528\,r_0\,R\,m^{2} - 756\,R^{2} + 288\,r_0^{2} 
- 464\,r_0\,R\,m^{2}\,L^{2} 
\nonumber \\ &&
 \qquad + 8\,r_0\,R\,m^{2}\,L^{4} + 720\,r_0\,R 
+ 1062\,R^{2}\,m^{2} + 20\,R^{2}\,L^{6} + 993\,R^{2}\,L^{2} 
- 281\,R^{2}\,L^{4} 
\nonumber \\ &&
 \qquad + 160\,m^{4}\,r_0^{2} - 260\,r_0\,R\,L^{2} 
- 104\,r_0^{2}\,L^{2} - 16\,R^{2}\,m^{4}\,L^{2} + 684\,R^{2}\,m^{4} 
\nonumber \\ &&
 \qquad + 102\,R^{2}\,m^{2}\,L^{4} - 8\,R^{2}\,m^{2}\,L^{6} 
- 388\,R^{2}\,m^{2}\,L^{2} + 560\,r_0\,R\,m^{4} + 8\,r_0^{2}\,L^{4} 
\nonumber \\ &&
 \qquad + 1872\,r_0^{2}\,m^{2} + 16\,r_0^{2}\,m^{2}\,L^{4} 
- 352\,r_0^{2}\,m^{2}\,L^{2})(u^{\phi})^2
\nonumber \\ &&
 \qquad /
(4\,{\cal L}^{(2)}\,(L^2-1)\,(L^2-4)\,(L^2-9))
\biggr] Y_{\ell m}^*(\theta_0,\,\phi_0) \,,
\cr
K_{\ell m}^{\rm S,H}(t,\,r)
&=& {2 \over L}\,\pi\,\mu \biggl[
{\displaystyle \frac {1}{192}} 
(12744\,r_0 - 9408\,R\,m^{2}\,L^{3} + 6912\,R\,m^{2}\,L 
+ 2688\,R\,m^{2}\,L^{5} - 120\,R\,L^{8} 
\nonumber \\ &&
 \qquad + 144\,r_0\,L^{8} + 144\,R\,L^{9} + 960\,r_0\,m^{4} 
- 288\,R\,m^{4}\,L^{2} - 192\,R\,m^{2}\,L^{7} + 120\,R\,m^{2}\,L^{6} 
\nonumber \\ &&
 \qquad - 144\,r_0\,m^{2}\,L^{6} - 648\,r_0\,m^{2} 
- 2172\,r_0\,L^{6} - 15402\,r_0\,L^{2} + 6462\,R\,m^{2}\,L^{2} 
\nonumber \\ &&
 \qquad - 1758\,R\,m^{2}\,L^{4} + 9438\,r_0\,L^{4} 
+ 16455\,R\,L^{2} + 1938\,R\,L^{6} + 9800\,R\,L^{5} - 14788\,R\,L^{3} 
\nonumber \\ &&
 \qquad - 9453\,R\,L^{4} - 3240\,R\,m^{2} + 7272\,R\,m^{4} + 7056\,R\,L 
- 2212\,R\,L^{7} - 7884\,R 
\nonumber \\ &&
 \qquad - 4140\,r_0\,m^{2}\,L^{2} + 1764\,r_0\,m^{2}\,L^{4})
(2\,L-1)\,(2\,L+1)\,(u^{\phi})^2
\nonumber \\ &&
 \qquad 
/({\cal L}^{(4)}\,(L^2-1)\,(L^2-4)
\,(L^2-9))
\mbox{} - {\displaystyle \frac {2\,i\,m\,\,T\,u^{\phi}}{r_0}}  
\nonumber \\ &&
 \qquad 
+ {\displaystyle \frac {1}{4}} \,{\displaystyle \frac {
(2\,r_0 + R - 8\,R\,L + 8\,R\,L^{3})\,M}
{r_0^{3}\,(L^2-1)}}  
+ {\displaystyle \frac {(2\,r_0 
- R + 2\,R\,L)}{r_0^{2}}} 
\biggr] Y_{\ell m}^*(\theta_0,\,\phi_0) \,,
\cr
G_{\ell m}^{\rm S,H}(t,\,r)
&=& {2 \over L}\,\pi\,\mu \biggl[
{\displaystyle \frac {1}{48}} 
( - 28296\,r_0\,R^{2} - 5184\,r_0^{3}\,m^{2} 
+ 30780\,R^{3} + 9792\,r_0^{2}\,R\,m^{2}\,L^{2} 
\nonumber \\ &&
 \qquad + 192\,r_0^{2}\,R\,m^{2}\,L^{6} 
- 384\,r_0^{2}\,R\,m^{2}\,L^{7} - 6048\,r_0^{2}\,R 
- 1104\,r_0^{3}\,L^{2} - 2544\,r_0^{3}\,L^{6} 
\nonumber \\ &&
 \qquad + 7488\,r_0^{3}\,L^{4} + 192\,r_0^{3}\,L^{8} 
+ 1920\,r_0^{3}\,m^{4} - 384\,r_0^{3}\,m^{2}\,L^{6} 
- 8064\,r_0^{3}\,m^{2}\,L^{2} 
\nonumber \\ &&
 \qquad + 4416\,r_0^{3}\,m^{2}\,L^{4} + 13416\,r_0^{2}\,R\,L^{2} 
- 15536\,r_0^{2}\,R\,L^{3} + 1560\,r_0^{2}\,R\,L^{6} 
\nonumber \\ &&
 \qquad - 2864\,r_0^{2}\,R\,L^{7} - 7680\,r_0^{2}\,R\,L^{4} 
+ 11872\,r_0^{2}\,R\,L^{5} + 32\,R^{3}\,L^{11} 
- 71343\,R^{3}\,L^{2} 
\nonumber \\ &&
 \qquad + 10047\,R^{3}\,L^{6} + 38356\,R^{3}\,L^{5} + 15650\,R^{3}\,L^{3} 
+ 39596\,R^{3}\,L^{4} - 13182\,R^{3}\,L^{7} 
\nonumber \\ &&
 \qquad - 5976\,R^{3}\,L^{8} - 104004\,R^{3}\,m^{2} + 688\,R^{3}\,L^{9} 
+ 35688\,R^{3}\,m^{4} + 496\,R^{3}\,L^{10} 
\nonumber \\ &&
 \qquad - 41544\,R^{3}\,L - 13824\,r_0\,R^{2}\,m^{2}\,L 
- 192\,r_0\,R^{2}\,m^{2}\,L^{8} + 18816\,r_0\,R^{2}\,m^{2}\,L^{3} 
\nonumber \\ &&
 \qquad + 384\,r_0\,R^{2}\,m^{2}\,L^{7} + 43266\,r_0\,R^{2}\,L^{2} 
- 9480\,r_0\,R^{2}\,m^{2}\,L^{4} - 4416\,r_0\,R^{2}\,m^{4}\,L^{2} 
\nonumber \\ &&
 \qquad - 1536\,r_0^{2}\,R\,m^{4}\,L^{2} 
- 5376\,r_0\,R^{2}\,m^{2}\,L^{5} + 2832\,r_0\,R^{2}\,m^{2}\,L^{6} 
- 11088\,r_0\,R^{2}\,m^{2}\,L^{2} 
\nonumber \\ &&
 \qquad - 18816\,r_0^{2}\,R\,m^{2}\,L^{3} 
+ 5376\,r_0^{2}\,R\,m^{2}\,L^{5} + 13824\,r_0^{2}\,R\,m^{2}\,L 
- 96\,r_0^{2}\,R\,L^{8} 
\nonumber \\ &&
 \qquad + 192\,r_0^{2}\,R\,L^{9} - 2592\,r_0^{2}\,R\,m^{2} 
+ 16704\,r_0^{2}\,R\,m^{4} + 6336\,r_0^{2}\,R\,L 
+ 8742\,r_0\,R^{2}\,L^{6} 
\nonumber \\ &&
 \qquad - 22624\,r_0\,R^{2}\,L^{5} + 53168\,r_0\,R^{2}\,L^{3} 
- 24816\,r_0\,R^{2}\,L^{4} + 3632\,r_0\,R^{2}\,L^{7} 
\nonumber \\ &&
 \qquad - 1584\,r_0\,R^{2}\,L^{8} + 62856\,r_0\,R^{2}\,m^{2} 
- 192\,r_0\,R^{2}\,L^{9} + 40464\,r_0\,R^{2}\,m^{4} 
+ 96\,r_0\,R^{2}\,L^{10} 
\nonumber \\ &&
 \qquad - 33984\,r_0\,R^{2}\,L - 4256\,R^{3}\,m^{2}\,L^{5} 
- 1560\,R^{3}\,m^{2}\,L^{6} + 108648\,R^{3}\,m^{2}\,L^{2} 
\nonumber \\ &&
 \qquad + 5248\,R^{3}\,m^{4}\,L - 20076\,R^{3}\,m^{2}\,L^{4} 
- 8416\,R^{3}\,m^{4}\,L^{2} + 128\,R^{3}\,m^{4}\,L^{3} 
\nonumber \\ &&
 \qquad + 12672\,R^{3}\,m^{2}\,L + 4496\,R^{3}\,m^{2}\,L^{3} 
+ 256\,R^{3}\,m^{4}\,L^{4} + 976\,R^{3}\,m^{2}\,L^{7} 
- 64\,R^{3}\,m^{2}\,L^{9} 
\nonumber \\ &&
 \qquad + 288\,R^{3}\,m^{2}\,L^{8} - 2784\,r_0^{2}\,R\,m^{2}\,L^{4} 
- 1728\,r_0^{3})(u^{\phi})^2
\nonumber \\ &&
 \qquad /
(r_0^{2}\,{\cal L}^{(4)}\,(L^2-1)\,(L^2-4)\,(L^2-9))
\biggr]  Y_{\ell m}^*(\theta_0,\,\phi_0) \,.
\label{eq:ScoeffH}
\end{eqnarray}

The gauge transformation from the harmonic gauge to the RW gauge is given by 
\begin{eqnarray}
M_{0\ell m}^{{\rm S,H} \to {\rm RW}}(t,\,r)
&=& {2 \over L}\,\pi\,\mu \biggl[
- T (72\,r_0 - 112\,R\,m^{2}\,L^{5} 
+ 392\,R\,m^{2}\,L^{3} - 288\,R\,m^{2}\,L - 120\,r_0\,m^{4} 
\nonumber \\ && \qquad
 + 8\,R\,m^{2}\,L^{7} + 4\,R\,m^{2}\,L^{6} + 48\,R\,m^{4}\,L^{2} 
+ 8\,r_0\,m^{2}\,L^{6} - 828\,r_0\,m^{2} 
- 8\,r_0\,L^{6} 
\nonumber \\ && \qquad
 - 314\,r_0\,L^{2} + 172\,R\,m^{2}\,L^{2} - 50\,R\,m^{2}\,L^{4} 
+ 106\,r_0\,L^{4} - 941\,R\,L^{2} - 20\,R\,L^{6} 
\nonumber \\ && \qquad
 + 277\,R\,L^{4} - 414\,R\,m^{2} - 612\,R\,m^{4} + 612\,R 
+ 344\,r_0\,m^{2}\,L^{2} - 100\,r_0\,m^{2}\,L^{4})(u^{\phi})^2
\nonumber \\ && \qquad
/(2\,{\cal L}^{(2)}\,(L^2-1)\,(L^2-4)\,(L^2-9))
\nonumber \\ && \qquad
 - i(8\,R^{2}\,L^{6}  + 16\,r_0\,R\,L^{5} - 46\,R^{2}\,L^{4} 
+ 8\,r_0\,R\,L^{4} + 16\,r_0^{2}\,L^{4} - 80\,r_0\,R\,L^{3} 
- 56\,r_0^{2}\,L^{2} 
\nonumber \\ && \qquad
 + 55\,R^{2}\,L^{2} - 52\,r_0\,R\,L^{2} 
+ 64\,r_0\,R\,L + 4\,R^{2} + 40\,R^{2}\,m^{2} + 80\,r_0\,R 
+ 16\,r_0\,R\,m^{2} 
\nonumber \\ && \qquad
 - 32\,r_0^{2}) \,m\,u^{\phi}
/(4\,r_0\,{\cal L}^{(2)}\,(L^2-1)\,(L^2-4)) 
\nonumber \\ && \qquad
 - {\displaystyle \frac {T 
\,( - R + 2\,r_0)\,M}{r_0^{3}\,{\cal L}^{(2)}}} 
\biggr] Y_{\ell m}^*(\theta_0,\,\phi_0) \,,
\cr
M_{1\ell m}^{{\rm S,H} \to {\rm RW}}(t,\,r)
&=& {2 \over L}\,\pi\,\mu \biggl[
{\displaystyle \frac {1}{96}}  
( - 21504\,r_0\,R\,L^{5} + 75264\,r_0\,R\,L^{3} 
- 72312\,r_0\,R\,L^{4} - 69984\,r_0\,R\,m^{2} 
\nonumber \\ && \qquad
 + 13932\,R^{2} - 21600\,r_0^{2} 
- 36096\,r_0\,R\,m^{2}\,L^{4} + 107136\,r_0\,R\,m^{2}\,L^{2} 
- 96\,L^{8}\,r_0^{2} 
\nonumber \\ && \qquad
 - 107568\,r_0\,R + 50850\,R^{2}\,L^{4} + 58845\,R^{2}\,L^{6} 
- 120528\,R^{2}\,m^{2} - 99315\,R^{2}\,L^{2} 
\nonumber \\ && \qquad
 - 3456\,m^{2}\,L\,R^{2} + 6240\,m^{2}\,L^{6}\,r_0\,R 
+ 384\,m^{4}\,L^{3}\,R^{2} - 254232\,R^{2}\,L 
+ 21084\,r_0\,L^{6}\,R 
\nonumber \\ && \qquad
 + 15744\,m^{4}\,L\,R^{2} + 1536\,m^{4}\,L^{2}\,r_0\,R 
- 28896\,m^{2}\,L^{5}\,R^{2} + 69936\,m^{2}\,L^{3}\,R^{2} 
\nonumber \\ && \qquad
 + 1536\,r_0\,L^{7}\,R - 27882\,L^{7}\,R^{2} 
+ 8064\,m^{4}\,r_0^{2} + 192\,r_0^{2}\,L^{9} 
+ 1488\,L^{9}\,R^{2} 
\nonumber \\ && \qquad
 - 3360\,L^{8}\,r_0\,R + 192\,L^{10}\,r_0\,R 
+ 144684\,r_0\,R\,L^{2} + 6336\,r_0^{2}\,L 
- 55296\,r_0\,R\,L 
\nonumber \\ && \qquad
 + 25944\,r_0^{2}\,L^{2} + 4080\,m^{2}\,L^{7}\,R^{2} 
+ 384\,m^{4}\,L^{4}\,R^{2} - 192\,R^{2}\,m^{2}\,L^{9} + 96\,R^{2}\,L^{11} 
\nonumber \\ && \qquad
 - 96\,R^{2}\,m^{2}\,L^{8} - 384\,r_0\,R\,m^{2}\,L^{8} 
+ 244086\,R^{2}\,L^{3} + 36444\,R^{2}\,L^{5} + 1872\,L^{10}\,R^{2} 
\nonumber \\ && \qquad
 - 23880\,L^{8}\,R^{2} + 1752\,r_0^{2}\,L^{6} 
- 10608\,r_0^{2}\,L^{4} + 11872\,r_0^{2}\,L^{5} 
- 15536\,r_0^{2}\,L^{3} 
\nonumber \\ && \qquad
 - 103680\,r_0^{2}\,m^{2} - 12096\,r_0^{2}\,m^{2}\,L^{4} 
- 18816\,r_0^{2}\,m^{2}\,L^{3} + 13824\,r_0^{2}\,m^{2}\,L 
\nonumber \\ && \qquad
 + 73728\,r_0^{2}\,m^{2}\,L^{2} + 84096\,r_0\,R\,m^{4} 
+ 9720\,R^{2}\,m^{2}\,L^{6} + 5376\,r_0^{2}\,m^{2}\,L^{5} 
\nonumber \\ && \qquad
 + 337824\,R^{2}\,m^{2}\,L^{2} - 115752\,R^{2}\,m^{2}\,L^{4} 
- 27168\,R^{2}\,m^{4}\,L^{2} + 248688\,R^{2}\,m^{4} 
\nonumber \\ && \qquad
 - 2864\,r_0^{2}\,L^{7} + 576\,r_0^{2}\,m^{2}\,L^{6} 
- 384\,r_0^{2}\,m^{2}\,L^{7} 
+ 2304\,m^{4}\,r_0^{2}\,L^{2})(u^{\phi})^2
\nonumber \\ && \qquad
/({\cal L}^{(4)}\,(L^2-1)\,(L^2-4)\,(L^2-9)) 
\biggr] Y_{\ell m}^*(\theta_0,\,\phi_0) \,,
\cr
M_{2\ell m}^{{\rm S,H} \to {\rm RW}}(t,\,r)
&=& {2 \over L}\,\pi\,\mu \biggl[
- {\displaystyle \frac {1}{96}}  
(11208\,r_0^{2}\,R\,L^{2} - 4768\,R^{3}\,m^{2}\,L^{5} 
+ 4416\,r_0^{3}\,m^{2}\,L^{4} - 8064\,r_0^{3}\,m^{2}\,L^{2} 
\nonumber \\ && \qquad
 - 384\,r_0^{3}\,m^{2}\,L^{6} + 7296\,r_0^{2}\,R\,L^{4} 
- 42120\,r_0\,R^{2} - 2864\,r_0^{2}\,R\,L^{7} 
- 3528\,r_0^{2}\,R\,L^{6} 
\nonumber \\ && \qquad
 - 15536\,r_0^{2}\,R\,L^{3} + 20544\,r_0^{
2}\,R\,m^{4} - 12960\,r_0^{2}\,R\,m^{2} + 192\,r_0^{2}\,R\,L^{9} 
\nonumber \\ && \qquad
 + 288\,r_0^{2}\,R\,L^{8} + 11872\,r_0^{2}\,R\,L^{5} 
+ 6336\,r_0^{2}\,R\,L + 9318\,r_0\,R^{2}\,L^{6} 
+ 68994\,r_0\,R^{2}\,L^{2} 
\nonumber \\ && \qquad
 + 96\,r_0\,R^{2}\,L^{10} - 6336\,r_0^{2}\,R\,m^{2}\,L^{2} 
+ 75792\,r_0\,R^{2}\,m^{4} + 192\,r_0\,R^{2}\,L^{9} 
\nonumber \\ && \qquad
 + 52488\,r_0\,R^{2}\,m^{2} - 1584\,r_0\,R^{2}\,L^{8} 
- 2096\,r_0\,R^{2}\,L^{7} - 384\,r_0^{2}\,R\,m^{2}\,L^{7} 
\nonumber \\ && \qquad
 - 576\,r_0^{2}\,R\,m^{2}\,L^{6} - 32688\,r_0\,R^{2}\,L^{4} 
+ 22096\,r_0\,R^{2}\,L^{3} + 1120\,r_0\,R^{2}\,L^{5} 
- 656\,R^{3}\,L^{9} 
\nonumber \\ && \qquad
 + 6048\,r_0^{2}\,R\,m^{2}\,L^{4} 
- 18816\,r_0^{2}\,R\,m^{2}\,L^{3} - 21312\,r_0\,R^{2}\,L 
+ 13824\,r_0^{2}\,R\,m^{2}\,L 
\nonumber \\ && \qquad
 + 5376\,r_0^{2}\,R\,m^{2}\,L^{5} 
+ 432\,r_0\,R^{2}\,m^{2}\,L^{2} - 192\,r_0\,R^{2}\,m^{2}\,L^{8} 
+ 2832\,r_0\,R^{2}\,m^{2}\,L^{6} 
\nonumber \\ && \qquad
 + 5376\,r_0\,R^{2}\,m^{2}\,L^{5} 
- 1536\,r_0^{2}\,R\,m^{4}\,L^{2} + 13824\,r_0\,R^{2}\,m^{2}\,L 
- 7488\,r_0\,R^{2}\,m^{4}\,L^{2} 
\nonumber \\ && \qquad
 - 10632\,r_0\,R^{2}\,m^{2}\,L^{4} - 64\,R^{3}\,m^{2}\,L^{9} 
+ 1040\,R^{3}\,m^{2}\,L^{7} + 768\,R^{3}\,m^{4}\,L^{4} 
+ 48\,R^{3}\,m^{2}\,L^{3} 
\nonumber \\ && \qquad
 + 47520\,R^{3}\,m^{2}\,L - 7168\,R^{3}\,m^{4}\,L^{3} 
- 6240\,R^{3}\,m^{4}\,L^{2} - 11868\,R^{3}\,m^{2}\,L^{4} 
\nonumber \\ && \qquad
 + 75712\,R^{3}\,m^{4}\,L - 384\,r_0\,R^{2}\,m^{2}\,L^{7} 
+ 18504\,R^{3}\,m^{2}\,L^{2} + 1992\,R^{3}\,m^{2}\,L^{6} 
\nonumber \\ && \qquad
 - 18816\,r_0\,R^{2}\,m^{2}\,L^{3} - 96\,R^{3}\,m^{2}\,L^{8} 
- 5184\,r_0^{3}\,m^{2} + 32\,R^{3}\,L^{11} - 24492\,R^{3}\,L^{5} 
\nonumber \\ && \qquad
 + 52594\,R^{3}\,L^{3} + 33004\,R^{3}\,L^{4} + 5426\,R^{3}\,L^{7} 
- 152\,R^{3}\,L^{8} - 27540\,R^{3}\,m^{2} 
\nonumber \\ && \qquad
 - 35448\,R^{3}\,m^{4} + 48\,R^{3}\,L^{10} - 30600\,R^{3}\,L 
+ 19980\,R^{3} - 50019\,R^{3}\,L^{2} - 5597\,R^{3}\,L^{6} 
\nonumber \\ && \qquad
 - 1104\,r_0^{3}\,L^{2} - 2544\,r_0^{3}\,L^{6} 
+ 7488\,r_0^{3}\,L^{4} + 192\,r_0^{3}\,L^{8} 
+ 1920\,r_0^{3}\,m^{4} 
\nonumber \\ && \qquad
 - 9504\,r_0^{2}\,R - 1728\,r_0^{3})(u^{\phi})^2
\nonumber \\ && \qquad
/({\cal L}^{(4)}\,(L^2-1)\,(L^2-4)\,(L^2-9)) 
\biggr] Y_{\ell m}^*(\theta_0,\,\phi_0) \,,
\cr
\Lambda_{\ell m}^{{\rm S,H} \to {\rm RW}}(t,\,r)
 &=& {2 \over L}\,\pi\,\mu \biggl[
-{\displaystyle \frac {1}{12}} \,i(48\,r_0\,R\,L^{5} 
- 240\,r_0\,R\,L^{3} + 72\,r_0\,R\,L^{4} 
- 288\,r_0\,R\,m^{2} + 4108\,R^{2} 
\nonumber \\ && \qquad
 + 1056\,r_0^{2} - 1488\,r_0\,R - 66\,R^{2}\,L^{4} 
+ 24\,R^{2}\,L^{6} + 1392\,R^{2}\,m^{2} - 1147\,R^{2}\,L^{2} 
\nonumber \\ && \qquad
 + 192\,R^{2}\,L + 84\,r_0\,R\,L^{2} 
+ 192\,r_0\,R\,L - 456\,r_0^{2}\,L^{2} - 240\,R^{2}\,L^{3} 
+ 48\,R^{2}\,L^{5} 
\nonumber \\ && \qquad
 + 48\,r_0^{2}\,L^{4} + 288\,r_0^{2}\,m^{2} 
- 48\,R^{2}\,m^{2}\,L^{2})r_0 \,m
\,(u^{\phi})^2
\nonumber \\ && \qquad
/({\cal L}^{(4)}\,(L^2-1)\,(L^2-4))
\biggr] \partial_{\theta}\,Y_{\ell m}^*(\theta_0,\,\phi_0) \,.
\label{eq:Gfunctions}
\end{eqnarray}

And then, the coefficients of the S part under the RW gauge are 
calculated as 
\begin{eqnarray}
h_{0\ell m}^{\rm S,RW}(t,\,r)
&=& {2 \over L}\,\pi\,\mu \biggl[
{\displaystyle \frac 
{4\,i\,T \,m\,r_0\,(L^{2} - 2)\,(u^{\phi})^2}
{{\cal L}^{(2)}\,(L^2-1)}}  
- (8\,r_0 - 6\,r_0\,m^{2} 
- 18\,r_0\,L^{2} 
\nonumber \\ && \qquad
 + 4\,r_0\,L^{4} - 4\,R + 16\,R\,L - 13\,R\,m^{2} 
- 7\,R\,L^{2} - 20\,R\,L^{3} + 2\,R\,L^{4} + 4\,R\,L^{5})u^{\phi}
\nonumber \\ && \qquad
/({\cal L}^{(2)}\,(L^2-1)\,(L^2-4))
\biggr] \partial_{\theta}\,Y_{\ell m}^*(\theta_0,\,\phi_0) 
\,,
\cr
h_{1\ell m}^{\rm S,RW}(t,\,r)
&=& {2 \over L}\,\pi\,\mu \biggl[
-{\displaystyle \frac {1}{3}} \,i( - 60\,r_0\,L^{3} 
+ 174\,r_0\,L^{2} + 48\,L\,r_0 - 792\,r_0 
+ 6\,r_0\,L^{4} + 12\,r_0\,L^{5} 
\nonumber \\ && \qquad
 - 216\,r_0\,m^{2} + 3380\,R - 881\,R\,L^{2} 
- 39\,R\,L^{4} - 24\,R\,m^{2}\,L^{2} + 984\,R\,m^{2} + 12\,R\,L^{6}) 
r_0 \,m\,(u^{\phi})^2 
\nonumber \\ && \qquad
/({\cal L}^{(4)}\,(L^2-1)\,(L^2-4)) 
\biggr] 
\partial_{\theta}\,Y_{\ell m}^*(\theta_0,\,\phi_0) \,,
\cr
H_{0\ell m}^{\rm S,RW}(t,\,r)
&=& {2 \over L}\,\pi\,\mu \biggl[
{\displaystyle \frac {1}{16}}  
( - 648\,r_0 + 1792\,R\,m^{2}\,L^{5} - 6272\,R\,m^{2}\,L^{3} 
+ 4608\,R\,m^{2}\,L - 40\,R\,L^{8} 
\nonumber \\ && \qquad
 + 48\,r_0\,L^{8} + 48\,R\,L^{9} + 1920\,r_0\,m^{4} 
- 128\,R\,m^{2}\,L^{7} - 56\,R\,m^{2}\,L^{6} - 608\,R\,m^{4}\,L^{2} 
\nonumber \\ && \qquad
 - 112\,r_0\,m^{2}\,L^{6} + 13104\,r_0\,m^{2} 
- 564\,r_0\,L^{6} + 2394\,r_0\,L^{2} 
- 1106\,R\,m^{2}\,L^{2} 
\nonumber \\ && \qquad
 + 550\,R\,m^{2}\,L^{4} + 930\,r_0\,L^{4} + 17457\,R\,L^{2} 
+ 902\,R\,L^{6} + 2520\,R\,L^{5} - 2316\,R\,L^{3} 
\nonumber \\ && \qquad
 - 6691\,R\,L^{4} + 6228\,R\,m^{2} + 9752\,R\,m^{4} + 432\,R\,L 
- 684\,R\,L^{7} - 10260\,R 
\nonumber \\ && \qquad
 - 4876\,r_0\,m^{2}\,L^{2} + 1388\,r_0\,m^{2}\,L^{4})
(u^{\phi})^2
\nonumber \\ && \qquad
/({\cal L}^{(2)}\,(L^2-1)\,(L^2-4)\,(L^2-9)) 
- {\displaystyle 
\frac {2\,i\,m \,T\,u^{\phi}}{r_0}} 
 \nonumber \\ && \qquad
 - {\displaystyle \frac {1}{16}} \,{\displaystyle \frac 
{ \,( - 62\,r_0 + 56\,r_0\,L^{2} 
+ 33\,R - 8\,R\,L - 36\,R\,L^{2} + 40\,R\,L^{3} - 32\,R\,L^{5})\,M}
{r_0^{3}\,{\cal L}^{(2)}\,(L^2-1)
}}  
\nonumber \\ && \qquad
 + {\displaystyle \frac { \,(2\,r_0 
- R + 2\,R\,L)}{r_0^{2}}} 
\biggr] Y_{\ell m}^*(\theta_0,\,\phi_0) \,,
\cr
H_{1\ell m}^{\rm S,RW}(t,\,r)
&=& {2 \over L}\,\pi\,\mu \biggl[
- T (360 - 742\,L^{2} 
- 69\,L^{6} + 375\,L^{4} - 56\,m^{2}\,L^{5} - 2\,m^{2}\,L^{6} 
- 404\,m^{2}\,L^{2} 
\nonumber \\ && \qquad
 + 4\,L^{8} + 64\,m^{2}\,L^{4} - 90\,m^{2} - 16\,m^{4}\,L^{2} 
- 296\,m^{4} - 144\,m^{2}\,L + 196\,m^{2}\,L^{3} + 4\,m^{2}\,L^{7}) 
(u^{\phi})^2
\nonumber \\ && \qquad
/({\cal L}^{(2)}\,(L^2-1)\,(L^2-4)\,(L^2-9)) 
\nonumber \\ && \qquad
- i\,m(4\,R\,L^{6} - 21\,R\,L^{4} + 19\,R\,L^{2} + 4\,R + 20\,R\,m^{2} 
+ 4\,r_0\,L^{5} - 2\,r_0\,L^{4} - 20\,r_0\,L^{3} 
\nonumber \\ && \qquad
 + 4\,r_0\,L^{2} + 16\,L\,r_0 + 16\,r_0 
+ 4\,r_0\,m^{2})u^{\phi}/(r_0\,{\cal L}^{(2)}\,(L^2-1)\,(L^2-4)) 
\nonumber \\ && \qquad
 + {\displaystyle \frac {4\,T \,L^{2}\,M}{r_0^{3}\,{\cal L}^{(2)}}} 
\biggr] Y_{\ell m}^*(\theta_0,\,\phi_0) \,,
\cr
H_{2\ell m}^{\rm S,RW}(t,\,r)
&=& {2 \over L}\,\pi\,\mu \biggl[
{\displaystyle \frac {1}{64}}  
( - 137592\,r_0 - 224\,R\,m^{2}\,L^{8} - 77056\,R\,m^{2}\,L^{5} 
+ 186496\,R\,m^{2}\,L^{3} 
\nonumber \\ && \qquad
 - 9216\,R\,m^{2}\,L + 4960\,R\,L^{10} - 448\,r_0\,m^{2}\,L^{8} 
- 512\,R\,m^{2}\,L^{9} - 73728\,L\,r_0 
\nonumber \\ && \qquad
 + 192\,R\,L^{11} + 192\,r_0\,L^{10} + 1024\,R\,m^{4}\,L^{3} 
+ 41984\,R\,m^{4}\,L + 1664\,R\,m^{4}\,L^{4} 
\nonumber \\ && \qquad
 + 2048\,m^{4}\,L^{2}\,r_0 - 62976\,R\,L^{8} 
- 3200\,r_0\,L^{8} + 5024\,R\,L^{9} + 100352\,r_0\,L^{3} 
\nonumber \\ && \qquad
 - 28672\,r_0\,L^{5} + 112128\,r_0\,m^{4} 
+ 10880\,R\,m^{2}\,L^{7} + 23696\,R\,m^{2}\,L^{6} - 74048\,R\,m^{4}\,L^{2} 
\nonumber \\ && \qquad
 + 6560\,r_0\,m^{2}\,L^{6} + 2048\,r_0\,L^{7} 
- 88128\,r_0\,m^{2} + 19228\,r_0\,L^{6} 
+ 163590\,r_0\,L^{2} 
\nonumber \\ && \qquad
 + 857898\,R\,m^{2}\,L^{2} - 286590\,R\,m^{2}\,L^{4} 
- 70298\,r_0\,L^{4} - 285729\,R\,L^{2} + 151486\,R\,L^{6} 
\nonumber \\ && \qquad
 + 107832\,R\,L^{5} + 643372\,R\,L^{3} + 153079\,R\,L^{4} 
- 312012\,R\,m^{2} + 663528\,R\,m^{4} 
\nonumber \\ && \qquad
 - 676656\,R\,L - 79764\,R\,L^{7} + 41364\,R 
+ 118908\,r_0\,m^{2}\,L^{2} - 34876\,r_0\,m^{2}\,L^{4})
(u^{\phi})^2
\nonumber \\ && \qquad
/({\cal L}^{(4)}\,(L^2-1)\,(L^2-4)\,(L^2-9)) 
- {\displaystyle \frac {2\,i\,m \,T\,u^{\phi}}
{r_0}} 
\nonumber \\ && \qquad
 + {\displaystyle \frac {1}{4}}\,
{\displaystyle \frac {(2\,r_0 
+ R - 8\,R\,L + 8\,R\,L^{3})\,M}{r_0^{3}\,(L^2-1)}}  
 + {\displaystyle \frac { \,(2\,r_0 
- R + 2\,R\,L)}{r_0^{2}}} 
\biggr] Y_{\ell m}^*(\theta_0,\,\phi_0) \,,
\cr
K_{\ell m}^{\rm S,RW}(t,\,r)
&=& {2 \over L}\,\pi\,\mu \biggl[
{\displaystyle \frac {1}{192}}  
( - 99144\,r_0 - 1056\,R\,m^{2}\,L^{8} - 61824\,R\,m^{2}\,L^{5} 
+ 112320\,R\,m^{2}\,L^{3} 
\nonumber \\ && \qquad
 - 62208\,R\,m^{2}\,L + 288\,R\,L^{10} - 1536\,r_0\,m^{2}\,L^{7} 
- 576\,r_0\,m^{2}\,L^{8} - 768\,R\,m^{2}\,L^{9} 
\nonumber \\ && \qquad
 + 25344\,L\,r_0 + 768\,r_0\,L^{9} + 576\,R\,L^{11} 
+ 576\,r_0\,L^{10} - 1152\,R\,m^{4}\,L^{4} 
\nonumber \\ && \qquad
 + 13056\,m^{4}\,L^{2}\,r_0 - 5184\,R\,L^{8} 
- 75264\,r_0\,m^{2}\,L^{3} + 21504\,r_0\,m^{2}\,L^{5} 
\nonumber \\ && \qquad
 + 55296\,m^{2}\,L\,r_0 - 9216\,r_0\,L^{8} 
- 9760\,R\,L^{9} - 62144\,r_0\,L^{3} + 47488\,r_0\,L^{5} 
\nonumber \\ && \qquad
 + 31296\,r_0\,m^{4} + 12480\,R\,m^{2}\,L^{7} 
+ 15504\,R\,m^{2}\,L^{6} + 26304\,R\,m^{4}\,L^{2} 
\nonumber \\ && \qquad
 + 9504\,r_0\,m^{2}\,L^{6} - 11456\,r_0\,L^{7} 
- 414072\,r_0\,m^{2} + 46932\,r_0\,L^{6} 
+ 170154\,r_0\,L^{2} 
\nonumber \\ && \qquad
 + 114210\,R\,m^{2}\,L^{2} - 68394\,R\,m^{2}\,L^{4} 
- 113478\,r_0\,L^{4} + 426969\,R\,L^{2} + 37578\,R\,L^{6} 
\nonumber \\ && \qquad
 - 202456\,R\,L^{5} + 406212\,R\,L^{3} - 171543\,R\,L^{4} 
+ 138024\,R\,m^{2} + 296856\,R\,m^{4} 
\nonumber \\ && \qquad
 - 253584\,R\,L + 59012\,R\,L^{7} - 335988\,R 
+ 296460\,r_0\,m^{2}\,L^{2} - 66708\,r_0\,m^{2}\,L^{4}) 
(u^{\phi})^2
\nonumber \\ && \qquad
/({\cal L}^{(4)}\,(L^2-1)\,(L^2-4)\,(L^2-9)) 
- {\displaystyle \frac {2\,i\,m \,T\,u^{\phi}}{r_0}} 
\nonumber \\ && \qquad
 + {\displaystyle \frac {1}{4}}{\displaystyle \frac { 
 \,(2\,r_0 + R - 8\,R\,L 
+ 8\,R\,L^{3})\,M}{r_0^{3}\,(L^2-1)}}  
 + {\displaystyle \frac { \,(2\,r_0 
- R + 2\,R\,L)}{r_0^{2}}} 
\biggr] 
Y_{\ell m}^*(\theta_0,\,\phi_0) \,.
\label{eq:RWSmetric}
\end{eqnarray}

\section{$\bm{\ell=0}$ and $\bm{1}$ modes}\label{app:L01}

In this Appendix, we derive the contributions to the self-force
in the RW gauge from the $\ell=0$ and $1$ modes. 
As discussed at the beginning of Sec.~\ref{sec:result},
and described in Eq.~(\ref{eq:Fldecomp}), although there is
no physical contribution from the $\ell=0$ and $\ell=1$
modes to the self-force in the rigorous sense,
since we can calculate the S part only locally in the
vicinity of the particle, its spherical extension inevitably
contaminates each $\ell$ mode with other $\ell$ modes.
Therefore, in particular, we have to take account of the
corrections from the $\ell=0$ and $\ell=1$ modes
to the self-force.

For the $\ell=0$ and $\ell=1$ odd modes, the RW gauge
condition is automatically satisfied, since
$h_0^{(e)}=h_1^{(e)}=G=0$ for $\ell=0$ and 
$h_2=0$ for $\ell=1$ odd modes.
An appropriate choice of gauge is then to consider the perturbation
under the retarded causal boundary condition in the harmonic gauge.
In fact, if we recall the gauge transformation equations
from the harmonic gauge to the RW gauge given by Eq.~(\ref{eq:ggtran}),
we see that all the gauge transformation generators for
$\ell=0$ and $\ell=1$ odd modes vanish.
Thus, no gauge transformation is needed for the S part
of these modes,
and our task is to find the exact solutions in the harmonic
gauge with the retarded boundary condition
 and perform the subtraction of the S part
under the harmonic gauge.

For the $\ell=1$ even mode, the RW gauge condition is non-trivial
and there is a gauge degree of freedom in
the RW gauge, reflecting the fact that it is a pure gauge mode
describes a shift of the center of mass
coordinates in the source-free case. On the other hand, the gauge
transformation of this mode from the harmonic gauge to
the RW gauge is uniquely
determined. Thus, to determine the self-force unambiguously in the
RW gauge, one first has to solve the perturbation equations in
the harmonic gauge exactly (under the retarded boundary condition),
transform the result to the RW
gauge, and perform the subtraction of the S part. However,
unfortunately, we were unable to solve for the $\ell=1$ even
mode in the harmonic gauge due to a
complicated structure of the perturbation equations
(i.e., in the form of coupled hyperbolic equations).
Thus, there remains a gauge ambiguity in the final result.
Nevertheless, in the Newton limit when the coordinates can be
defined globally, we can resolve the gauge ambiguity and
give a definite meaning to the resulting self-force.

To summarize, the regularized self-force in the RW gauge
is expressed as
\begin{eqnarray}
F^{\rm R,RW} &=& 
 \sum_{\ell \geq 2}
 \left(F^{\rm full,RW}_{\ell} -F^{\rm S,RW,Ap}_{\ell}\right)
 +\delta F^{\rm RW}_{\ell=0,1} \,, 
\end{eqnarray}
where
\begin{eqnarray}
\delta F^{\rm RW}_{\ell=0,1}
=\delta F^{\rm RW}_{\ell=0}+\delta F^{\rm RW}_{\ell=1{\rm (odd)}} 
+\delta F^{\rm RW}_{\ell=1{\rm (even)}}
\end{eqnarray}
with
\begin{eqnarray}
 \delta F^{\rm RW}_{\ell=0} &=& 
 F^{\rm full,H}_{\ell=0} -F^{\rm S,H,Ap}_{\ell=0}
 \,,
\nonumber \\
 \delta F^{\rm RW}_{\ell=1{\rm (odd)}} &=& 
 F^{\rm full,H}_{\ell=1{\rm (odd)}} -F^{\rm S,H,Ap}_{\ell=1{\rm (odd)}}
 \,, 
\nonumber\\
 \delta F^{\rm RW}_{\ell=1{\rm (even)}} &=& 
 F^{\rm full,RW}_{\ell=1{\rm (even)}} -F^{\rm S,RW,Ap}_{\ell=1{\rm (even)}} \,,
\end{eqnarray}
where there remains a gauge ambiguity in
$\delta F^{\rm RW}_{\ell=1{\rm (even)}}$. 

\subsection{$\bm{\ell=0}$ mode}

First, we consider the $\ell=0$ mode of the full metric perturbation.
It is noted that the $\ell=0$ mode consists of only the even parity
part and all the derivatives of $Y_{00}$ vanish.
As noted above, this mode satisfies
the RW gauge condition $h_0^{(e)}=h_1^{(e)}=G=0$
automatically. So, the appropriate choice
of gauge is the harmonic gauge under the retarded boundary condition.
To find the exact solution in this gauge, we consider a gauge 
transformation of the exact solution found by Zerilli.

This mode represents the perturbation in the total mass of the system
and was analyzed by Zerilli.
For the $\ell=0$ mode, there are two gauge degrees of
freedom. The choice made by Zerilli is 
\begin{eqnarray}
H_1^{{\rm full,Z}}(t,r)=K^{{\rm full,Z}}(t,r)=0 \,.
\end{eqnarray}
which we call the Zerilli (Z) gauge and denote the quantities in it by 
the superscript ${\rm Z}$.
In the case of a circular orbit, the $\ell=0$ mode metric perturbation is 
solved to be
\begin{eqnarray}
H_2^{{\rm full,Z}}(t,r)&=&{a \over r-2M}\Theta(r-r_0) \,,
\cr
H_0^{{\rm full,Z}}(t,r)&=&
a\left[
{1 \over r_0-2M}\Theta(r_0-r)
+{1 \over r-2M}\Theta(r-r_0) \right] \,.
\end{eqnarray}
Here we imposed the boundary condition that the black hole
mass is unperturbed and the perturbation satisfies the
asymptotic flatness. Note that the Zerilli gauge is singular 
in the sense that the metric has a discontinuity at $r=r_0$.
The constant $a$ is given by 
\begin{eqnarray}
a=2(4\pi)^{1/2} \mu u^t \left(1-{2\,M \over r_0}\right) \,.
\end{eqnarray}
Note that the $\ell=0$ mode is independent of time.
So we may write $H_2^{{\rm full,Z}}(t,r)=H_2^{{\rm full,Z}}(r)$.

Now we consider the gauge transformation from the above Z gauge 
to the harmonic gauge. 
The equations for the gauge transformation are
formally written as 
\begin{eqnarray}
\xi_{\mu;\nu}{}^{\nu} &=& \bar{h}_{\mu\nu}^{\rm Z}{}^{;\nu}
\,, \cr
\bar{h}_{\mu\nu} &=& h_{\mu\nu}
-{1\over 2}g_{\mu\nu} h_{\alpha}{}^{\alpha} \,.
\label{eq:ZtoH}
\end{eqnarray}
Detailed discussions on the gauge transformation 
to the harmonic gauge are given in~\cite{Sago:2002fe}. 

We set the gauge transformation generator $\xi_\mu$ as
\begin{eqnarray}
\{\xi_{\mu}^{{\rm Z \rightarrow H}}\}
=\{M_0(r)Y_{00}(\theta,\,\phi), M_1(r)Y_{00}(\theta,\,\phi), 0,0\} \,.
\end{eqnarray}
In the circular case, 
the $\ell=0$ mode of Eq.~(\ref{eq:ZtoH}) is explicitly 
written down as 
\begin{eqnarray}
&&
\left[{d^2 \over dr^2}+{2\over r}{d \over dr} \right]M_0(r)=0\,,
\label{eq:M0}
\\
&&\left[{ r-2 \,M \over r} {d^2 \over dr^2}
+{2 \over r} {d \over dr} 
-{2(r-2\,M) \over r^3}\right]M_1(r) 
= S(r) \,, 
\label{eq:M1}
\end{eqnarray}
where
\begin{eqnarray}
S(r) &=& 
4\pi \frac{r^3}{(r-2\,M)^2}A_{00}^{(0)}(r) 
+{M \over r (r-2\,M)} H_0^{{\rm Z}}(r)
+{ 2\,r-3\,M \over r (r-2\,M)}H_2^{{\rm Z}}(r) 
\nonumber \\ 
&=&
{a  \over 2 (r_0-2 \,M)} \delta(r_0-r)
+{a \,M \over  r (r-2 \,M)(r_0-2\, M)} \Theta(r_0-r)
\nonumber \\ 
&&+{2\, a (r-M) \over r(r-2 \,M)^2} \Theta(r-r_0)
\,.
\end{eqnarray}
Since $M_0$ is independent of the source, we set it to zero
in accordance with the retarded boundary condition.
Thus we focus on the equation for $M_1$.

We employ the Green function method to solve Eq.~(\ref{eq:M1}). 
Two independent homogeneous solutions are easily obtained as 
\begin{eqnarray}
M_1^{{\rm (homo,1)}} &=& {1 \over r(r-2\,M)} \,,
\cr 
M_1^{{\rm (homo,2)}} &=& {r^2 \over r-2\,M} \,. 
\end{eqnarray}
Using the above homogeneous solutions, we construct
the other two independent solutions
$M_1^{\rm in}$ and $M_1^{\rm out}$ which are regular 
at the event horizon and infinity, respectively.
We find
\begin{eqnarray}
M_1^{\rm in} &=& -8\,M^3 M_1^{{\rm (homo,1)}} + M_1^{{\rm (homo,2)}} 
={r^2 + 2\,M\,r + 4\,M^2 \over r} \,, 
\nonumber\\
M_1^{\rm out} &=& M_1^{{\rm (homo,1)}}= {1 \over r(r-2\,M)} \,.
\end{eqnarray}
Then the Green function is derived as
\begin{eqnarray}
G(r,r')&=& {1 \over W}\left[
M_1^{\rm in}(r)M_1^{\rm out}(r')\Theta(r'-r)
+M_1^{\rm out}(r)M_1^{\rm in}(r')\Theta(r-r')\right] \,;
\nonumber\\
W &=& (r-2\,M)^2\left[
M_1^{\rm in}(r)\partial_r M_1^{\rm out}(r)
-M_1^{\rm out}(r)\partial_r M_1^{\rm in}(r)\right] 
= -3 \,,
\end{eqnarray}
and $M_1$ is given by
\begin{eqnarray}
M_1(r) = \int_{2M}^{\infty} G(r,r') r'(r'-2\,M)S(r') dr' \,.
\end{eqnarray}
Although the integral can be performed without any approximation, 
we only show the result to 1PN order,
\begin{eqnarray}
M_1(r) &=& \left[
-{5\, a \over 6}{r \over r_0} 
-{a \over 6}{(13 \,r_0+6\, r) M \over r_0^2}
\right] \Theta(r_0-r) 
\nonumber \\ 
&& + \left[
-{a \over 6}{6\,r^2-r_0^2 \over r^2} 
-{a \over 6}{(30\, r^2-9 r \,r_0-2 r_0^2) M \over r^3}
\right] \Theta(r-r_0) \,.
\end{eqnarray}

The metric perturbation transforms under the above gauge 
transformation as 
\begin{eqnarray}
H_0^{{\rm H}}(r) &=& H_0^{{\rm Z}}(r)+{2\,M \over r^2}M_1(r) 
\,, \cr
H_1^{{\rm H}}(r) &=& -{d \over dr} M_0(r)+{2\,M \over r(r-2\, M)} M_0(r) \,, \cr
H_2^{{\rm H}}(r) &=& H_2^{{\rm Z}}(r)
-2\left(1-{2\,M \over r}\right)
\left({d \over dr} M_1(r) +{M \over r(r-2\,M)}M_1(r)\right) 
\,, \cr
K^{{\rm H}}(r) &=& -{2(r-2\,M)\over r^2}M_1(r) 
\,.
\end{eqnarray}
Note that we have $H_1=0$ because $M_0=0$.
(It may be noted that $H_1$ does not contribute
to the force for a circular orbit even if it is non-zero.)
Then the metric perturbation in the harmonic gauge is found as
\begin{eqnarray}
H_0^{{\rm H}} &=& \left[
a{1 \over r_0} 
+{a \over 3}{(6\, r -5\, r_0) M \over r\, r_0^2}
\right] \Theta(r_0-r) 
\nonumber \\ 
&& + \left[
a{1 \over r} 
+{a \over 3}{ r_0^2\, M \over r^4}
\right] \Theta(r-r_0) \,, \cr
H_2^{{\rm H}} &=& \left[
{5\, a \over 3}{1 \over r_0} 
+{a \over 3}{(6\, r -5 \,r_0) M \over r\, r_0^2}
\right] \Theta(r_0-r) 
\nonumber \\ 
&& + \left[
{a \over 3}{3\, r^2+2 \,r_0^2 \over r^3} 
-{a \over 3}{ (18\, r^2-18\, r\, r_0-r_0^2) M \over r^4}
\right] \Theta(r-r_0) \,, \cr
K^{{\rm H}} &=& \left[
{5\, a \over 3}{1 \over r_0} 
+a{(r_0 + 2\, r) M \over r\, r_0^2}
\right] \Theta(r_0-r) 
\nonumber \\ 
&& + \left[
{a \over 3}{6\, r^2 -r_0^2 \over r^3} 
+3 a{(2\, r - r_0) M \over r^3}
\right] \Theta(r-r_0) \,,
\end{eqnarray}
and the full force is calculated as 
\begin{eqnarray}
F^r_{{\rm full,H}}(\ell =0)
=\left[ {7\,\mu^2\, M \over r_0^3}\right] \,\Theta(r_0-r)
+ \left[-{\mu^2 \over r_0^2}
+{9\,\mu^2 \,M \over 2\, r_0^3}\right] \,\Theta(r-r_0) \,.
\end{eqnarray}

Next, we consider the S part of the metric perturbation.
Its harmonic coefficients are given in 
Eqs.~(\ref{eq:harmonicS}).
Only the harmonic coefficients $H_0$, $H_2$ and
$K$ remain for the $\ell=0$ mode.
To 1PN order, we have
\begin{eqnarray}
H_{0}^{\rm S,H}(r)
&=& \sqrt{4\,\pi}\,\mu\,\biggl\{
\biggl[ 3\,{\displaystyle \frac {M}
{r_0^{2}}}  + {\displaystyle \frac {2 }{r_0}}  \biggr]\Theta(r_0-r)
 + \biggl[ {\displaystyle \frac
{( - 5\,R + 3\,r_0)\,M}{r_0^{3}}}
+ {\displaystyle \frac {2\,(r_0 - R)}{r_0^{2}}} \biggr] \Theta(r-r_0)
\biggr\}
 \,,
\cr
H_{2}^{\rm S,H}(r)
&=&
\sqrt{4\,\pi}\,\mu\,\biggl\{
\biggl[ - {\displaystyle \frac {M}
{r_0^{2}}}  + {\displaystyle \frac {2 }{r_0}}  \biggr]\Theta(r_0-r)
 + \biggl[ - {\displaystyle \frac
{(R + r_0)\,M}{r_0^{3}}}  + {\displaystyle \frac
{2\,(r_0 - R)}{r_0^{2}}}  \biggr] \Theta(r-r_0)
\biggr\}
\,,
\cr
K^{\rm S,H}(r)
&=&
\sqrt{4\,\pi}\,\mu\,\biggl\{
\biggl[ {\displaystyle \frac {M}
{r_0^{2}}}  + {\displaystyle \frac {2 }{r_0}}  \biggr]\Theta(r_0-r)
 + \biggl[ {\displaystyle \frac
{( - 3\,R + r_0)\,M}{r_0^{3}}}
+ {\displaystyle \frac {2\,(r_0 - R)}
{r_0^{2}}}  \biggr] \Theta(r-r_0)
\biggr\}
 \,.
\end{eqnarray}
The S-force in the harmonic gauge 
is calculated as 
\begin{eqnarray}
F^r_{{\rm S,H}}(\ell =0) &=& \left[
{33\,\mu^2 \,M \over 8\, r_0^3} \right] \,\Theta(r_0-r) 
+\left[-{\mu^2 \over r_0^2}
+{13\,\mu^2 \,M \over 8\, r_0^3}\right] \,\Theta(r-r_0) 
\,.
\end{eqnarray}

{}From the above results,
we obtain the contribution of the $\ell=0$ force as 
\begin{eqnarray}
\delta F^{r}_{{\rm RW}}(\ell =0) &=& \delta F^{r}_{{\rm H}}(\ell =0)
\nonumber \\ &=& F^r_{{\rm full,H}}(\ell =0) - F^r_{{\rm S,H}}(\ell =0)
\nonumber \\ &=& {23\,\mu^2\, M \over 8\, r_0^3} \,.
\end{eqnarray}

\subsection{$\bm{\ell=1}$ odd parity mode}

The $\ell=1$ odd mode represents the angular momentum 
perturbation added to the system. It also satisfies the 
odd parity RW gauge condition $h_2=0$ automatically.
Therefore, as in the $\ell=0$ case,
we look for the exact solution in the harmonic gauge
with the retarded boundary condition.

The full metric perturbation consists of the two 
components, $h_{0}^{\rm full}$ and $h_{1}^{\rm full}$. 
These were also solved by Zerilli.
There is one gauge degree of freedom, and we may put
$h_1=0$. The appropriate boundary condition is that
the black hole angular momentum is unperturbed and
the perturbation is well-behaved at infinity.
Then we find
\begin{eqnarray}
h_0^{{\rm Z}}(t,r)=
\left(b {r^2 \over r_0^3}\Theta(r_0-r)
+b {1\over r}\Theta(r-r_0)\right)\delta_{0,m} \,,
\end{eqnarray}
where $b$ is given by 
\begin{eqnarray}
b=2\sqrt{4 \pi \over 3} \mu \,u^{\phi} \,r_0^2 \,.
\end{eqnarray}
Note that only the $m=0$ mode is non-zero, and it
is time-independent. 

Next, we consider the gauge transformation to the 
harmonic gauge. We set
\begin{eqnarray}
\xi_\mu=\Lambda_m^{{\rm Z}\to{\rm H}}(r)
\left(0,0,-{1\over\sin\theta}\partial_\phi Y_{1m}(\theta,\,\phi)\,,
\sin\theta\partial_\theta Y_{1m}(\theta,\,\phi)\right).
\end{eqnarray}
The equation for $\Lambda_m^{{\rm Z}\to{\rm H}}$ becomes
\begin{eqnarray}
\left[-\left(1-{2M\over r}\right)^{-1}\partial_t^2
+\partial_r\left(1-{2M\over r}\right)\partial_r
-{2\over r^2}\right]\Lambda_m^{{\rm Z}\to{\rm H}}(r)=0\,.
\end{eqnarray}
This is a source-free hyperbolic equation. So,
with the retarded boundary condition, we find
$\Lambda_m^{{\rm Z}\to{\rm H}}=0$, that is,
the Zerilli gauge is equivalent to the harmonic gauge
with the retarded boundary condition.
The full force is then calculated as
\begin{eqnarray}
F^{r {\rm (odd)}}_{{\rm full,H}}(\ell =1) &=& 
\left[
-{4\,\mu^2 M \over r_0^3} \right] \,\Theta(r_0-r) 
+\left[{2\,\mu^2 M \over r_0^3} \right] \,\Theta(r-r_0)  \,.
\end{eqnarray}

The harmonic coefficients of the S part are given as
\begin{eqnarray}
h_{0\,1m}^{\rm S,H}(r)
&=& -\sqrt{{4\,\pi \over 3}} \,\mu \biggl\{ \biggl[
 - {\displaystyle \frac {8}{9}} \,(4\,R + 2\,r_0)\,u^{\phi}
\biggr] \Theta(r_0-r)
\nonumber \\ &&
+\biggl[
 - {\displaystyle \frac {8}{9}} \,(2\,r_0 - 2\,R)\,u^{\phi}
\biggr] \Theta(r-r_0)
\biggr\}
\,\delta_{0,m} \,,
\cr
h_{1\,1m}^{\rm S,H}(r)
&=& 0 \,,
\end{eqnarray}
and the S-force is obtained as 
\begin{eqnarray}
F^{r {\rm (odd)}}_{{\rm S,H}}(\ell =1) &=& \left[
-{4\,\mu^2 M \over r_0^3} \right] \,\Theta(r_0-r) 
+\left[{2\,\mu^2 M \over r_0^3} \right] \,\Theta(r-r_0) 
\,.
\end{eqnarray}

Subtracting the S part from the full force, 
we find
\begin{eqnarray}
\delta F^{r {\rm (odd)}}_{{\rm RW}}(\ell =1) 
&=& \delta F^{r {\rm (odd)}}_{{\rm H}}(\ell =1)
\nonumber \\ &=& 
F^{r {\rm (odd)}}_{{\rm full,H}}(\ell =1) 
- F^{r {\rm (odd)}}_{{\rm S,H}}(\ell =1) \cr
&=& 0 \,.
\end{eqnarray}
Thus, our spherical extension turns out to be accurate enough
to reproduce the correct $\ell=1$ odd mode up to 1PN order.

\subsection{$\bm{\ell=1}$ even parity mode}

The $\ell=1$ even mode represents essentially a gauge mode
that describes a shift of the center of momentum of the system.
The coefficient $G$ is absent from the beginning,
while there is no loss in the gauge freedom. 
Hence there remains one degree of gauge freedom in the RW gauge.
As mentioned at the beginning of this Appendix,
to fix the gauge completely, it is necessary 
to solve the perturbation equations in the harmonic
gauge with the retarded boundary condition, and to perform
the gauge transformation to the RW gauge. However, 
because the perturbation equations become complicated,
coupled hyperbolic equations in the harmonic gauge,
we were unable to solve for this mode. Here, we therefore
give up fixing the gauge unambiguously,
but solve the perturbation equations in the RW gauge,
imposing a gauge condition by hand.

To look for an exact solution in the RW gauge,
following Zerilli,
we choose $K=0$ in addition to $h_0^{(e)}=h_1^{(e)}=0$.
Let us also call it the Zerilli gauge.
The harmonic coefficients for the full metric perturbation in the
Zerilli gauge are given by
\begin{eqnarray}
H_{2\,1m}^{{\rm full,Z}}(t,r)&=&{1 \over (r-2M)^2}\,f_m(t)\,\Theta(r-r_0) \,,
\cr
H_{1\,1m}^{{\rm full,Z}}(t,r)&=&-{r \over (r-2M)^2}
\,\partial_t f_m(t)\,\Theta(r-r_0) \,,
\cr
H_{0\,1m}^{{\rm full,Z}}(t,r)&=&{1 \over 3(r-2M)^2}
\left(f_m(t)+{r^3 \over M} \,\partial_t^2 f_m(t) \right)
\Theta(r-r_0) \,,
\end{eqnarray}
where
\begin{eqnarray}
f_m(t) = 8\,\pi\,\mu\,u^t
{(r_0-2\,M)^2 \over r_0}Y_{1 m}^*(\theta_0(t),\,\phi_0(t)) \,,
\end{eqnarray}
and we have imposed the boundary condition that
the perturbation is regular at horizon. It may be noted that 
although the $\ell=1$ even mode is locally a pure gauge, it is not 
so in the global sense because of the regularity at the horizon. 
Note that the $m=0$ components vanish because the orbit is on
the equatorial plane.
It is also noted that $H_1^{\rm full,Z}$ and $H_2^{\rm full,Z}$
are discontinuous at $r=r_0$, while $H_0^{\rm full,Z}$ is 
continuous because $\partial_t^2f_m=-\Omega^2 f_m=-(M/r_0^3)f_m$ for
$m=\pm1$, and the force depends only on $H_0^{\rm full,Z}$.
The full force in this gauge is derived as 
\begin{eqnarray}
F^{r {\rm (even)}}_{{\rm full,Z}}(\ell =1) &=& \left[-{3\,\mu^2 \over r_0^2}
-{3\,\mu^2 \,M \over 2\, r_0^3}\right] \,\Theta(r-r_0) \,.
\end{eqnarray}

The coefficient $H_{0\,1m}^{{\rm full,Z}}$ in the
above behaves as $\sim r$ at infinity. Without violating
the RW gauge condition, it is possible to remove this singular
behavior. Namely, we consider a gauge transformation,
\begin{eqnarray}
H_{01 m}^{\rm full,RW}(t,\,r)
&=& H_{01 m}^{\rm full,Z}(t,\,r)
+{2\,r \over r-2\,M}
\left[ \partial_t
\,M_{01 m}^{{\rm full,Z} \to {\rm RW}}(t,\,r)
-{M(r-2\,M) \over r^3}
M_{11 m}^{{\rm full,Z} \to {\rm RW}}(t,\,r)\right] \,,
\nonumber\\
H_{11 m}^{\rm full,RW}(t,\,r)
&=& H_{11 m}^{\rm full,Z}(t,\,r)
+
\left[ \partial_t
\,M_{11 m}^{{\rm full,Z} \to {\rm RW}}(t,\,r)
+\partial_r \,M_{01 m}^{{\rm full,Z} \to {\rm RW}}(t,\,r)
-{2\,M \over r(r-2\,M) }
M_{01 m}^{{\rm full,Z} \to {\rm RW}}(t,\,r)\right] \,,
\nonumber\\
H_{21 m}^{\rm full,RW}(t,\,r)
&=& H_{21 m}^{\rm full,Z}(t,\,r)
+{2(r-2\,M) \over r}
\left[ \partial_r
\,M_{11 m}^{{\rm full,Z} \to {\rm RW}}(t,\,r)
+{M \over r(r-2\,M)}
M_{11 m}^{{\rm full,Z} \to {\rm RW}}(t,\,r)\right] \,, 
\nonumber\\
K_{1 m}^{{\rm full,RW}}(t,\,r) &=&
{2 \over r^2} 
\left(
2(r-2\,M) M_{11 m}^{{\rm full,Z} \to {\rm RW}}(t,\,r)
-M_{21 m}^{{\rm full,Z} \to {\rm RW}}(t,\,r)
\right)\,,
\nonumber \\ 
h_{0~1m}^{(e){\rm full,RW}}(t,r)&=&0=
-M_{01 m}^{{\rm full,Z} \to {\rm RW}}(t,\,r)
- \partial_t  M_{21 m}^{{\rm full,Z} \to {\rm RW}}(t,\,r)
\,,
\nonumber\\
h_{1~1m}^{(e){\rm full,RW}}(t,r)&=&0=
-M_{11 m}^{{\rm full,Z} \to {\rm RW}}(t,\,r)
- r^2 \,\partial_r  \left(
{M_{21 m}^{{\rm full,Z} \to {\rm RW}}(t,\,r)
\over r^2}\right)
\,.
\end{eqnarray}
As a solution of the above gauge equations that makes
the metric perturbation regular at infinity, 
we choose
\begin{eqnarray}
M_{01 m}^{{\rm full,Z} \to {\rm RW}}(t,\,r) &=&
{i\,r \over 6\,m\,\Omega\,r_0^3} f_m(t) \,, \cr
M_{11 m}^{{\rm full,Z} \to {\rm RW}}(t,\,r) &=& 
{1 \over 6\,m^2\,\Omega^2\,r_0^3} f_m(t) \,, \cr
M_{21 m}^{{\rm full,Z} \to {\rm RW}}(t,\,r) &=& 
{r \over 6\,m^2\,\Omega^2\,r_0^3} f_m(t) \,.
\label{eq:even1gtr}
\end{eqnarray}

By the above gauge transformation, the $r$-component of the force
changes by
\begin{eqnarray}
\delta F^{r {\rm (even)}}_{{\rm full,Z}\rightarrow{\rm RW}}(\ell =1) &=& 
\sum_{m=-1}^1 \mu \biggl(
{2\,M\,(r_0-2\,M) \over r_0^4} M_{11 m}^{{\rm full,Z} \to {\rm RW}}(t_0,r_0)
+{2\,i\,m\,M\,\Omega \over (r_0-3\,M) r_0} M_{01 m}^{{\rm full,Z} \to {\rm RW}}(t_0,r_0)
\nonumber \\ && \qquad \qquad
-{(r_0-2\,M) \over (r_0-3\,M)} \,\partial_t^2 M_{11 m}^{{\rm full,Z} \to {\rm RW}}
(t_0,r_0)
+{2\,M \over r_0(r_0-3\,M)} \,\partial_t M_{01 m}^{{\rm full,Z} \to {\rm RW}}(t_0,r_0)
\nonumber \\ && \qquad \qquad
-{i\,m\,\Omega(r_0-2\,M) \over (r_0-3\,M)} 
\,\partial_r M_{01 m}^{{\rm full,Z} \to {\rm RW}}(t_0,r_0)
-{M (r_0-2\,M)^2 \over (r_0-3\,M) r_0^3} 
\,\partial_r M_{11 m}^{{\rm full,Z} \to {\rm RW}}(t_0,r_0)
\nonumber \\ && \qquad \qquad
-{i\,\Omega\,m(r_0-2\,M) \over (r_0-3\,M)} 
\,\partial_t M_{11 m}^{{\rm full,Z} \to {\rm RW}}(t_0,r_0)
\biggr) Y_{1 m}(\theta_0,\,\phi_0) 
\,.
\end{eqnarray}
So, to 1PN order, we find 
\begin{eqnarray}
\delta F^{r {\rm (even)}}_{{\rm full,Z}\rightarrow{\rm RW}}(\ell =1) &=& 
\sum_{m=-1}^1 \mu 
\left[ {(r_0-2\,M)^2 \over 2\,r_0^4\,(r_0-3\,M)}f_m(t) \right] 
Y_{1 m}(\theta_0,\,\phi_0) 
\nonumber \\ &=& {3\,\mu^2 \over r_0^2} - {21\,\mu^2\,M \over 2\,r_0^3} \,.
\end{eqnarray}
Thus, the full force in this RW gauge is given by
\begin{eqnarray}
F^{r {\rm (even)}}_{{\rm full,RW}}(\ell =1) &=& \left[
{3\,\mu^2 \over r_0^2} - {21\,\mu^2\,M \over 2\,r_0^3}
\right] \,\Theta(r_0-r) 
+ \left[
-{12\,\mu^2 \,M \over r_0^3}\right] \,\Theta(r-r_0) \,.
\label{eq:ffulleven}
\end{eqnarray}

It may be noted that, at Newtonian order, the $r$ coordinate of the
Zerilli gauge, in which the metric inside the orbit is unperturbed,
corresponds to placing the black hole at $r=0$, while the
gauge transformation that regularizes the asymptotic behavior at
infinity makes $r$ the radial coordinate measured in the center of mass
coordinate system. In other words, $r_0$ in the
Zerilli gauge gives the relative distance between the black hole
and the particle, while $r_0$ after the transformation gives 
the distance from the center of mass to the particle.
This explains the Newtonian part of the change in the force,
$3\mu^2/r_0^2$. In this sense, the gauge freedom is under control
at Newtonian order.

Now we turn to the S part. The harmonic
coefficients in the harmonic gauge are given by
\begin{eqnarray}
H_{0\,1 m}^{\rm S,H}(t,\,r)
&=& {\displaystyle \frac {4}{3}} \,\pi \,\mu \biggl\{ \biggl[
{\displaystyle \frac {2\,r_0 + 2\,R}{r_0^{2}}}
 + {\displaystyle \frac {3\,M\,R}{r_0^{3}}}  -
{\displaystyle \frac {2\,i\,T\,m\,u^{\phi}}{r_0}}  +
{\displaystyle \frac {2}{9}} \,(2\,r_0\,m^{2} + 9\,R + R
\,m^{2} + {\displaystyle \frac {27}{2}} \,r_0)\,
(u^{\phi})^2
\biggr] \Theta(r_0-r)
\nonumber \\ &&
+\biggl[
{\displaystyle \frac {2\,
r_0 - 4\,R}{r_0^{2}}}  - {\displaystyle \frac {3
\,M\,R}{r_0^{3}}}  - {\displaystyle \frac {2\,i\,T\,m\,u^{\phi}}
{r_0}}  + {\displaystyle \frac {2}{9}} \,( -
{\displaystyle \frac {63}{2}} \,R + 2\,r_0\,m^{2} + R\,m
^{2} + {\displaystyle \frac {27}{2}} \,r_0)\,(u^{\phi})^2
\biggr] \Theta(r-r_0)
\biggr\}
\nonumber \\ &&
\times
Y_{1 m}^*(\theta_0,\,\phi_0) \,,
\cr
H_{1\,1 m}^{\rm S,H}(t,\,r)
&=& {\displaystyle \frac {4}{3}} \,\pi \,\mu \biggl\{ \biggl[
4\,{\displaystyle \frac {T\,M}{r_0^{3}}}  + {\displaystyle
\frac {16}{9}} \,i\,m\,u^{\phi} - {\displaystyle \frac {16}{9}
} \,({\displaystyle \frac {3}{2}}  - m)\,({\displaystyle \frac
{3}{2}}  + m)\,(u^{\phi})^2\,T
\biggr] \Theta(r_0-r)
\nonumber \\ &&
+\biggl[
4\,{\displaystyle \frac {T\,M}{r_0^{3}}}  + {\displaystyle
\frac {16}{9}} \,i\,m\,u^{\phi} - {\displaystyle \frac {16}{9}
} \,({\displaystyle \frac {3}{2}}  - m)\,({\displaystyle \frac {3}{2}}
+ m)\,(u^{\phi})^2\,T
\biggr] \Theta(r-r_0)
\biggr\}
Y_{1 m}^*(\theta_0,\,\phi_0) \,,
\cr
H_{2\,1 m}^{\rm S,H}(t,\,r)
&=& {\displaystyle \frac {4}{3}} \,\pi \,\mu \biggl\{ \biggl[
{\displaystyle \frac {2\,r_0 + 2\,R}{r_0^{2}}}
 + {\displaystyle \frac {3\,M\,R}{r_0^{3}}}  -
{\displaystyle \frac {2\,i\,T\,m\,u^{\phi}}{r_0}}  -
{\displaystyle \frac {2}{9}} \,(9\,R - 2\,r_0\,m^{2} - R
\,m^{2} + {\displaystyle \frac {9}{2}} \,r_0)\,(u^{\phi})^2
\biggr] \Theta(r_0-r)
\nonumber \\ &&
+\biggl[
{\displaystyle \frac {2\,r_0 - 4\,R}{r_0^{2}}}
 - {\displaystyle \frac {3\,M\,R}{r_0^{3}}}  -
{\displaystyle \frac {2\,i\,T\,m\,u^{\phi}}{r_0}}  -
{\displaystyle \frac {2}{9}} \,({\displaystyle \frac {9}{2}} \,
r_0 - 2\,r_0\,m^{2} - {\displaystyle \frac {9}{2}} \,R
- R\,m^{2})\,(u^{\phi})^2
\biggr] \Theta(r-r_0)
\biggr\}
\nonumber \\ &&
\times
Y_{1 m}^*(\theta_0,\,\phi_0) \,,
\cr
h_{0\,1 m}^{\rm (e) S,H}(t,\,r)
&=& {\displaystyle \frac {4}{3}} \,\pi \,\mu \biggl\{ \biggl[
{\displaystyle \frac {{\displaystyle \frac {8}{9}} \,i\,m\,(
{\displaystyle \frac {9}{4}} \,R^{2} + 4\,r_0\,R + 2\,
r_0^{2})\,u^{\phi}}{r_0}}  + {\displaystyle
\frac {8}{9}} \,m^{2}\,T\,(4\,R + 2\,r_0)\,(u^{\phi})^2
\biggr] \Theta(r_0-r)
\nonumber \\ &&
+\biggl[
{\displaystyle \frac {{\displaystyle \frac {8}{9}} \,i\,m\,(
 - 2\,r_0\,R + {\displaystyle \frac {9}{4}} \,R^{2} + 2\,
r_0^{2})\,u^{\phi}}{r_0}}  + {\displaystyle
\frac {8}{9}} \,m^{2}\,T\,(2\,r_0 - 2\,R)\,(u^{\phi})^2
\biggr] \Theta(r-r_0)
\biggr\}
\nonumber \\ &&
\times Y_{1 m}^*(\theta_0,\,\phi_0) \,,
\cr
h_{1\,1 m}^{\rm (e) S,H}(t,\,r)
&=& {\displaystyle \frac {4}{3}} \,\pi \,\mu \biggl\{ \biggl[
- {\displaystyle \frac {32}{81}} m
^{2}\,( - {\displaystyle \frac {9}{4}} \,R^{2} + 2\,r_0^{2}
+ r_0\,R)\,(u^{\phi})^2
\biggr] \Theta(r_0-r)
\nonumber \\ &&
+\biggl[
- {\displaystyle \frac {32}{81}}m
^{2}\,( - {\displaystyle \frac {9}{4}} \,R^{2} + 2\,r_0^{2}
+ r_0\,R)\,(u^{\phi})^2
\biggr] \Theta(r-r_0)
\biggr\} Y_{1 m}^*(\theta_0,\,\phi_0) \,,
\cr
K_{1 m}^{\rm S,H}(t,\,r)
&=& {\displaystyle \frac {4}{3}} \,\pi \,\mu \biggl\{ \biggl[
{\displaystyle \frac {2\,r_0 + 2\,R}{r_0^{2}}}
 + {\displaystyle \frac {3\,M\,R}{r_0^{3}}}  -
{\displaystyle \frac {2\,i\,T\,m\,u^{\phi}}{r_0}}  +
{\displaystyle \frac {2}{9}} \,(2\,r_0\,m^{2} +
{\displaystyle \frac {9}{2}} \,r_0 + R\,m^{2})\,
(u^{\phi})^2
\biggr] \Theta(r_0-r)
\nonumber \\ &&
+ \biggl[
{\displaystyle \frac {2\,r_0 - 4\,R}{r_0^{2}}}
 - {\displaystyle \frac {3\,M\,R}{r_0^{3}}}  -
{\displaystyle \frac {2\,i\,T\,m\,u^{\phi}}{r_0}}  +
{\displaystyle \frac {2}{9}} \,( - {\displaystyle \frac {27}{2}}
\,R +2\,r_0\,m^{2} + R\,m^{2} + {\displaystyle
\frac {9}{2}} \,r_0)\,(u^{\phi})^2
\biggr] \Theta(r-r_0)
\biggr\}
\nonumber \\ &&
\times Y_{1 m}^*(\theta_0,\,\phi_0) \,,
\end{eqnarray}
We transform the above to the RW gauge, as discussed in
Section~\ref{sec:dir}.
Since $G$ is absent from the beginning,
Eqs.(\ref{eq:GGT4}) that give the 
gauge transformation from the harmonic gauge to the RW gauge
are simplified as
\begin{eqnarray}
M_{2\,1m}^{{\rm S,H}\to{\rm RW}}(t,\,r)=0
\,,\quad
M_{0\,1 m}^{{\rm S,H} \to {\rm RW}}(t,\,r)
=-h_{0\,1 m}^{{\rm (e)S,H}}(t,\,r) 
\,,\quad
M_{1\,1 m}^{{\rm S,H} \to {\rm RW}}(t,\,r)
=-h_{11 m}^{{\rm (e)S,H}}(t,\,r)
\,.
\end{eqnarray}
The resulting harmonic coefficients in the RW gauge are 
expressed as those given in Eqs.~(\ref{eq:GTeven}) 
except for the gauge functions $M_0$ and $M_1$ that are now given by
the above equations. From these, we find
\begin{eqnarray}
H_{01 m}^{\rm S,RW}(t,\,r)
&=& {\displaystyle \frac {4}{3}} \,\pi \,\mu \biggl\{ \biggl[
{\displaystyle \frac {2\,r_0 + 2\,R}{r_0^{2}}}
 + {\displaystyle \frac {3\,M\,R}{r_0^{3}}}  -
{\displaystyle \frac {2\,i\,T\,m\,u^{\phi}}{r_0}}
\nonumber \\ && \quad  +
{\displaystyle \frac {2}{9}} \,(-14\,r_0\,m^{2} + 9\,R - 31\,R\,m^{2}
+ {\displaystyle \frac {27}{2}} \,r_0)\,
(u^{\phi})^2
\biggr] \Theta(r_0-r)
\nonumber \\ &&
+\biggl[
{\displaystyle \frac {2\,
r_0 - 4\,R}{r_0^{2}}}  - {\displaystyle \frac {3
\,M\,R}{r_0^{3}}}  - {\displaystyle \frac {2\,i\,T\,m\,u^{\phi}}
{r_0}}
\nonumber \\ && \quad + {\displaystyle \frac {2}{9}} \,\left( -
{\displaystyle \frac {63}{2}} \,R - 14\,r_0\,m^{2} + 17\,R\,m^{2}
+ {\displaystyle \frac {27}{2}} \,r_0\right)\,(u^{\phi})^2
\biggr] \Theta(r-r_0)
\biggr\}
\nonumber \\ &&
\times
Y_{1 m}^*(\theta_0,\,\phi_0) \,,
\cr
H_{11 m}^{\rm S,RW}(t,\,r)
&=& {\displaystyle \frac {4}{3}} \,\pi \,\mu \biggl\{ \biggl[
4\,{\displaystyle \frac {T\,M}{r_0^{3}}}  - {\displaystyle
\frac {8}{9}}\,\left( 2\,r_0+{9\over 2}\,R \right) \,i\,m\,u^{\phi}
- {\displaystyle \frac {8}{9}}
\,\left({\displaystyle \frac {9}{2}}  + 2\,m^2\right)\,(u^{\phi})^2\,T
\biggr] \Theta(r_0-r)
\nonumber \\ &&
+\biggl[
4\,{\displaystyle \frac {T\,M}{r_0^{3}}}  + {\displaystyle
\frac {8}{9}}\,\left( 4\,r_0-{9\over 2}\,R \right) \,i\,m\,u^{\phi}
- {\displaystyle \frac {8}{9}}
\,\left({\displaystyle \frac {9}{2}}  - 4\,m^2\right)\,(u^{\phi})^2\,T
\biggr] \Theta(r-r_0)
\biggr\}
\nonumber \\ &&
\times
Y_{1 m}^*(\theta_0,\,\phi_0) \,,
\cr
H_{21 m}^{\rm S,RW}(t,\,r)
&=& {\displaystyle \frac {4}{3}} \,\pi \,\mu \biggl\{ \biggl[
{\displaystyle \frac {2\,r_0 + 2\,R}{r_0^{2}}}
 + {\displaystyle \frac {3\,M\,R}{r_0^{3}}}  -
{\displaystyle \frac {2\,i\,T\,m\,u^{\phi}}{r_0}}
\nonumber \\ && \quad  -
{\displaystyle \frac {2}{9}} \,(9\,R - 2\,r_0\,m^{2} + 15\,R
\,m^{2} + {\displaystyle \frac {9}{2}} \,r_0)\,(u^{\phi})^2
\biggr] \Theta(r_0-r)
\nonumber \\ &&
+\biggl[
{\displaystyle \frac {2\,r_0 - 4\,R}{r_0^{2}}}
 - {\displaystyle \frac {3\,M\,R}{r_0^{3}}}  -
{\displaystyle \frac {2\,i\,T\,m\,u^{\phi}}{r_0}}
\nonumber \\ && \quad  -
{\displaystyle \frac {2}{9}} \,({\displaystyle \frac {9}{2}} \,
r_0 - 2\,r_0\,m^{2} - {\displaystyle \frac {9}{2}} \,R
+ 15\, R\,m^{2})\,(u^{\phi})^2
\biggr] \Theta(r-r_0)
\biggr\} Y_{1 m}^*(\theta_0,\,\phi_0) \,,
\cr
K_{1 m}^{\rm S,RW}(t,\,r)
&=& {\displaystyle \frac {4}{3}} \,\pi \,\mu \biggl\{ \biggl[
{\displaystyle \frac {2\,r_0 + 2\,R}{r_0^{2}}}
 + {\displaystyle \frac {3\,M\,R}{r_0^{3}}}  -
{\displaystyle \frac {2\,i\,T\,m\,u^{\phi}}{r_0}}
\nonumber \\ && \quad  +
{\displaystyle \frac {2}{9}} \,(2\,r_0\,m^{2} +
{\displaystyle \frac {9}{2}} \,r_0 + R\,m^{2})\,
(u^{\phi})^2
\biggr] \Theta(r_0-r)
\nonumber \\ &&
+ \biggl[
{\displaystyle \frac {2\,r_0 - 4\,R}{r_0^{2}}}
 - {\displaystyle \frac {3\,M\,R}{r_0^{3}}}  -
{\displaystyle \frac {2\,i\,T\,m\,u^{\phi}}{r_0}}
\nonumber \\ && \quad  +
{\displaystyle \frac {2}{9}} \,( - {\displaystyle \frac {27}{2}}
\,R + 2\,r_0\,m^{2} + \,R\,m^{2} + {\displaystyle
\frac {9}{2}} \,r_0)\,(u^{\phi})^2
\biggr] \Theta(r-r_0)
\biggr\} Y_{1 m}^*(\theta_0,\,\phi_0) \,,
\end{eqnarray}
We note that only $H_1^{\rm S,RW}$ is discontinuous at $r=r_0$.
However, as mentioned before, the force depends only on 
$H_0^{\rm S,RW}$ and $K^{\rm S,RW}$ which are continuous.
We obtain the S-force as
\begin{eqnarray}
F^{r {\rm (even)}}_{{\rm S,RW}}(\ell =1) &=& \left[
{\mu^2 \over r_0^2}+{21\,\mu^2 M \over 8\, r_0^3} \right] \,\Theta(r_0-r)
+\left[-{2\,\mu^2 \over r_0^2}+{9\,\mu^2 M \over 8\, r_0^3}
\right] \,\Theta(r-r_0)
\,.
\end{eqnarray}

Subtracting the above from the full force (\ref{eq:ffulleven}), we
find
\begin{eqnarray}
\delta F^{r {\rm (even)}}_{{\rm RW}}(\ell =1) 
&=& F^{r {\rm (even)}}_{{\rm full,RW}}(\ell =1) 
- F^{r {\rm (even)}}_{{\rm S,RW}}(\ell =1)
\nonumber \\ &=& {2\,\mu^2 \over r_0^2} - {105\,\mu^2 M \over 8\, r_0^3} \,.
\label{eq:dfeven}
\end{eqnarray}
We note that the Newtonian term, $2\mu^2/r_0^2$, is
precisely the correction to the force at $O(\mu^2)$
when $r_0$ is the distance from the center
of mass to the location of the particle.

If we recall the fact that both $H_1^{\rm full,Z}$ and $H_2^{\rm full,Z}$
are discontinuous at $r=r_0$ 
and the gauge transformation from the Zerilli gauge 
to a RW gauge given by Eq.~(\ref{eq:even1gtr})
does not change the discontinuity, while only $H_1^{\rm S,RW}$
is discontinuous for the S part, 
we see that the RW gauge we adopted
to obtain the full force is different from the RW gauge 
for the S part obtained by the transformation from the harmonic gauge.
Fortunately, however, because the force depends only on $H_0$ 
(and $K$) for circular orbits, and its discontinuity structure at $r=r_0$
happens to be the same in both gauges, the resulting force
(\ref{eq:dfeven}) turns out to contain no discontinuity.
Furthermore, as discussed above, the correct Newtonian
force is recovered at $O(\mu^2)$.
It is not clear if this desirable property holds because the
orbit is circular or because only the 1PN order correction
is considered. If this happens to be no longer the case
for general orbits, it will be necessary to find a 
gauge transformation that remedies the discrepancy.
In any case, except for the correction at Newtonian order,
the gauge ambiguity remains in the final result, and its
resolution is left for future work.

\section{$\bm{m}$-summation of tensor harmonics}\label{app:M}

In this Appendix,
we summarize the formulas for summing over $m$-modes of 
the tensor harmonics for arbitrary $\ell$.
Specifically, the $m$-sum we need to evaluate takes the form,
\begin{eqnarray}
\sum_{m=-\ell}^\ell m^N|Y_{\ell m}(\pi/2,0)|^2\,,
\label{eq:msumform}
\end{eqnarray}
where $N$ is a non-negative integer. To perform the summation, we
introduce the generating function,
\begin{eqnarray}
\Gamma_{\ell}(z)
 =\sum_{m=-\ell}^{\ell} e^{mz} |Y_{\ell m}(\pi/2,0)|^2 \,.
\end{eqnarray}
Then the sum (\ref{eq:msumform}) may be evaluated as
$\lim\limits_{z\to0}\partial_z^N\Gamma_{\ell}(z)$.
The above function is calculated as 
\begin{eqnarray}
\Gamma_{\ell}(z) &=& {2\ell+1\over 4\pi}e^{\ell z}
{}_2F_1 \left({1\over 2},\,-\ell;\,1;\,1-e^{-2z}\right) \,,
\label{eq:GMG}
\end{eqnarray}
where ${}_2F_1$ is the hypergeometric function. 
This can be easily expanded to an arbitrary order of $z$.
For example, to $O(z^6)$, we have
\begin{eqnarray}
\Gamma_{\ell}(z) &=& {2\ell+1\over 4\pi}
\Biggl\{1+\left({\ell \,(\ell+1) \over 2}\right){1\over 2}z^2
+\left({\ell \,(\ell+1)(3 \ell^2+3\ell-2) \over 8}\right)
{1\over 4!}z^4
\nonumber \\ && \qquad \qquad
+\left({\ell \,(\ell+1)(5\ell^4+10\ell^3-5\ell^2-10\ell+8) \over 16}
\right){1\over 6!}z^6
+O(z^8)\Biggr\} \,.
\end{eqnarray}

In the cases of the vector and tensor harmonics,
it is necessary to evaluate the $m$-sum of the form,
\begin{eqnarray}
\sum_{m=-\ell}^\ell m^N|\partial_{\theta} Y_{\ell m}(\pi/2,0)|^2\,.
\label{eq:mtsumform}
\end{eqnarray}
We introduce the generating function,
\begin{eqnarray}
\Delta_{\ell}(z)
 =\sum_{m=-\ell}^{\ell} e^{mz} |\partial_{\theta} Y_{\ell m}(\pi/2,0)|^2 \,.
\end{eqnarray}
This is expressed in terms of a hypergeometric function as
\begin{eqnarray}
\Delta_{\ell}(z) &=&
{2\ell+1 \over 4\pi^2} e^{(\ell-1)z}
{\Gamma(\ell+1/2)\Gamma(3/2) \over \Gamma(\ell)}
{}_2F_1 \left({3\over 2},\,-\ell+1;\,-\ell+{1\over 2};\,e^{-2z}\right) \,.
\end{eqnarray}
The sum (\ref{eq:mtsumform}) is evaluated by taking the
derivatives of the above generating function. 
Expanding in powers of $z$, the $m$-sum (\ref{eq:mtsumform}) is
calculated as
\begin{eqnarray}
\Delta_{\ell}(z) &=& {2\ell+1 \over 4\pi}
\Biggl\{
\left({\ell\,(\ell + 1) \over 2}\right)
+ \left({\ell\,(\ell + 1)\,(\ell - 1)\,(\ell + 2) \over 8}\right)
\,{1\over 2}z^{2}
\nonumber \\ && \qquad \qquad
+ \left({\ell\,(\ell + 1)\,(\ell - 1)\,(\ell + 2)
\,(\ell^{2} + \ell - 4)\over 16}\right)\,{1 \over 4!} z^{4}
\nonumber \\ && \qquad \qquad
+ \left({\ell\,(\ell + 1)\,(\ell - 1)\,(\ell + 2)
\,(5\,\ell^{4} + 10\,\ell^{3} - 45\,\ell^{2} - 50\,\ell + 136)
\over 128}\right)\,{1\over 6!}z^{6} + O(z^8)
\Biggr\}
\,.
\end{eqnarray}

\end{appendix}



\begin{thebibliography}{ederf}

\bibitem{LIGO} LIGO web page: http://www.ligo.caltech.edu/

\bibitem{GEO} GEO600 web page: http://www.geo600.uni-hannover.de/

\bibitem{TAMA} TAMA300 web page: http://tamago.mtk.nao.ac.jp/

\bibitem{VIRGO} VIRGO web page: http://www.virgo.infn.it/

\bibitem{LISA} LISA web page: http://lisa.jpl.nasa.gov/

\bibitem{DECIGO}
N.~Seto, S.~Kawamura and T.~Nakamura,
Phys.\ Rev.\ Lett.\  {\bf 87}, 221103 (2001)
[arXiv:astro-ph/0108011].

\bibitem{MSSTT} For a review, see e.g., Y.~Mino, M.~Sasaki, M.~Shibata,
T.~Tagoshi and T.~Tanaka, Prog.\ Theor.\ Phys.\ Suppl.\ {\bf 128}, 1 (1997).

\bibitem{Detweiler:2002mi}
S.~Detweiler and B.~F.~Whiting,
Phys.\ Rev.\ D {\bf 67}, 024025 (2003)
[arXiv:gr-qc/0202086].


\bibitem{MNS1}
Y.~Mino, H.~Nakano and M.~Sasaki,
Prog.\ Theor.\ Phys.\  {\bf 108}, 1039 (2002)
[arXiv:gr-qc/0111074].

\bibitem{Barack:2001gx}
L.~Barack, Y.~Mino, H.~Nakano, A.~Ori and M.~Sasaki,
Phys.\ Rev.\ Lett.\  {\bf 88}, 091101 (2002)
[arXiv:gr-qc/0111001].

\bibitem{Barack:2002mh}
L.~Barack and A.~Ori,
Phys.\ Rev.\ D {\bf 66}, 084022 (2002) 
[arXiv:gr-qc/0204093].

\bibitem{Barack:2002bt}
L.~Barack and A.~Ori,
Phys.\ Rev.\ D {\bf 67}, 024029 (2003) 
[arXiv:gr-qc/0209072].

\bibitem{Barack:2001ph}
L.~Barack and A.~Ori,
Phys.\ Rev.\ D {\bf 64}, 124003 (2001)
[arXiv:gr-qc/0107056].

\bibitem{Barack:2002ku}
L.~Barack and C.~O.~Lousto,
Phys.\ Rev.\ D {\bf 66}, 061502 (2002)
[arXiv:gr-qc/0205043].

\bibitem{Chrzanowski:wv}
P.~L.~Chrzanowski,
Phys.\ Rev.\ D {\bf 11}, 2042 (1975).

\bibitem{Ori:2002uv}
A.~Ori,
Phys.\ Rev.\ D {\bf 67}, 124010 (2003) 
[arXiv:gr-qc/0207045].

\bibitem{Barack:2002mh2}
L.~Barack and A.~Ori,
Phys.\ Rev.\ Lett. {\bf 90}, 111101 (2003)
[arXiv:gr-qc/0212103].

\bibitem{reaction}
Y.~Mino, M.~Sasaki and T.~Tanaka,
Phys.\ Rev.\ D {\bf 55}, 3457 (1997)
[arXiv:gr-qc/9606018].

\bibitem{QuiWal}
T.~C.~Quinn and R.~M.~Wald,
Phys.\ Rev.\ D {\bf 56}, 3381 (1997);
Phys.\ Rev.\ D {\bf 60}, 064009 (1999).

\bibitem{RegWhe} T. Regge and J. A. Wheeler, Phys. Rev. {\bf 108}, 1063 (1957).

\bibitem{Zer} F. Zerilli, Phys.\ Rev.\ D {\bf 2}, 2141 (1970).

\bibitem{Teukolsky:1973ha}
S.~A.~Teukolsky,
Astrophys.\ J.\  {\bf 185}, 635 (1973).

\bibitem{Sago:2002fe}
N.~Sago, H.~Nakano and M.~Sasaki,
Phys.\ Rev.\ D {\bf 67}, 104017 (2003)
[arXiv:gr-qc/0208060].

\bibitem{Jhingan:2002kb}
S.~Jhingan and T.~Tanaka,
Phys.\ Rev.\ D {\bf 67}, 104018 (2003)
[arXiv:gr-qc/0211060].

\bibitem{Minopc} The idea of calculating the
self-force in the RW gauge was originally due to Y. Mino
(private communication).

\bibitem{Mano2}
S.~Mano, H.~Suzuki and E.~Takasugi,
Prog.\ Theor.\ Phys.\ {\bf 96}, 549 (1996)
[arXiv:gr-qc/9605057].

\bibitem{BHPC} Black hole perturbation club (M.~Sasaki, T.~Tanaka, S.~Jhingan, 
H.~Nakano, N.~Sago and W.~Hikida), in preparation.

\bibitem{Fujita} R.~Fujita et al., in preparation.

\bibitem{Minon} Y.~Mino, Phys.\ Rev.\ D {\bf 67}, 084027 (2003)
[arXiv:gr-qc/0302075].

\bibitem{psireg} T.~Tanaka et al., in preparation.

\bibitem{SaNa} N.~Sago and H.~Nakano, in preparation.

\bibitem{chandra}
S.~Chandrasekhar, Proc. R. Soc. (London) {\bf A343},
289 (1975);

{\it Mathematical Theory of Black Holes\/}
(Oxford University Press, 1983), \\
M.~Sasaki and T.~Nakamura,
Phys.\ Lett.\ A {\bf 89}, 68 (1982).

\end{thebibliography}
\end{document}